\documentclass[12pt]{iopart}

\usepackage{iopams} 
\usepackage{graphicx}
\usepackage{pdflscape}

\begin{document}

\title[]{Non-classical behaviour of scalar wave fields in any state of spatial coherence and point sources in the phase-space representation}

\author{Rom\'an Casta\~neda$^{1*}$, Gustavo Ca\~nas-Cardona$^1$ and Herbert Vinck-Posada$^2$}

\address{$^1$Physics School, Universidad Nacional de Colombia Sede Medell\'in, A.A. 3840, Medell\'in - Colombia\\
$^2$Physics Department, Universidad Nacional de Colombia sede Bogot\'a, Bogot\'a - Colombia}
\ead{rcastane@unal.edu.co}

\begin{abstract}
 
In the framework of the spatial coherence wavelets, different features of the first-order spatial coherence (Young's interference) are analyzed by calculating the 
corresponding marginal power spectrum, a close related quantity to the classical Wigner Distribution Function (WDF) of the optical field. The consideration of the radiant 
and virtual point sources evidences some effects, conventionally attributed to non-classical correlations of light, although such type of the correlations are not explicitly 
included in the model. Specifically, a light state is produced that has similar morphology to the WDF of the well-known quantum Sch\"odinger cat state.

\end{abstract}

\maketitle

\section{{\bf INTRODUCTION}}

The advances on cavity quantum electrodynamics (CQED) and on semi-conductor quantum electrodynamics (SCQED), reached in the last three decades, allowed developing devices 
that approach the predictions of the quantum theory of the matter-radiation interaction to the experimental results. in this context, the manipulation of the quantum states
of light is a subject of growing interest, with topics as the production of only-one photon sources \cite{Article1,Article2,Article3}, the analysis of the quantum
interference mechanisms like the Schr\"odinger cat states of light \cite{Article4} and those derived from the matter-radiation interaction \cite{Article5} for instance. They
are promising topics for technological applications in quantum computation and information processing \cite{Article6}, quantum teleportation \cite{Article7} and quantum
opto-electronic system \cite{Article8}-\cite{Article10}.\\

\noindent
In spite of the accepted quantum nature of topics as the Schr\"odinger cat states, a classical approach to them seems to be possible by adding novel considerations to the 
phase-space representation of the optical wave-field, like the spatial coherence wavelets emitted by both radiant and virtual point sources \cite{Article11}. It could 
revaluate the actual limits of the classical theories and the real grounds of the physical behavior of light. For instance, the analysis presented in this paper suggest
that the Schr\"odinger cat states are actually originated by the spatial coherence state of the light.\\

\noindent
The more recent description of the properties and behavior of optical fields in arbitrary state of spatial coherence has been proposed in the framework of their phase-space
representation \cite{Book2}. In this context, the spatial coherence wavelets and the marginal power spectrum \cite{Article12} have been introduced as field descriptors,
in such a way that the propagation of power and correlation features of the field is depicted in terms of wavelets emitted by source pairs gathered in classes determined
by the pair separations \cite{Article13}.\\

\noindent
This description is compatible to the second order spatial coherence theory of the optical field \cite{Book3}, and provides new insight about the interference and the 
diffraction phenomena, as discussed in the present work. For instance, the marginal power spectrum provides a ray-map with two kinds of rays, i.e.{\it the carrier rays}
along which the radiant energy of the field is transported, and the {\it modulating $0-\pi$ rays} which propagate positive and negative energies, responsible for the
interference phenomenon \cite{Article14}. Additionally to the conventional geometrical definition of the rays \cite{Book4}, the carrier rays are radiometric quantities, 
whose energy does not depend on the spatial coherence state of the field but it is recordable by squared modulus detectors, while the positive and negative energies 
associated to the modulating $0-\pi$ rays are determined by the spatial coherence state of the field but they are not detectable by such type of detectors, i.e. their 
existence is inferred from the redistribution of the power across the interference and the diffraction patterns \cite{Article14}.\\

\noindent
Carrier and modulating $0-\pi$ rays are emitted by two sets of point sources of different nature\cite{Article11}, i.e. {\it radiant point sources} for the first kind of rays,
and {\it virtual point sources} responsible for the non-recordable modulating $0-\pi$ rays. These sets of point sources are optically disjoint, so that they can be regarded
as allocated over two distinct layers of the space, which eventually are bringing together onto the same plane in specific arrangements. In such cases, dual point sources
are induced by the coincidence of radiant and virtual sources at the same point on the plane. The union of the sets of radiant and virtual point sources provides a unified
structure of point sources for the field. Consequently, carrier and modulating $0-\pi$ rays emitted by the dual sources share the same paths in space.\\

\noindent
This model is successfully applied in the description of diffraction of light with arbitrary states of spatial coherence and the Young's experiment with diffraction effects
\cite{Article11}. Two potential uses make it attractive, i.e. the developing of the robust simulation algorithms and of new strategies for optical information processing 
based on the manipulation of the virtual sources of the second layer of the space, instead of the radiant sources of the first layer as conventional. Both uses will be 
explored in next  papers. Furthermore, this model reveals some unexpected features of the field, morphologically similar to the quantum squeezed states and the Schr\"odinger 
cat states, without appealing to explicit quantum constrains.

\section{{\bf THE TWO-LAYER STRUCTURE OF THE SPACE}}

Let us suppose a stationary scalar wave field in any state of spatial coherence, on propagation between the aperture plane (AP) and the observation plane (OP), at a distance 
$z$ to each other in the Fresnel-Fraunhofer domain. Let us assume that the centre and difference coordinates $(\boldsymbol{\xi}_{A},\boldsymbol{\xi}_{D})$ and 
$\mathbf{(r_{A},r_{D})}$ univocally specify the positions of any pair of points at the AP and the OP respectively, i.e. $(\boldsymbol{\xi}_{A}\pm\boldsymbol{\xi}_{D})/2$ 
and $\mathbf{(\boldsymbol{\xi}_{A}\pm\boldsymbol{\xi}_{D})/2}$ \cite{Article12}. The phase space representation of this field is provided by the spatial coherence wavelets 
\cite{Article12}, denoted as

\begin{eqnarray}\label{eq1}
\textsf{\textbf{W}}(\mathbf{r}_{A}+\mathbf{r}_{D}/2,\mathbf{r}_{A}-\mathbf{r}_{D}/2;\boldsymbol{\xi}_{A})=\textsf{\textbf{S}}(\boldsymbol{\xi}_{A},
\mathbf{r}_{A})exp\left(-i\frac{k}{z}\boldsymbol{\xi}_{A} \cdot \mathbf{r}_{D}\right),
\end{eqnarray}\\
\noindent
where $k=2\pi/\lambda$, with $\lambda$ the wavelength of the field, and 

\begin{eqnarray}\label{eq2}
\fl \textsf{\textbf{S}}(\boldsymbol{\xi}_{A},\mathbf{r}_{A})=\int_{AP}\textit{W}(\boldsymbol{\xi}_{A}+\boldsymbol{\xi}_{D}/2,\boldsymbol{\xi}_{A}-\boldsymbol{\xi}_{D}/2)
exp\left(i\frac{k}{z}\boldsymbol{\xi}_{A}\cdot\boldsymbol{\xi}_{D}\right)exp\left(-i\frac{k}{z}\boldsymbol{\xi}_{D}\cdot\mathbf{r}_{A}\right)d^{2}\xi_{D}
\end{eqnarray}\\
\noindent
is a Wigner distribution function with energy units, called the marginal power spectrum \cite{Article12}, with 

\begin{eqnarray}\label{eq3}
\fl \textit{W}(\boldsymbol{\xi}_{A}+\boldsymbol{\xi}_{D}/2,\boldsymbol{\xi}_{A}-\boldsymbol{\xi}_{D}/2)&=&\mu(\boldsymbol{\xi}_{A}+\boldsymbol{\xi}_{D}/2,\boldsymbol{\xi}_{A}
-\boldsymbol{\xi}_{D}/2)\sqrt{S(\boldsymbol{\xi}_{A}+\boldsymbol{\xi}_{D}/2)}\\\nonumber
&\times&t(\boldsymbol{\xi}_{A}+\boldsymbol{\xi}_{D}/2)\sqrt{S(\boldsymbol{\xi}_{A}-\boldsymbol{\xi}_{D}/2)}t^{*}(\boldsymbol{\xi}_{A}-\boldsymbol{\xi}_{D}/2)
\end{eqnarray}\\
\noindent
the cross-spectral density \cite{Book3} of the field at the AP, $\mu(\boldsymbol{\xi}_{A}+\boldsymbol{\xi}_{D}/2,\boldsymbol{\xi}_{A}-\boldsymbol{\xi}_{D}/2)=
\arrowvert\mu(\boldsymbol{\xi}_{A}+\boldsymbol{\xi}_{D}/2,\boldsymbol{\xi}_{A}-\boldsymbol{\xi}_{D}/2)\arrowvert exp\left[i \alpha(\boldsymbol{\xi}_{A}+
\boldsymbol{\xi}_{D}/2,\boldsymbol{\xi}_{A}-\boldsymbol{\xi}_{D}/2)\right]$ the complex degree of spatial coherence \cite{Book3}, $\textit{S}(\boldsymbol{\xi}_{A}
\pm\boldsymbol{\xi}_{D}/2)$ the power distribution  of the field at the AP and $t(\boldsymbol{\xi}_{A}\pm\boldsymbol{\xi}_{D}/2)=\arrowvert t(\boldsymbol{\xi}_{A}
\pm\boldsymbol{\xi}_{D}/2)\arrowvert exp\left[i\phi(\boldsymbol{\xi}_{A}\pm\boldsymbol{\xi}_{D}/2)\right]$  the complex transmission of this plane. The marginal power 
spectrum provides the phase-space diagram or ray-map of the scalar wave field in any state of spatial coherence \cite{Article12}.\\

\noindent
The superposition of the spatial coherence wavelets yields the cross-spectral density of the field at the OP \cite{Article12}, i.e.

\begin{eqnarray}\label{eq4a}
\textit{W}(\mathbf{r}_{A}+\mathbf{r}_{D}/2,\mathbf{r}_{A}-\mathbf{r}_{D}/2)&=&\left(\frac{1}{\lambda z}\right)^{2}exp\left(i\frac{k}{z}\mathbf{r}_A
\cdot\mathbf{r}_{D}\right)\\ \nonumber &\times& 
\int_{AP}\textsf{\textbf{W}}(\mathbf{r}_{A}+\mathbf{r}_{D}/2,\mathbf{r}_{A}-\mathbf{r}_{D}/2;\boldsymbol{\xi}_{A})d^{2}\xi_{A}                                                                                                           
\end{eqnarray}\\
\noindent
whose evaluation for $\mathbf{r}_{D}=0$ is the power spectrum recorded at this plane, i.e.

\begin{eqnarray}\label{eq4b}
\fl S(\mathbf{r}_{A})=\textit{W}(\mathbf{r}_{A},\mathbf{r}_{A})=\left(\frac{1}{\lambda z}\right)^{2}\int_{AP}\textsf{\textbf{W}}(\mathbf{r}_{A},\mathbf{r}_{A};
\boldsymbol{\xi}_{A})d^{2}\xi_{A}=\left(\frac{1}{\lambda z}\right)^{2}\int_{AP}\textsf{\textbf{S}}(\boldsymbol{\xi}_{A},\mathbf{r}_{A})d^{2}\xi_{A}
\end{eqnarray}\\
\noindent
Because of the definition of the marginal power spectrum in equation (\ref{eq2}), the superposition of spatial coherence wavelets in equation (\ref{eq4a}) can explain the
interference and diffraction phenomena of light, which are close related to the lowest order of spatial coherence and are observed as modulations on the power spectrum 
in equation (\ref{eq4b}) recorded at the OP \cite{Article14}. However, it is not enough to predict phenomena due to spatial coherence properties of higher order.

\subsubsection{{\bf Radiant and virtual point  sources.}}

It is useful to introduce the dimensionless function $1\equiv C \delta(\boldsymbol{\xi}_{D})+[1-C \delta(\boldsymbol{\xi}_{D})]$ \cite{Article12} with 
$\delta(\boldsymbol{\xi}_{D})$ the Dirac's delta and $C$ a constant that assures the dimensionless character of the function and the accomplishment of the conservation
law of the total energy, into the integral in equation (\ref{eq2}) in order to express the marginal power spectrum as 
$\textsf{\textbf{S}}(\boldsymbol{\xi}_{A},\mathbf{r}_{A})=\textsf{\textbf{S}}_{ind}(\boldsymbol{\xi}_{A},\mathbf{r}_{A})+\textsf{\textbf{S}}_{pairs}(\boldsymbol{\xi}_{A},
\mathbf{r}_{A})$, with 

\begin{eqnarray}\label{eq5a}
\textsf{\textbf{S}}_{ind}(\boldsymbol{\xi}_{A},\mathbf{r}_{A})=C S(\boldsymbol{\xi}_{A})\arrowvert t(\boldsymbol{\xi}_{A})\arrowvert^{2}\geq 0
\end{eqnarray}\\
\noindent
and

\begin{eqnarray}\label{eq5b}
\fl \textsf{\textbf{S}}_{pairs}(\boldsymbol{\xi}_{A},\mathbf{r}_{A})&=& 2\int\limits_{AP, \boldsymbol{\xi}_{D}\neq0}\arrowvert \mu(\boldsymbol{\xi}_{A}+
\boldsymbol{\xi}_{D}/2,\boldsymbol{\xi}_{A}-\boldsymbol{\xi}_{D}/2) \arrowvert \sqrt{S(\boldsymbol{\xi}_{A}+\boldsymbol{\xi}_{D}/2)}\\ &\times& \arrowvert 
t(\boldsymbol{\xi}_{A}+\boldsymbol{\xi}_{D}/2)\arrowvert \sqrt{S(\boldsymbol{\xi}_{A}-\boldsymbol{\xi}_{D}/2)}\arrowvert t(\boldsymbol{\xi}_{A}-\boldsymbol{\xi}_{D}/2)
\arrowvert \nonumber \\ &\times& cos(\frac{k}{z}\boldsymbol{\xi}_{D}\cdot\mathbf{r}_{A}-\frac{k}{z}\boldsymbol{\xi}_{A}\cdot\boldsymbol{\xi}_{D}-\alpha
(\boldsymbol{\xi}_{A}+\boldsymbol{\xi}_{D},\boldsymbol{\xi}_{A}-\boldsymbol{\xi}_{D}/2)-\Delta\phi)d^{2}\xi_{D} \nonumber
\end{eqnarray}\\
\noindent
where equation (\ref{eq3}) was taken into account with $W(\boldsymbol{\xi}_{A},\boldsymbol{\xi}_{A})=S(\boldsymbol{\xi}_{A})\arrowvert t(\boldsymbol{\xi}_{A})\arrowvert^{2}$ 
the power distribution that emerges from the AP, $\mu(\boldsymbol{\xi}_{A},\boldsymbol{\xi}_{A})=1$ and $\Delta\phi=\phi(\boldsymbol{\xi}_{A}+
\boldsymbol{\xi}_{D}/2)-\phi(\boldsymbol{\xi}_{A}-\boldsymbol{\xi}_{D}/2)$. The cosine function in equation (\ref{eq5b}) results from considering the two degrees of freedom 
in direction of each separation vector $\boldsymbol{\xi}_{D}$. Accordingly, $\textsf{\textbf{S}}(\boldsymbol{\xi}_{A},\mathbf{r}_{A})$ determine the flux of radiant energy 
provided by each individual centre of secondary disturbance at the AP onto any point at the OP, while $\textsf{\textbf{S}}_{pairs}(\boldsymbol{\xi}_{A},\mathbf{r}_{A})$ 
determines positive and negative energies due to the correlation between the pairs of centres of secondary disturbance, placed within  the spatial coherence support centred 
at $\boldsymbol{\xi}_{A}$ on the AP, onto any point at the OP. The term {\it structured spatial coherence support} denotes the region around any point on the AP determined
by the spatial coherence state of the wave, where pairs of emitters are included depending on specific correlation properties established in detail in \cite{Article11}. The 
positive and negative energies of $\textsf{\textbf{S}}_{pairs}(\boldsymbol{\xi}_{A},\mathbf{r}_{A})$ cannot be interpreted as a radiant flux \cite{Article14}. Nevertheless, 
they play a crucial role in describing interference and diffraction from the point of view of the energy transport between the AP and the OP \cite{Article14}. Indeed, 
equations (\ref{eq4b}),(\ref{eq5a}) and (\ref{eq5b}) yield the power spectrum at any point on the OP, i.e.

\begin{eqnarray}\label{eq6}
 S(\mathbf{r}_{A})&=&S_{ind}(\mathbf{r}_{A})+S_{pairs}(\mathbf{r}_{A})\nonumber \\ &=& \left(\frac{1}{\lambda z}\right)^{2}\int_{AP}[\textsf{\textbf{S}}_{ind}
(\boldsymbol{\xi}_{A},\mathbf{r}_{A})+\textsf{\textbf{S}}_{pairs}(\boldsymbol{\xi}_{A},\mathbf{r}_{A})]d^{2}\xi_{A}
\end{eqnarray}\\
\noindent
with $S_{ind}(\mathbf{r}_{A})\geq0$ the radiant power provided  by the individual  centres of secondary disturbance, and $S_{pairs}(\mathbf{r}_{A})$ the modulating power, 
which can take on positive and negative values, and therefore has a different physical meaning as the radiant power. The condition $S(\mathbf{r}_{A})\geq0$ imposes 
$S_{ind}(\mathbf{r}_{A})\geq\arrowvert S_{pairs}(\mathbf{r}_{A})\arrowvert$, and therefore $S(\mathbf{r}_{A})\leq 2 S_{ind}(\mathbf{r}_{A})$ stands, i.e. 
$S_{pairs}(\mathbf{r}_{A})$ modulates $S_{ind}(\mathbf{r}_{A})$ determining the power spectrum at any point of the OP, a behavior presented in interference and diffraction 
light. Furthermore, the conservation law of the total energy of the field can be expressed of as \cite{Article12}

\begin{eqnarray}\label{eq7}
\fl \int_{OP}S(\mathbf{r}_{A})d^{2}r_{A}=\int_{AP}S(\boldsymbol{\xi}_{A})\arrowvert t(\boldsymbol{\xi}_{A})\arrowvert^{2}d^{2}\xi_{A}=\left(\frac{1}{\lambda z}\right)^{2}
\int_{OP}\int_{AP}\textsf{\textbf{S}}(\boldsymbol{\xi}_{A},\mathbf{r}_{A})d^{2}\xi_{A}d^{2}r_{A},
\end{eqnarray}\\
\noindent
where

\begin{eqnarray}\label{eq8a}
\fl \int_{OP}S(\mathbf{r}_{A})d^{2}r_{A}=\int_{OP}[S_{ind}(\mathbf{r}_{A})+S_{pairs}(\mathbf{r}_{A})]d^{2}r_{A} \nonumber\\\hspace{0.3cm}=
\left(\frac{1}{\lambda z}\right)\int_{OP}\int_{AP}[\textsf{\textbf{S}}_{ind}(\boldsymbol{\xi}_{A},\mathbf{r}_{A})+\textsf{\textbf{S}}_{pairs}
(\boldsymbol{\xi}_{A},\mathbf{r}_{A}) ]d^{2}\xi_{A}d^{2}r_{A},
\end{eqnarray}\\
\noindent
with

\begin{eqnarray}\label{eq8b}
\fl \left( \frac{1}{\lambda z}\right)^{2}\int_{OP}\int_{AP}\textsf{\textbf{S}}_{ind}(\boldsymbol{\xi}_{A},\mathbf{r}_{A})d^{2}\xi_{A}d^{2}r_{A}=\int_{OP}S_{ind}
(\mathbf{r}_{A})d^{2}r_{A}=\int_{AP}S(\boldsymbol{\xi}_{A})\arrowvert t(\boldsymbol{\xi}_{A})\arrowvert^{2}d^{2}\xi_{A}
\end{eqnarray}\\
\noindent
and

\begin{eqnarray}\label{eq8c}
\left(\frac{1}{\lambda z}\right)^{2}\int_{OP}\int_{AP}\textsf{\textbf{S}}_{pairs}(\boldsymbol{\xi}_{A},\mathbf{r}_{A})d^{2}\xi_{A}d^{2}r_{A}=
\int_{OP}S_{pairs}(\mathbf{r}_{A})d^{2}r_{A}=0
\end{eqnarray}\\
\noindent
Equations (\ref{eq8b}) and (\ref{eq8c}) imply that $C=(\lambda z)^{2}/\int_{OP}d^{2}r_{A}$ and

\begin{eqnarray}\label{eq9}
\fl \int_{OP} cos\left( \frac{k}{z}\boldsymbol{\xi}_{D}\cdot\mathbf{r}_{A}-\frac{k}{z}\boldsymbol{\xi}_{A}\cdot\mathbf{r}_{D}-\alpha(\boldsymbol{\xi}_{A}+
\boldsymbol{\xi}_{D}/2,\boldsymbol{\xi}_{A}-\boldsymbol{\xi}_{D}/2)-\Delta\phi\right)d^{2}r_{A}=0.
\end{eqnarray}\\
\noindent
Accordingly, the modulating power $S_{pairs}(\mathbf{r}_{A})$ is not recordable by squared modulus detectors placed at the OP, but its modulation effects are revealed as a 
redistribution of the radiant power emitted at the AP onto the OP. This is the meaning of interference and diffraction from the point of view of the energy transport 
between the AP and the OP \cite{Article14}.\\

\noindent
The results above suggest that an optical field in any state of spatial coherence can be thought as emitted by the following types of point sources, placed at positions  
$\boldsymbol{\xi}_{A}$ on two distinct layers of the ordinary space, as conceptually depicted in Fig. \ref{fig1}:

\begin{itemize}
 \item {\it Radiant emitters} gathered on the first layer, which are responsible for the emission of the radiant energy of the field, i.e. the energy which is recorded by 
detectors at the OP. They are pure individual sources in the sense that their emissions, described by equation (\ref{eq5a}) , do not depend from correlations with neighbour 
emitters.
\begin{figure}[h]
 \begin{center}
  \includegraphics[width=0.6\textwidth,angle=0]{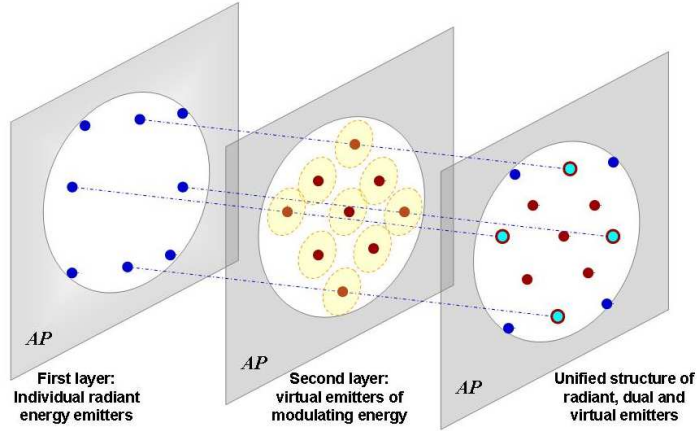}
\caption{Conceptual representation of radiant, virtual and dual point emitters (the dots at each layer). The extended circles at the second layer represent the spatial 
coherence support related to each virtual emitter, and the dots with thick line represent the dual emitters.}\label{fig1}
 \end{center}
\end{figure}
\item {\it Virtual emitters} gathered on the second layer, they emit the positive and negative modulating energies. They are virtual in the sense that their contributions 
are not directly  recordable by square modulus detectors but through the redistribution of the power spectrum at the OP that they produce. For such detectors, the second
layer is actually vacuumed. The emission of the point virtual emitters is described by equation (\ref{eq5b}) and depends on the correlation between all the pairs of centres 
of secondary disturbance placed within the spatial coherence support centred at the position of the corresponding point virtual emitter.
\end{itemize}

\begin{figure}[h]
 \begin{center}
  \includegraphics[width=0.6\textwidth]{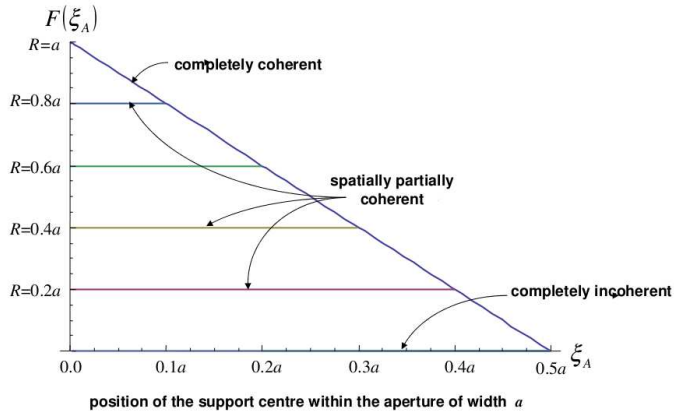}
 \end{center}
\caption{Growth in the number of pairs within the structured support of spatial coherence, as a function of the correlated pairs and the position of the support centre at 
the AP. Each horizontal line indicates the region of the AP in which all the supports have the same number of pairs. The oblique profile indicates the region near the 
aperture edge (at $\xi_{A}=10$ in arbitrary units) in which the number of pairs within any support diminish and the diminishing rate. There are no pairs within the support 
centred at the surrounding of the aperture edge no matter the spatial coherence state of the field.}\label{fig2}
\end{figure}

\noindent
In this two-layer structure of the ordinary space, two correlated radiant emitters of the first layer turn on a virtual emitter in the second layer, located in the middle 
point of the distance between them. A consequence:

\begin{itemize}
\item The distribution of radiant emitters on the first layer must be discrete, even regarding the set of maximal density of such emitters, because there should be place 
between consecutive correlated radiant emitters for allocating the turned-on virtual ones.
\item In contrast, the set of maximal density of virtual emitters of a scalar wave field approaches to continuum by the lowest-order of spatial coherence.
\item Only radiant emitters will be placed just at the edge of apertures at the AP.
\item Radiant and virtual emitters can be placed at the same positions $\boldsymbol{\xi}_{A}$ of the ordinary space (although in separate layers) if the correlation extends 
beyond the consecutive pairs.
\item The strengths provided by all the radiant sources are identical, except if the diffracted wavefront is no uniform at the AP; while those provided by the virtual 
sources depend of the size of the corresponding supports (Fig. \ref{fig2}), because each virtual source emits the addition of the contributions due to all the pairs of 
secondary disturbance within its support. So, only the virtual sources associated to supports of the same shape and size emit the same strength.
\end{itemize}

\noindent
Thus, the complete distribution of radiant and virtual point sources of the scalar wave field will be obtained when projecting the two layers of the space onto the same 
plane, as depicted in \ref{fig1}. Such distribution will contain a further type of point sources, i.e. {\it dual emitters} conformed by radiant and virtual point sources 
placed at the same position.

\noindent
From this point of view, fully spatially incoherent scalar wave fields propagate only through the first layer of the ordinary space and are emitted only by radiant point 
sources (i.e. such field  are unable to turn on virtual sources, and therefore they neither contain dual sources). Partially (or completely) spatial coherent fields use the 
both layers of the ordinary space in their propagation, because they are emitted by both the radiant and the virtual point sources (and then, by virtual emitters).\\

\noindent
Such structure of radiant and virtual point sources for the optical field, distributed onto two different layers of the ordinary space constitutes a novel description of the 
spatial coherence properties of the lowest-order of the field. It allows understanding the diffraction and interference of scalar optical fields in any state of spatial 
coherence not as a result of geometrical features (i.e. the effect of the aperture edges at the AP and the phase difference between superimposed wavefronts at the OP 
respectively) but as a result of the same radiative feature, i.e. the emission of positive and negative modulating energies by virtual point emitters distributed within the 
whole aperture at the AP.
 
\section{YOUNG'S EXPERIMENT}

\begin{center}
\begin{figure}[h]
\begin{tabular}{c c}
  \includegraphics[width=0.4\textwidth{}]{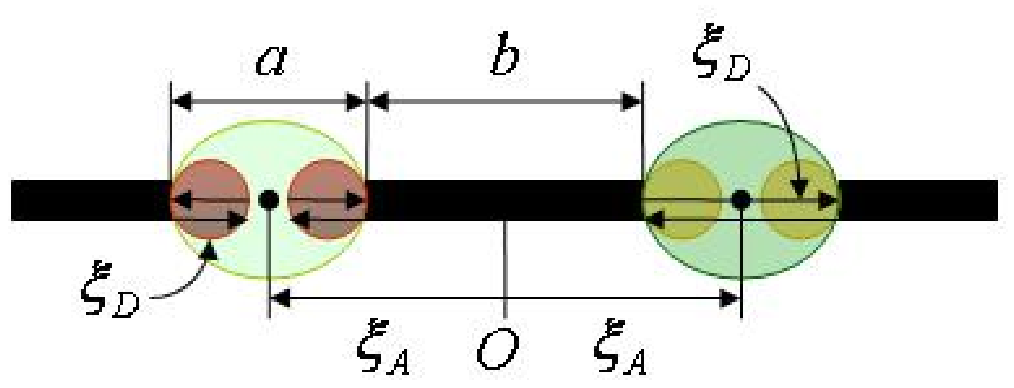}\label{fig3a} &  \includegraphics[width=0.4\textwidth{}]{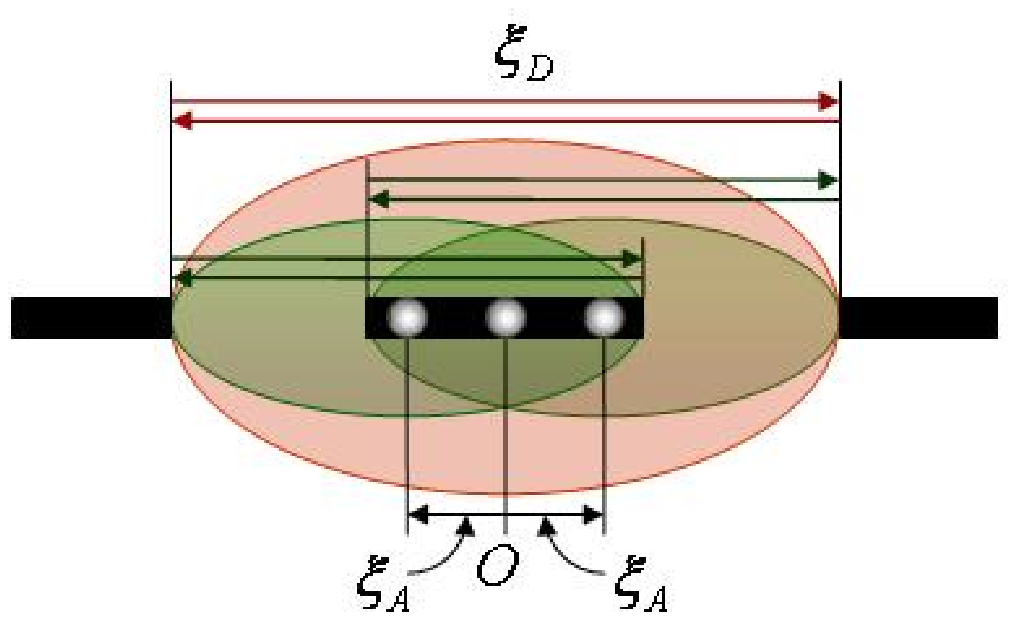}\label{fig3b}\\
  \textbf{a} & \textbf{b}
\end{tabular}
\caption{One-dimensional double slit mask for Young's experiment with diffraction effects. Spatial coherence supports for a) the diffraction of light through the slits and 
b) for the interference between the contributions of the slits.}\label{fig3}
\end{figure}
\end{center}

It is one-dimensionally performed by arranging a double slit mask at the AP, whose parameters are depicted in Fig. \ref{fig3}. For mathematical simplicity and without lack 
of generality, let us assume diffraction of an uniform scalar wave field in Fraunhofer domain by each slit, i.e. negligible phase term $\frac{k}{2}\xi_{A}\xi_{D}$ and 
constant power ${S}_{0}$ over the slits. The ray-map for this experiment can be expressed as $\textsf{\textbf{S}}(\xi_{A},x_{A})=\textsf{\textbf{S}}_{dif}
(\xi_{A}-(b+a)/2,x_{A})+\textsf{\textbf{S}}_{dif}(\xi_{A}+(b+a)/2,x_{A})+\textsf{\textbf{S}}_{int}(\xi_{A},x_{A})$, with

\begin{eqnarray}\label{eq10a}
\fl \textsf{\textbf{S}}_{dif}(\xi_{A},x_{A})&=&[CS_{0}+2S_{0}\int_{\xi_{D}\neq0}^{a-2\arrowvert\xi_{A}\arrowvert}\arrowvert \mu(\xi_{A}+\xi_{D}/2,\xi_{A}-{\xi}_{D}/2)
\arrowvert\\\nonumber &\times& cos\left(\frac{k}{z}\xi_{D}x_{A}-\alpha(\xi_{A}+\xi_{D}/2,\xi_{A}+\xi_{D}/2)\right)d^{2}\xi_{D}]rect(\xi_{A}/a)
\end{eqnarray}\\
\noindent
where $rect(\xi_{A}/a)=1$ for $\arrowvert\xi_{A}\arrowvert\leq{a/2}$ and equal to null otherwise, describing the diffraction of the scalar wave field through any of the 
slits; and 

\begin{eqnarray}\label{eq10b}
\textsf{\textbf{S}}_{int}(\xi_{A},x_{A})&=&2S_{0}rect(\xi/a)\int_{b+2\arrowvert\xi_{A}\arrowvert}^{2a+b-2\arrowvert\xi_{A}\arrowvert}\arrowvert
\mu(\xi_{A}+\xi_{D}/2,\xi_{A}-\xi_{D}/2)\arrowvert\\\nonumber &\times& cos\left(\frac{k}{z}\xi_{D}x_{A}-\alpha(\xi_{A}+\xi_{D}/2,\xi_{A}-\xi_{D}/2)\right)d\xi_{D}
\end{eqnarray}\\
\noindent
describing the interference between the contributions provided by the both slits.

\noindent
It is worth noting that there are both radiant and virtual emitters associated to $\textsf{\textbf{S}}_{dif}(\xi_{A},x_{A})$ while there are only virtual emitters associated 
to $\textsf{\textbf{S}}_{int}(\xi_{A},x_{A})$ for $b>0$. Indeed,

\begin{itemize}
 \item $\textsf{\textbf{S}}_{dif}(\xi_{A},x_{A})=\textsf{\textbf{S}}_{dif}^{(rad)}(\xi_{A})+\textsf{\textbf{S}}_{dif}^{(virt)}(\xi_{A},x_{A})$ holds, with 
  $\textsf{\textbf{S}}_{dif}^{(rad)}(\xi_{A})=CS_{0}rect(\xi_{A}/a)$ and $\textsf{\textbf{S}}_{dif}^{(virt)}(\xi_{A},x_{A})=\textsf{\textbf{S}}_{dif}(\xi_{A},x_{A})
  -CS_{0}rect(\xi_{A}/a)$.
 \item $\textsf{\textbf{S}}_{int}(\xi_{A},x_{A})=\textsf{\textbf{S}}_{int}^{(virt)}(\xi_{A},x_{A})$ because $b+2\arrowvert\xi_{A}\arrowvert>0$ holds for $b>0$
\end{itemize}

\noindent
From equation (\ref{eq4b}) follows $S(x_{A})=S_{rad}(x_{A})+S_{virt}(x_{A})=S_{dif}^{(rad)}(x_{A})+S_{dif}^{(virt)}(x_{A})+S_{int}^{(virt)}(x_{A})$, where

\begin{eqnarray}\label{eq11a}
S_{rad}(x_{A})=S_{dif}^{(rad)}=\left(\frac{1}{\lambda z} \right)2aCS_{0}\geq0
\end{eqnarray}\\
\noindent
is the uniform contribution of the radiant emitters within both diffracting slits onto the power spectrum at any point $x_{A}$ on the OP, and 
$S_{virt}(x_{A})=S_{dif}^{(virt)}(x_{A})+S_{int}^{(virt)}(x_{A})$ the contribution of the virtual emitters related to both the diffraction of the field through any slit and 
the interference between the contributions of both slits, onto the power spectrum at any point $x_{A}$ on the OP, i.e.

\begin{eqnarray}\label{eq11b}
S_{dif}^{(virt)}(x_{A})&=&\left(\frac{1}{\lambda z}\right)^{2}\int_{-(a+b/2)}^{-b/2}\textsf{\textbf{S}}_{dif}^{(virt)}(\xi_{A}+(b+a)/2,x_{A})d\xi_{A}\\\nonumber 
&+& \left(\frac{1}{\lambda z} \right)^{2}\int_{b/2}^{a+b/2}\textsf{\textbf{S}}_{dif}^{(virt)}(\xi_{A}-(b+a)/2,x_{A})d\xi_{A}
\end{eqnarray}\\
\noindent
and

\begin{eqnarray}\label{eq11c}
 S_{int}^{(virt)}(x_{A})=\left(\frac{1}{\lambda z} \right)^{2}\int_{-a/2}^{a/2}\textsf{\textbf{S}}_{int}^{(virt)}(\xi_{A},x_{A})d\xi_{A}
\end{eqnarray}\\
\noindent
More insight about the diffraction and interference involved in the Young's experiment with scalar wave fields in any state of spatial coherence is reached by considering 
separately the terms (\ref{eq11a}) to (\ref{eq11c}) of the power spectrum at the OP. Indeed, according to equation (\ref{eq7}), the total radiant energy is only determined 
by equation (\ref{eq11a}) independently of the spatial coherence state of the wave field, i.e.$\int_{OP}S(x_{A})dx_{A}=\int_{OP}S_{rad}(x_{A})dx_{A}=S_{dif}^{(rad)}
\int_{OP}dx_{A}=2aS_{0}$. Furthermore, the terms (\ref{eq11b}) and (\ref{eq11c}) can take on positive and negative values, depending on the spatial coherence state of the 
field. The positive values of $S_{dif}^{(virt)}(x_{A})$ determines the main maximum of the diffraction envelope of the power spectrum at the OP, while the negative values 
extend over the region outside of the main maximum, determining the points of null power of the diffraction envelope, its side lobes and its convergence to null in that 
region. In contrast, positive and negative values of $S_{int}^{(virt)}(x_{A})$ strongly oscillates within the diffraction envelope, without exceeding in magnitude the value 
of the diffraction envelope.\\

\noindent
Table 1 illustrates the ideas above. Figures in the first column on the left are the ray maps (marginal power spectra) produced by double slit masks, whose transmission 
function $t(\xi)$ is depicted on the top of each figure (i.e. slit width $a$ and distance $b$ between the consecutive slit edges). Diffraction and interference 
contributions, $\textsf{\textbf{S}}_{dif}(\xi_{A},x_{A})$ and $\textsf{\textbf{S}}_{int}(\xi_{A},x_{A})$ respectively, are delimited by white bars in each ray map. The 
coordinate origin for both $\xi_{A}$ and $x_{A}$ is at the middle point of each map. The profiles of the power spectra that emerge just from the masks at the AP and that 
arrive onto the OP are shown in the second column from the left and in the last column on the right respectively. Their energies are recordable by square modulus 
detectors placed at the corresponding planes. The profiles of the diffraction and interference contributions to the power spectra at the OP are also sketched in the third 
and fourth columns from the left respectively, and the diffraction contribution includes the energies provided by both the radiant and the virtual emitters placed within
the slits.\\

\noindent
Rows 1 to 5 show the changes due to the increasing of the slit width $a$ by maintaining the distance between the slit centres unchanged, in the diffraction of a completely 
spatially coherent wave field (i.e. $\arrowvert\mu(\xi_{A}+\xi_{A}/2,\xi_{A}-\xi_{D}/2)\arrowvert=1$; it is also assumed that $\alpha(\xi_{A}+\xi_{D}/2,\xi_{A}-\xi_{D}/2)=0$ 
for simplicity and without lack of generality). According to equations (\ref{eq10a}) and (\ref{eq10b}), the diffraction contribution of each slit to the ray map takes the 
form

\begin{eqnarray}\label{eq12a}
\textsf{\textbf{S}}_{dif}(\xi_{A},x_{A})=S_{0}a\frac{sin\left[\frac{k}{z}(a-2\arrowvert\xi_{A}\arrowvert) x_{A}\right]}{\frac{ka}{z}x_{A}}rect(\xi_{A}/a)
\end{eqnarray}\\
\noindent
while the interference contribution due to both slits will be given by

\begin{eqnarray}\label{eq12b}
\fl \textsf{\textbf{S}}_{int}(\xi_{A},x_{A})=2S_{0}rect(\xi_{A}/a)\left[\frac{sin[\frac{k}{z}(2a+b-2\arrowvert\xi_{A}\arrowvert)x_{A}]}{\frac{k}{z}x_{A}}
-\frac{sin[\frac{k}{z}(b+2\arrowvert\xi_{A}\arrowvert)x_{A}]}{\frac{k}{z}x_{A}}\right]
\end{eqnarray}\\
\noindent
The integration of equations (\ref{eq12a}) and (\ref{eq12b}) following equations (\ref{eq11a}) to (\ref{eq11c}) gives the power spectrum produced by the Young's experiment 
at the OP.\\

\noindent
Row 1 is corresponding to the mask with the narrowest slits that enclose few radiant and virtual emitters.Taking into account that $\arrowvert\xi_{A}\arrowvert\leqslant a/2$ 
and $a$ is very small and equations (\ref{eq12a}) and (\ref{eq12b}), the following approaches are applicable: $rect(\xi_{A}/a)\approx\delta(\xi_{A})$, 
$\textsf{\textbf{S}}_{dif}(\xi_{A},x_{A})\approx S_{0}a\left[1-\frac{1}{6}(\frac{ka}{z})^{2}x_{A}^{2}\right]\delta(\xi_{A})$ and $\textsf{\textbf{S}}_{int}
(\xi_{A},x_{A})\approx4aS_{0}\left[1-\frac{4}{6}(\frac{ka}{z})^{2}x_{A}^{2}\right]cos(\frac{kb}{z}x_{A})\delta(\xi_{A})$, with $\delta(\xi_{A})$ a Dirac's delta. 
Thus, the ray map on row 1 is described by

\begin{eqnarray}\label{eq13} 
\fl \textsf{\textbf{S}}(\xi_{A},x_{A})\approx
aS_{0}\left\{\delta\left(\xi_{A}-\frac{b+a}{2}\right)+\delta\left(\xi_{A}+\frac{b+a}{2}\right)+4cos\left(\frac{kb}{z}x_{A}\right)\delta(\xi_{A})\right\}\\ \nonumber
\fl -\frac{a}{6}\left(\frac{ka}{z}\right)^{2}S_{0}x_{A}^{2}\left\{\delta\left(\xi_{A}-\frac{b+a}{2}\right)+\delta\left(\xi_{A}+\frac{b+a}{2}\right)
+16cos\left(\frac{kb}{z}x_{A}\right)\delta(\xi_{A})\right\}
\end{eqnarray}\\
\noindent
The term on the first line of this expression is corresponding to the ray map for a Young experiment with delta-like slits, which enclose only radiant emitters, and then 
turn on only virtual emitters only for interference. The term on the second line describes the change of the ray map due to a small growth in slit width, which increases 
the number of enclosed radiant emitters that turn on some virtual emitters for diffraction. It introduces a parabolic disturbance on the term on the first line, along the 
$x_{A}$-axis of the OP. The virtual emitters within the slits increase $S_{dif}(x_{A})$ in the middle region at the AP and decrease it at the pattern edges by relative small 
values. The correlations between the radiant emitters in both slits turn on the virtual emitters of $S_{int}(x_{A})$, whose positive and negative modulating energies are 
cosine-like distributed on the OP, but with parabolic increments at the middle region and decrements at the pattern edges. Therefore, $S(x_{A})$ will be a high contrasted 
cosine-like pattern of fringes with a parabolic disturbance, which can be minimized if the observation region on the OP is small enough, i.e.

\begin{eqnarray}
 \fl S(x_{A})\approx2aS_{0}\left[1+2cos\left(\frac{kb}{z}x_{A}\right)\right]-\frac{aS_{0}}{3}\left(\frac{ka}{z}\right)^{2}x_{A}^{2}
\left[1+8cos\left(\frac{kb}{z}x_{A}\right)\right]\\\nonumber\approx2aS_{0}cos^{2}\left(\frac{kb}{2z}x_{A}\right)-\frac{aS_{0}}{3}\left(\frac{ka}{z}\right)^{2}x_{A}^{2}
\left[1+8cos\left(\frac{kb}{z}x_{A}\right)\right]
\end{eqnarray}\\
\noindent
As the slit width increases, more radiant emitters are included within, so that more virtual sources are turned on for both $S_{dif}(x_{A})$ and $S_{int}(x_{A})$. 
Accordingly, 1) the wave field that emerges just from the slits will exhibit oscillations, 2) both the increments and decrements of $S_{dif}(x_{A})$ becomes more significant, 
in such a way that $S_{dif}(x_{A})$ acquires a main maximum in a delimited region around the middle point of the OP and decreasing side lobes. The bigger the slits the 
narrower the main maximum. It is worth noting that $S_{dif}(x_{A})\geq0$ and that $S_{dif}(x_{A})=S(x_{A})$ for completely spatially incoherent wave fields. 3) In addition, 
the magnitudes of the positive and negative modulating energies of $S_{int}(x_{A})$ significantly grow within the region of the main maximum of $S_{dif}(x_{A})$ and 
appreciable diminish outside this region, in such a way that $S(x_{A})$ will be a high contrasted cosine-like pattern of fringes modulated by the profile of $S_{dif}(x_{A})$. 
These features can be appreciated in rows 2 to 5.\\

\noindent
It is worth noting that:
\begin{itemize}
 \item The distribution of the positive and negative modulating energies of $S_{int}(x_{A})$ resembles the Schr\"odinger cat state between mixed states 
of a quantum system, i.e. because of its spatial coherence state, the scalar wave field at both slits of the Young's mask mixes through its propagation to the OP in a 
similar way as the quantum cat state. This mixture is realized in the following terms: pairs of radiant emitters placed within different apertures of the mask are able
to turn on a virtual point source in between, depending on the correlation between their emissions. Instead of radiant energy, the virtual point source emit positive and 
negative modulating energies, whose distribution in space must fulfill condition (\ref{eq9}), i.e. the strengths of such energies must be equal and, if a certain amount of
modulating positive energy is emitted along a given path, then the same amount of modulating negative energy must be emitted along other path. Furthermore, phase of the 
complex degree of spatial coherence $\alpha(\xi_{A}+\xi_{D}/2,\xi_{A}-\xi_{D}/2)$ and/or the transmission phase difference at pairs of points on the different apertures of 
the mask $\bigtriangleup\phi$ in the cosine function of equation (\ref{eq9}) can be used in order to shift the distribution of the modulating energies. For instance, a shift by 
$\pi$ means that positive energies are turned in negative ones and vice versa. This behavior is compatible with the description of the mentioned Schr\"odinger cat state,
although it was obtained without appealing to quantum premises, but only by describing the spacial coherence state of the wave-field in terms of the structured spatial 
coherence supports, a classical concept to which the layer of virtual point sources is associated. These novel concepts seem to reveal that the Schr\"odinger cat state 
actually results from the spatial coherence state of the wave-field, even in the classical context as in the Young's experiment for instance. This state is formalized by 
equation (\ref{eq10b}) in this case, that clearly shows the dependence of the state strength and distribution on the magnitude and on the phase of the complex degree of
spatial coherence of the wave- field respectively.
 \item The region of the main maximum of $S_{dif}(x_{A})$ is corresponding to the region enclosed by the dotted rectangle in the ray map (the whole ray map in row 1 is 
enclosed by the rectangle); it also concentrates the significant positive and negative modulating energies of $S_{int}(x_{A})$. Therefore, the bigger the slits the narrower 
such region of the ray map. This feature resembles the quantum squeezing as discussed below.
\end{itemize}

\begin{figure}[h]
 \begin{center}
   \includegraphics[width=0.6\textwidth]{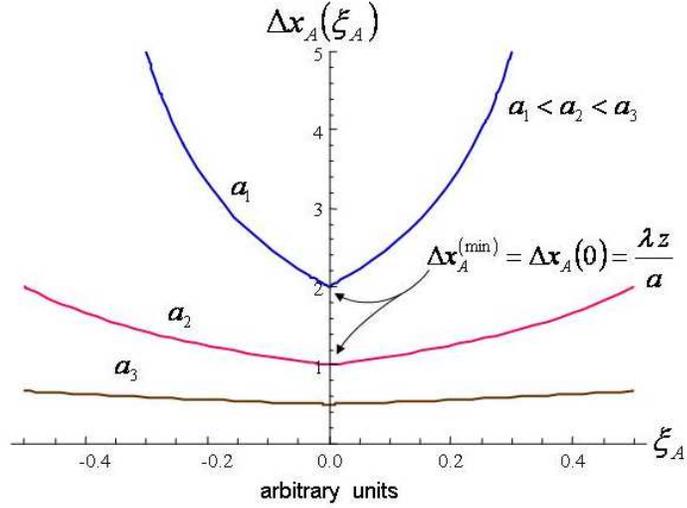}
 \end{center}
\caption{Support of the main maximum of the diffraction pattern by a slit as function of the slit width
and its growth with the coordinate $\xi_{A}$.}\label{fig4}
\end{figure}

\noindent
Profiles in columns 3 to 5 from the left of Table 1 point out that the region that concentrates the significant values of both the radiant and the modulating energies of 
the power spectrum at the OP is corresponding to the support of the main maximum of the diffraction pattern. It is delimited by the first zeroes of equation (\ref{eq12a}), 
i.e. $x_{A}^{(0)}(\xi_{A})=\pm0.5\frac{\lambda z}{a-2\arrowvert\xi_{A}\arrowvert}$, so that $\bigtriangleup x_{A}(\xi_{A})=\frac{\lambda z}{a-2\arrowvert\xi_{A}\arrowvert}$ 
(Fig. \ref{fig4}), whose minimum value is $\bigtriangleup x_{A}^{(min)}=\bigtriangleup x_{A}(0)=\frac{\lambda z}{a}$, i.e. $\bigtriangleup x_{A}(\xi_{A})\geq
\frac{\lambda z}{a}$. Thus, the smaller the slit width the bigger the support of the main maximum of diffraction; in addition, the smaller the slit width the faster the 
growth of the support size in the surrounding of $\xi_{A}=0$.\\

\noindent
Now, let us consider the ray-tracing from the AP to the OP as resulting from an ensemble of statistical realizations, each one consisting in tracing all the rays 
(both carrier and modulating) that travel for the AP to the OP at a given instant. The addition of the energies of all the rays arriving to the same point $x_{A}$ on the OP 
gives the power spectrum at that point. Two radiant sources emitting at the same time, can turn on a virtual source at this time depending on the correlation between their 
emissions. So, the probability of such activation will be close to one for fully spatially coherent wave fields and close to zero for fully spatially incoherent fields. 
Therefore, the ray emissions can randomly fluctuate from a realization to the next.\\

\noindent
In this sense, diffraction virtual sources cannot be turned on within ideal delta-like slits, but interference virtual sources do, at the middle point between the slits, 
for instance. Accordingly:

\begin{itemize}
 \item There are up to three well defined starting points at the AP for all rays of any realization produced by a mask with ideal delta-like slits, i.e. one at each slit 
for the unique radiant point source, and one in the middle point between the slits for the unique virtual point source. The radiant sources at the slits emit only carrier 
rays belonging to $\textsf{\textbf{S}}_{dif}(\xi_{A},x_{A})$, and the virtual source emits only $\it{0-\pi}$ modulating rays belonging to 
$\textsf{\textbf{S}}_{int}(\xi_{A},x_{A})$.
\item The power spectrum at the OP will be an interference pattern of cosine-like fringes without diffraction modulation, which extends over a broad region on the OP.
\end{itemize}

\noindent
As the slit width increases, more radiant sources are included within the slits, and therefore:

\begin{itemize}
 \item More starting points are enclosed in each slit, corresponding to the positions of both the radiant and the active virtual sources, associated to the diffraction. 
The same average number of carrier and $\it{0-\pi}$ modulating rays are emitted from the corresponding starting point within the slits if the wave field on the mask is 
uniform and its spatial coherence state over each slit is the same.
 \item Because of the emission of diffraction $\it{0-\pi}$ modulating rays, each slit provides a diffraction modulation on the power spectrum at the OP in each realization. 
Such modulations will be statistically identical if the slits have the same geometry. The modulation is smooth if few point sources are enclosed within the slits, i.e. if 
the slits are very narrow (figures on the rows 1 and 2 of the Table 1), and becomes stronger as the slit width increases, concentrating the energy of the power spectrum in 
the main maximum of the diffraction pattern.
 \item By illuminating great enough slits by high spatially coherent scalar wave fields, a significant amount of virtual sources will be turned on within the slits in each 
realization, whose modulating $\it{0-\pi}$ rays will produce a regular sequence of points of zero energy in the power spectrum, that delimits the diffraction maximum and 
determine the decreasing side lobes of the same size. The positions of the energy zeroes depend on the slit width (i.e. the distance between them is inversely proportional 
to the slit width, as shown by figures in rows 3 to 5).
 \item Scalar wave fields of low spatial coherence are able to turn on only few virtual sources within the slits in each realization, so that the energy of the power 
spectrum is also concentrated in the main maximum of the diffraction pattern, but they cannot completely annihilate the energy at any point of the OP, as depicted in 
figures \ref{fig13} to \ref{fig15}.
 \item The additional radiant emitters enclosed by the slits can turn on interference virtual emitters at new positions between the slits in each realization. So, starting 
points for interference $\it{0-\pi}$ rays should be regarded within a region of the same size as the slits, centred at the middle point between the slits.
 \item By high spatially coherent scalar wave fields, the strength of the positive and negative modulating energies emitted by the virtual emitters in the region centre is 
greater than that emitted by the virtual emitters at the region edge. The same feature occurs with the diffraction virtual emitters within the slits. Consequently, the 
interference modulating energies are modulated in a similar fashion as the radiant energy does by the diffraction modulating energies, which produces interference patterns 
of diffraction modulated cosine fringes, as depicted in figures on the rows 3 to 5 of Table 1. 
\end{itemize}

\noindent
According to the reasoning above, the expression $a\bigtriangleup x_{A}(0)=\lambda z$, obtained for high spatially coherent wave fields, points out that the area 
$a\bigtriangleup x_{A}(0)$ of the rectangles centred on the $\xi_{A}$-axis in both the diffraction and the interference regions of the ray map remains unchanged no matter 
the slit size. Thus, the greater the slit width the stronger the diffraction modulation and therefore the narrower the main diffraction maximum, where the radiant energy 
concentrates and the modulating energies have the higher strengths. This behavior resembles the squeezing effect in parameters of quantum systems involved in uncertainty 
principles (as position and momentum, for instance), in which the broadening of the uncertainty of any of the parameters causes the squeezing on the uncertainty of the other 
one.

\section{CONCLUSIONS}

Interference results are both qualitatively and quantitatively reproduced in the framework of the spatial coherence wavelets. The marginal power spectrum of the field,
a close related quantity to its classical WDF, produces similar features to the WDF of a cat state of light with ``quantum interference' that are not explicitly included 
in the model. In other words, it is remarkable that from only classical premises this model is able to predict quantum-like behaviors of light. In order to perform it, is
necessary to adopt the hypothesis of the existence of virtual point sources associated to the (classical) correlation properties of the field. Such unexpected result suggests
that this ``non-classical' behavior is actually related to the spatial coherence state of the light, even in the classical context.

\section*{Acknowledgement}

Authors are very grateful to Ing. Fis. Hernan Muñoz-Ossa for the development of the algorithm to calculate the ray-maps in the phase-space, and to the DIME-Universidad
Nacional de Colombia Sede Medell\'in for the corresponding financial support.

\newpage
\begin{landscape}

\textbf{TABLE 1}\\
		\\

 \begin{center}
  \begin{tabular}{|c|c|c|c|c|}\hline
  ${S}(\xi_{A},x_{A})$ & $S(\xi_{A})$ & $S_{dif}(x_{A})=S_{dif}^{(rad)}+S_{dif}^{(virt)}(x_{A})$ & $S_{int}(x_{A})=S_{int}^{(virt)}(x_{A})$
  & $S(x_{A})=S_{dif}(x_{A})+S_{int}(x_{A})$ \\ 

  \textbf{(arbitrary units)} & \textbf{(arbitrary units)} & \textbf{(arbitrary units)} & \textbf{(arbitrary units)} & \textbf{(arbitrary units)} \\ \hline

  \includegraphics[width=0.23\textwidth]{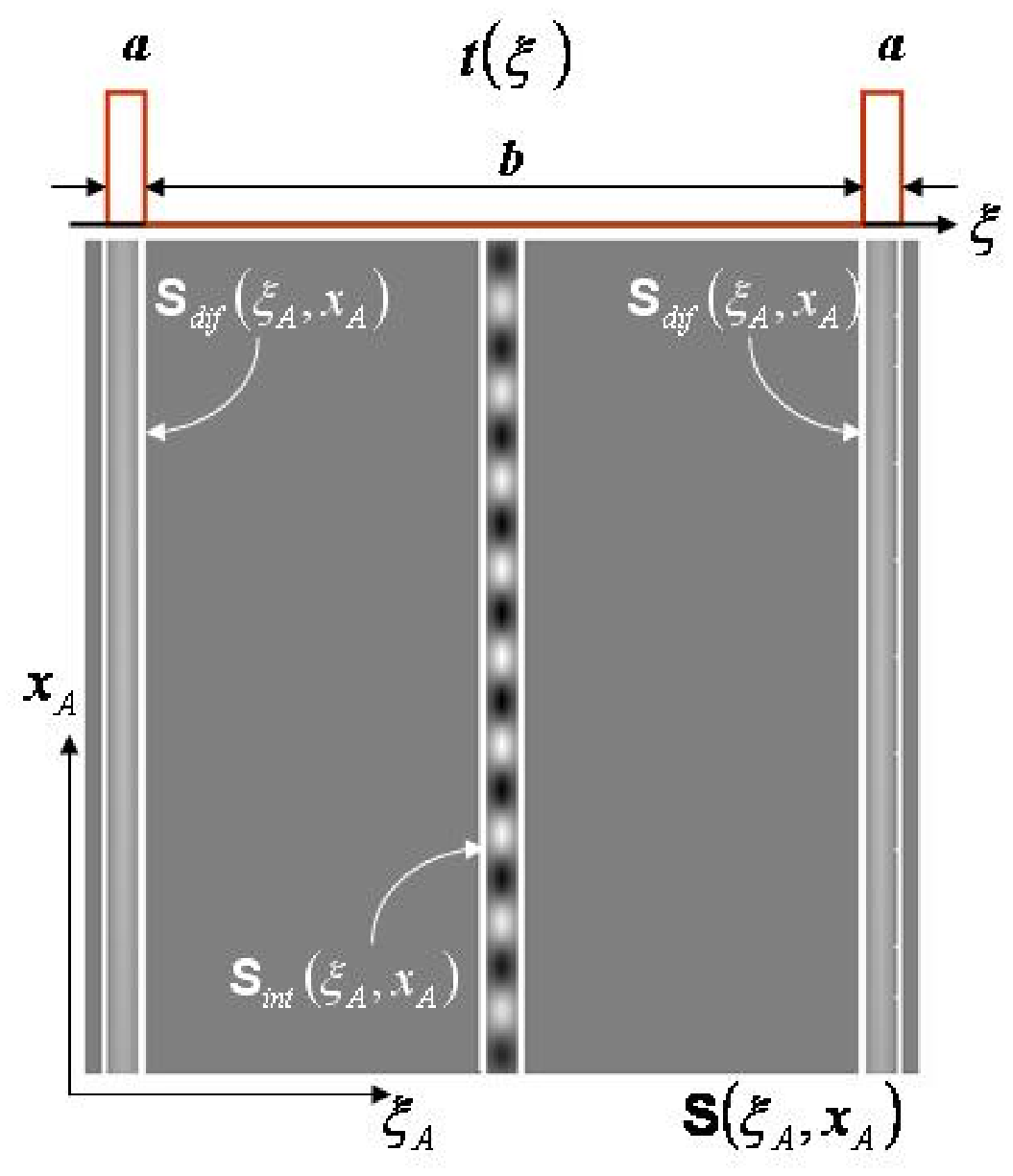}\label{fig5} & \includegraphics[width=0.25\textwidth]{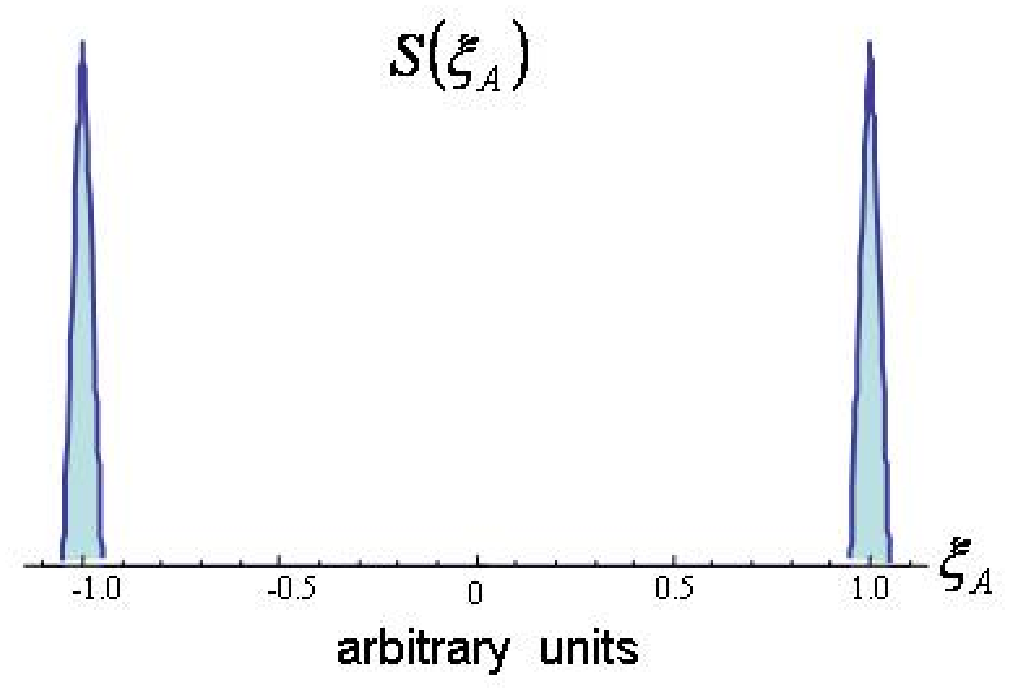} & \includegraphics[width=0.25\textwidth]{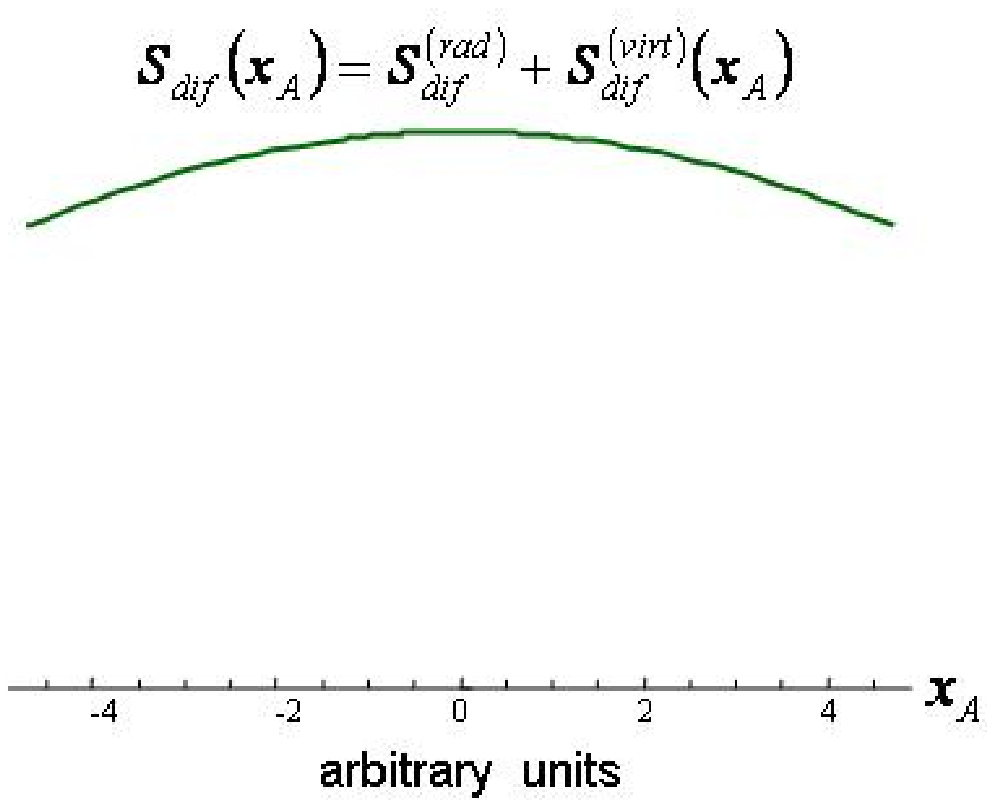}
  & \includegraphics[width=0.25\textwidth]{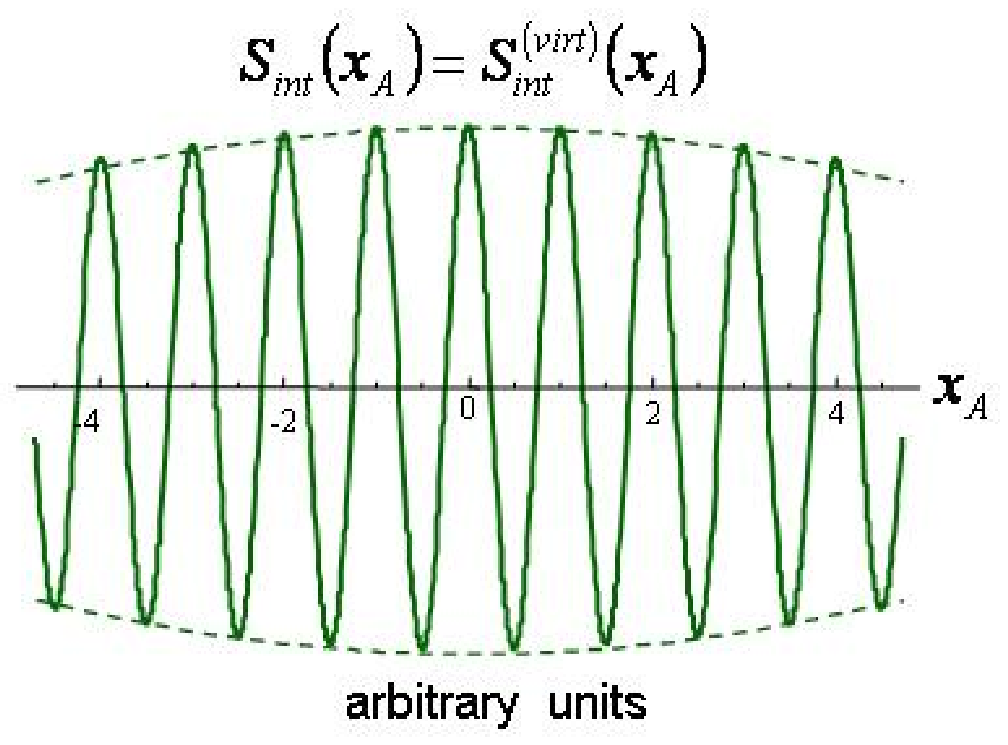} & \includegraphics[width=0.25\textwidth]{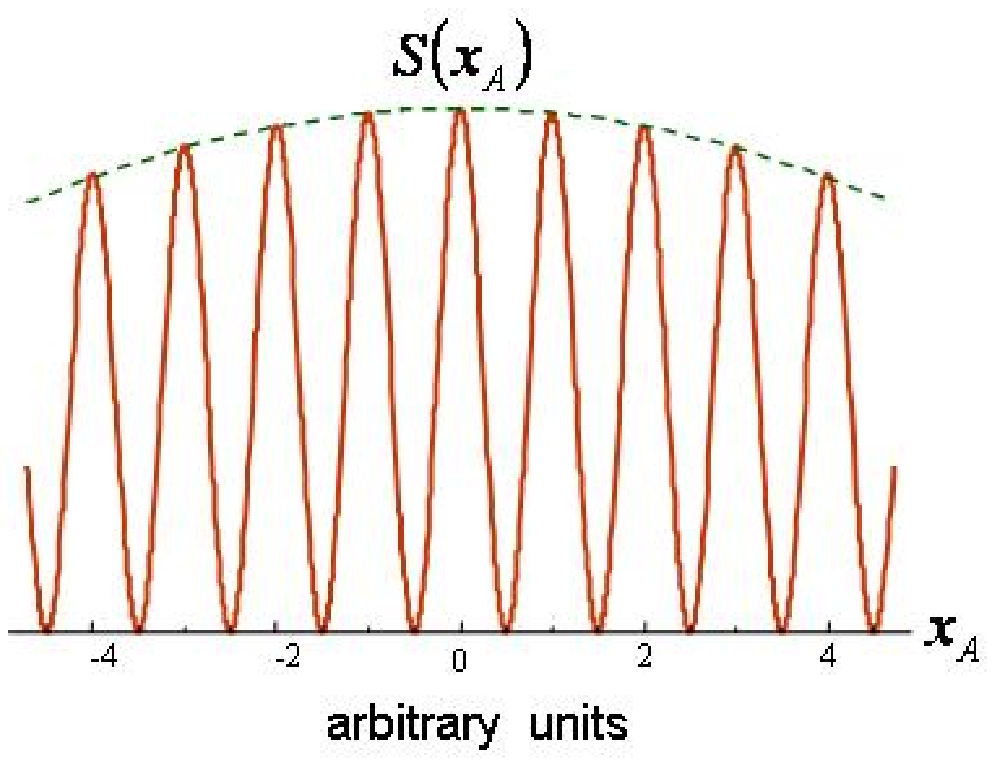} \\ 
  
  \textbf{(1)}& & & & \\ \hline 

  \includegraphics[width=0.23\textwidth]{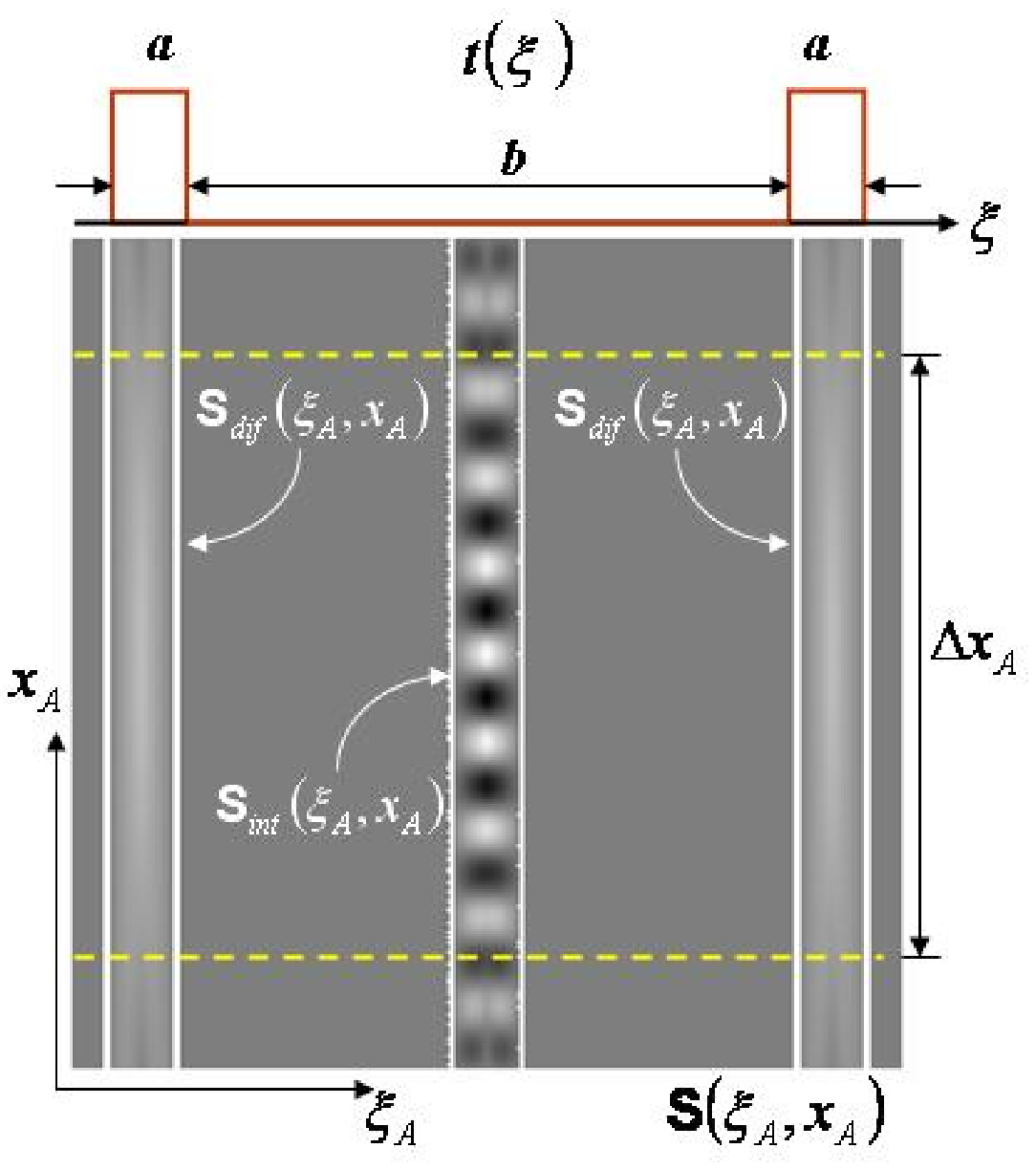}\label{fig6} & \includegraphics[width=0.25\textwidth]{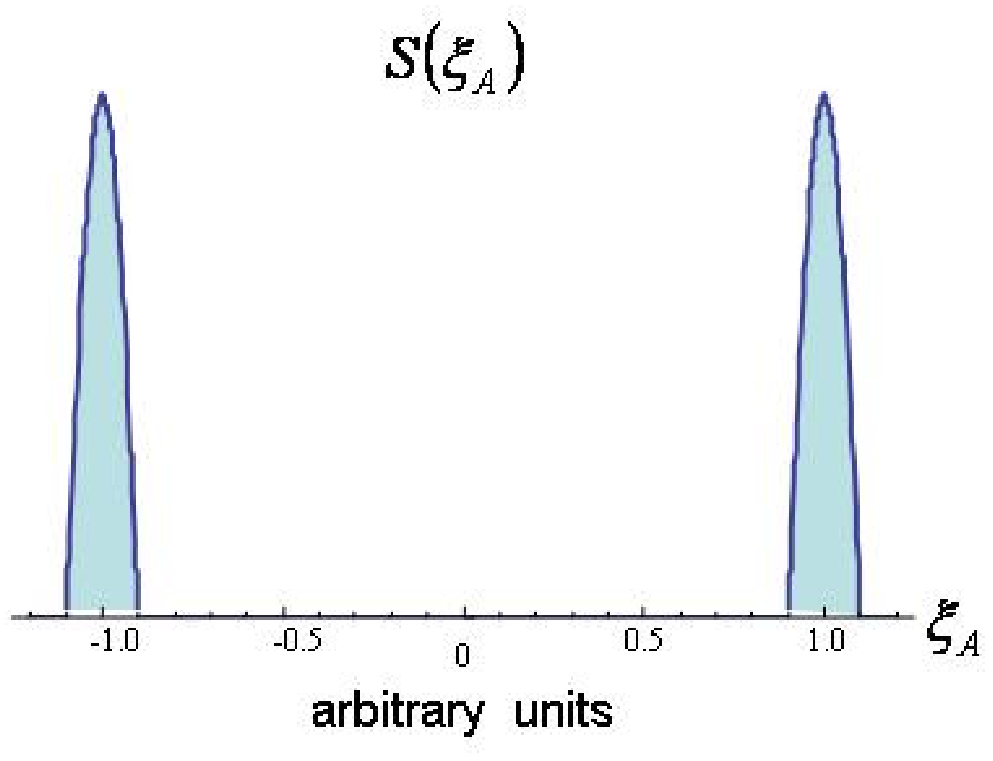} & \includegraphics[width=0.25\textwidth]{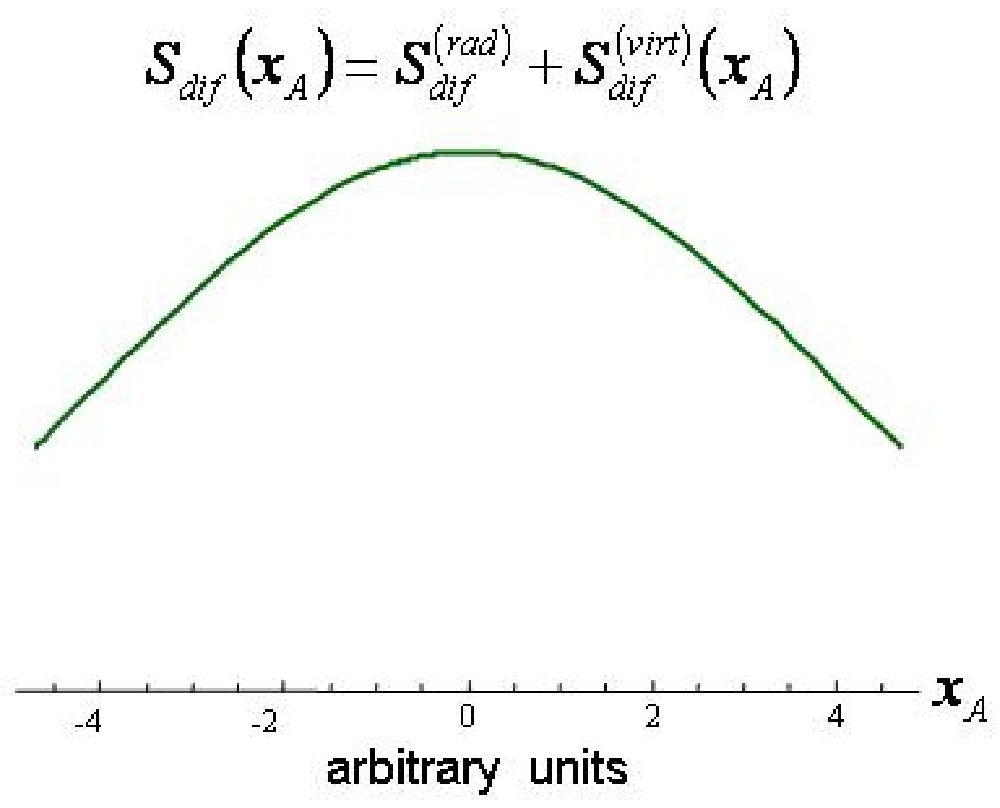} 
  & \includegraphics[width=0.25\textwidth]{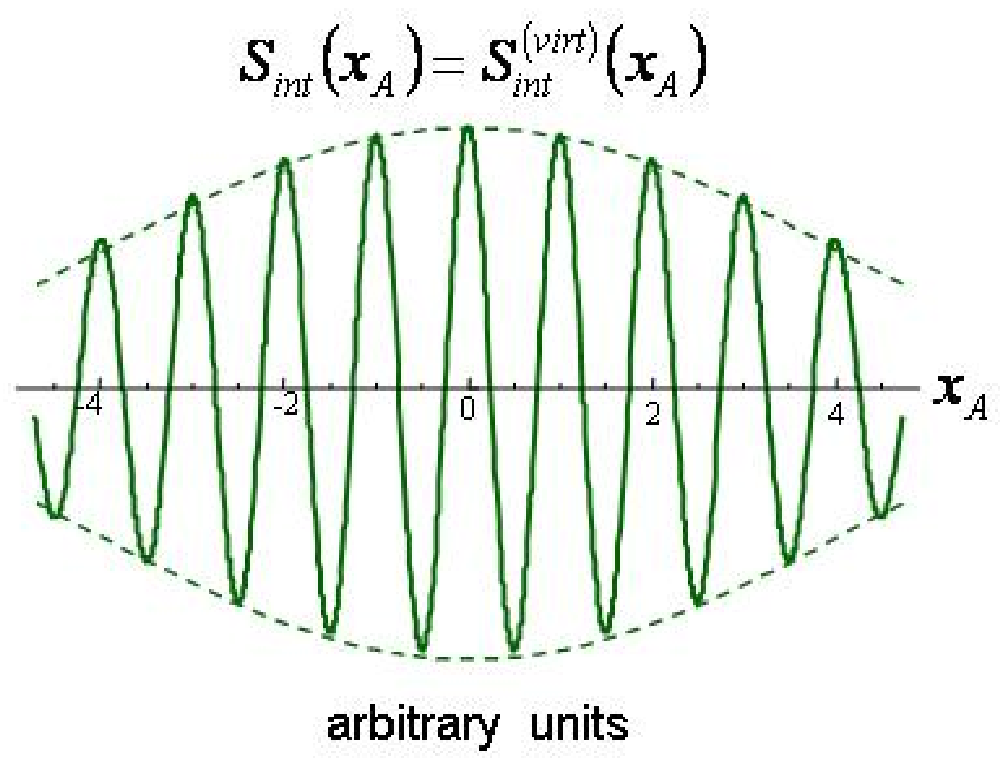} & \includegraphics[width=0.25\textwidth]{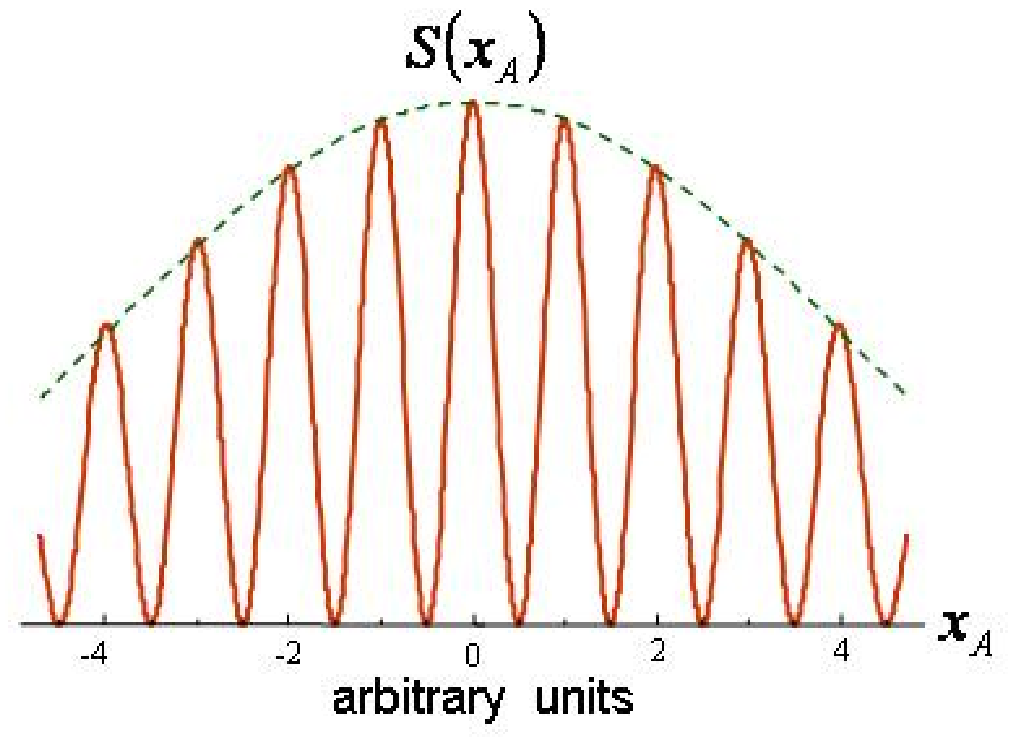}\\ 

  \textbf{(2)}& & & & \\ \hline 

  \end{tabular}
 \end{center} 
\end{landscape}

\newpage
\begin{landscape}
 \begin{center}
  \begin{tabular}{|c|c|c|c|c|}\hline
  \includegraphics[width=0.23\textwidth]{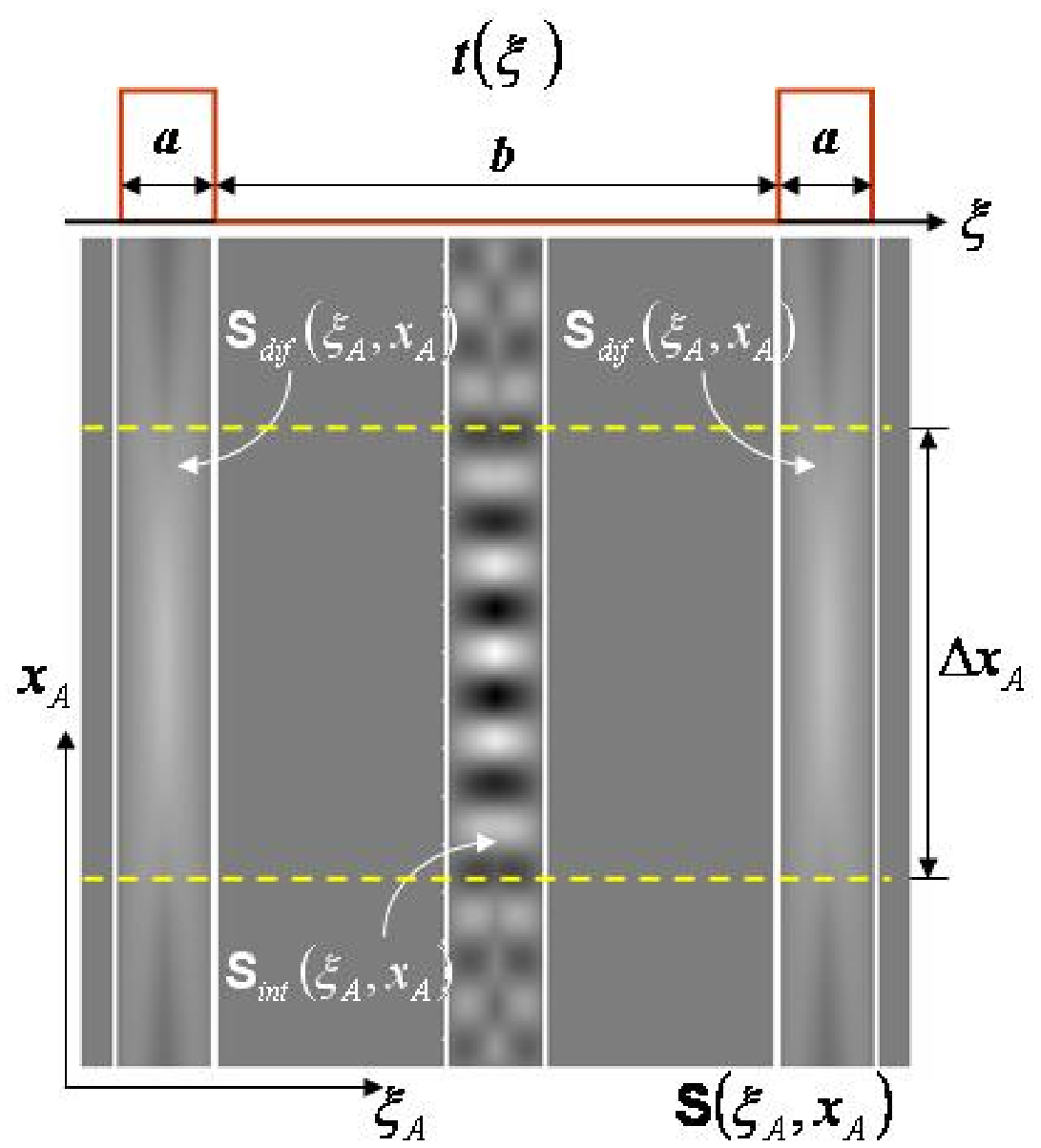}\label{fig7} & \includegraphics[width=0.25\textwidth]{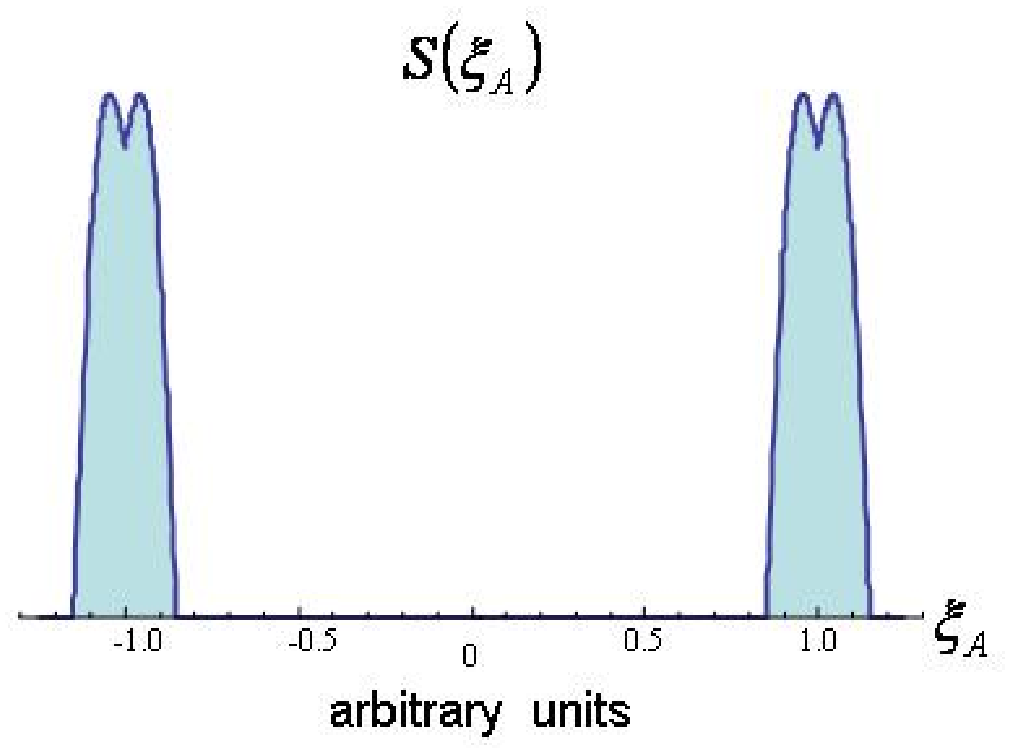} & \includegraphics[width=0.25\textwidth]{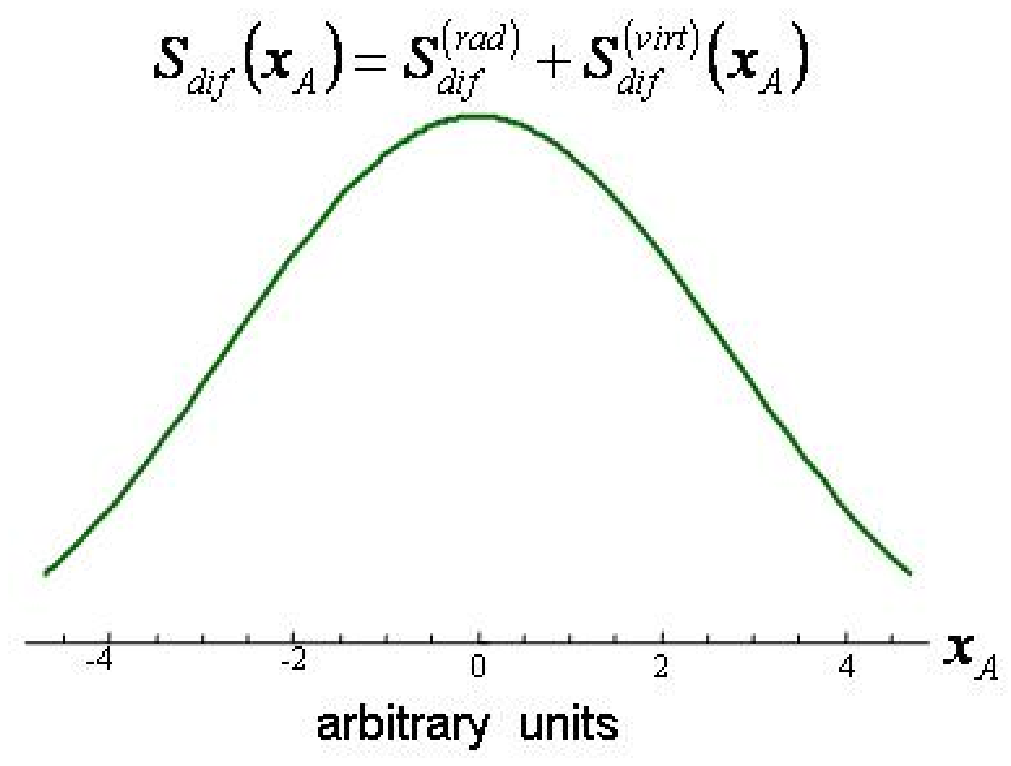} 
  & \includegraphics[width=0.25\textwidth]{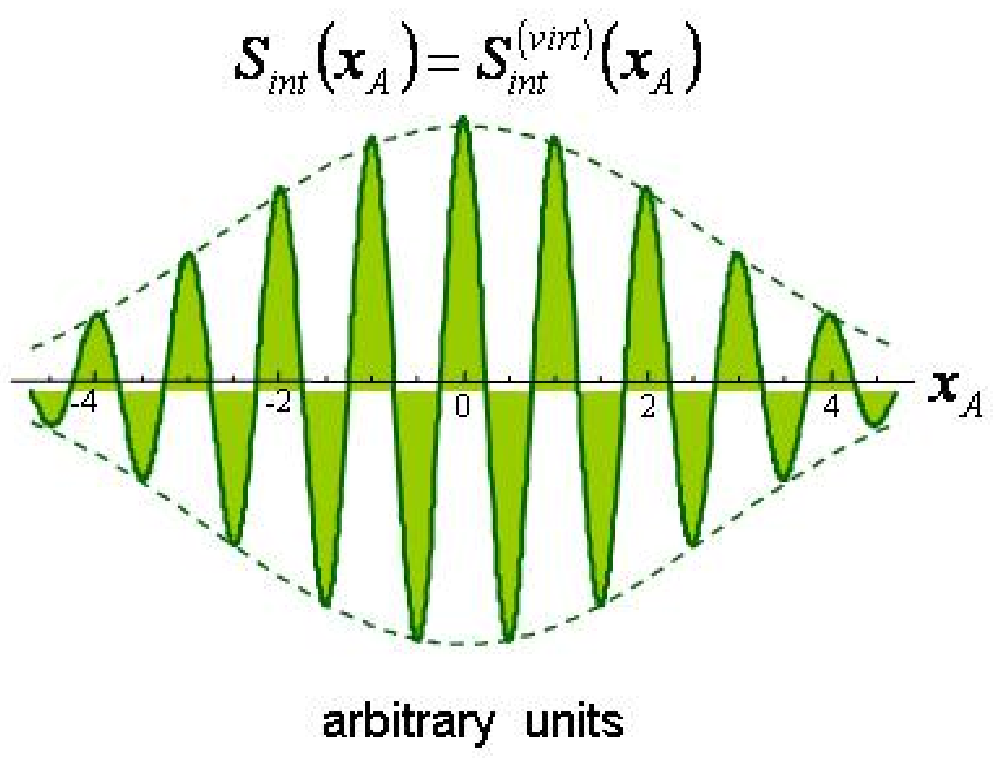} & \includegraphics[width=0.25\textwidth]{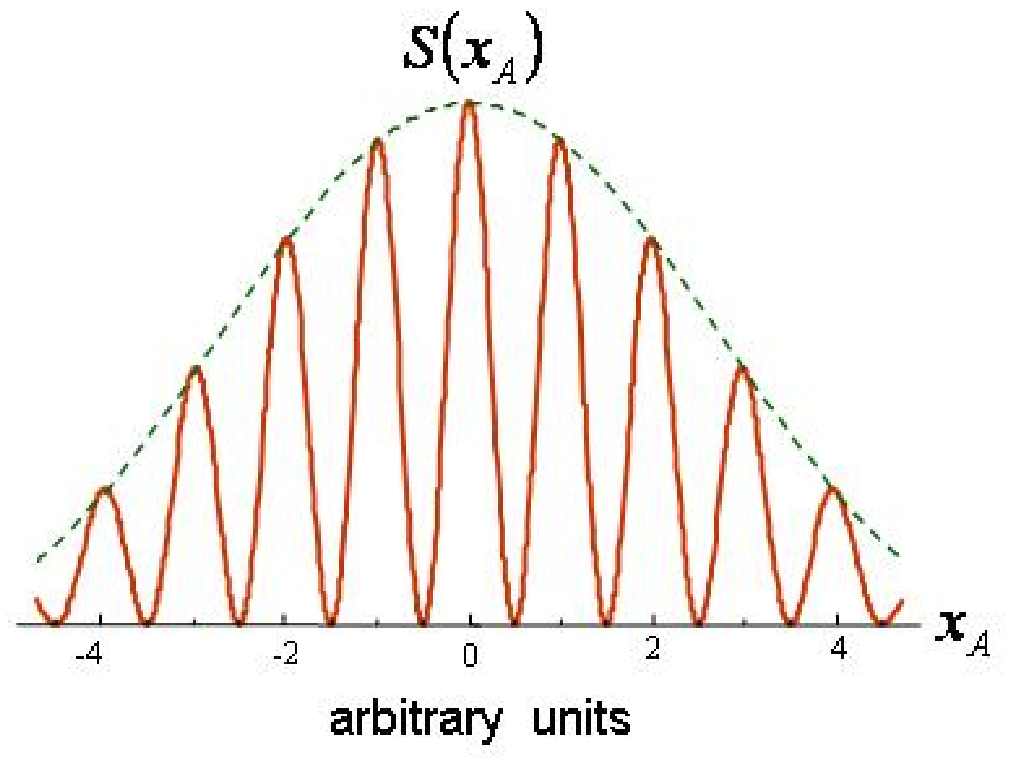}\\ 
  
  \textbf{(3)}& & & &\\ \hline 

  \includegraphics[width=0.23\textwidth]{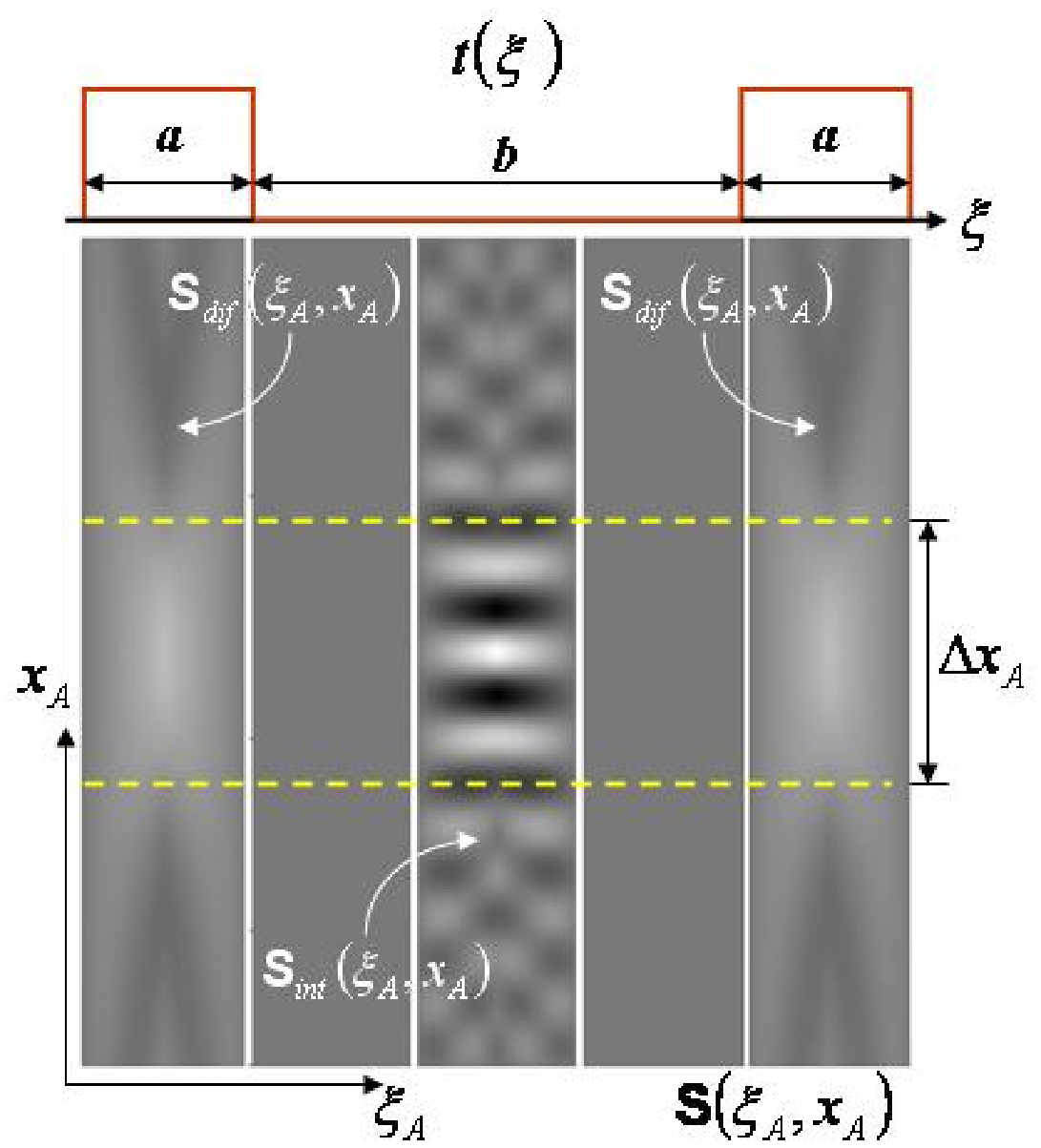}\label{fig8} & \includegraphics[width=0.25\textwidth]{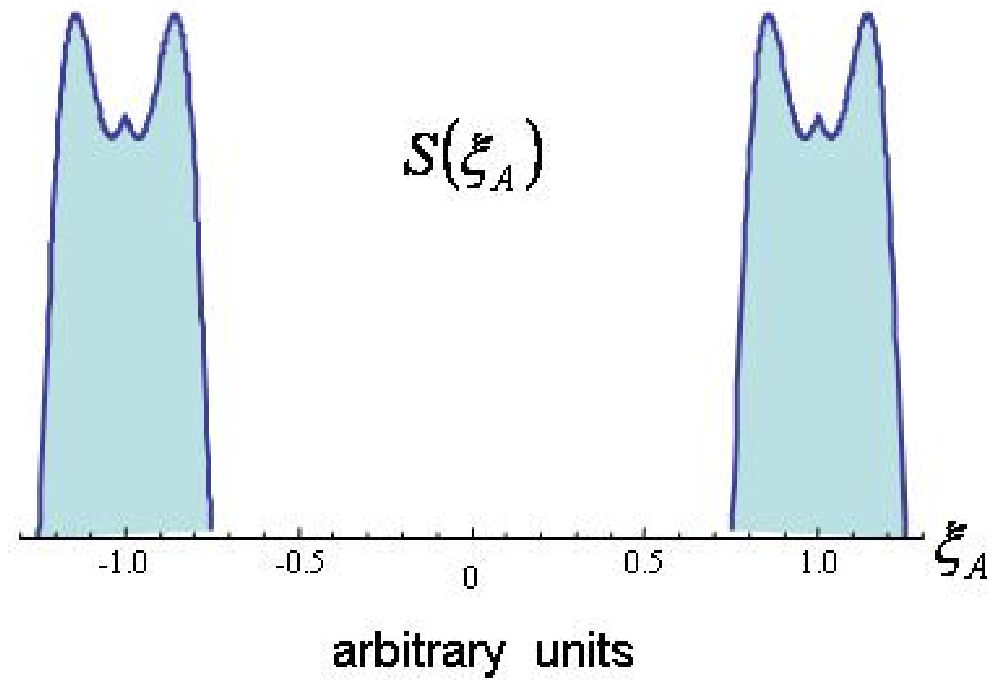} & \includegraphics[width=0.25\textwidth]{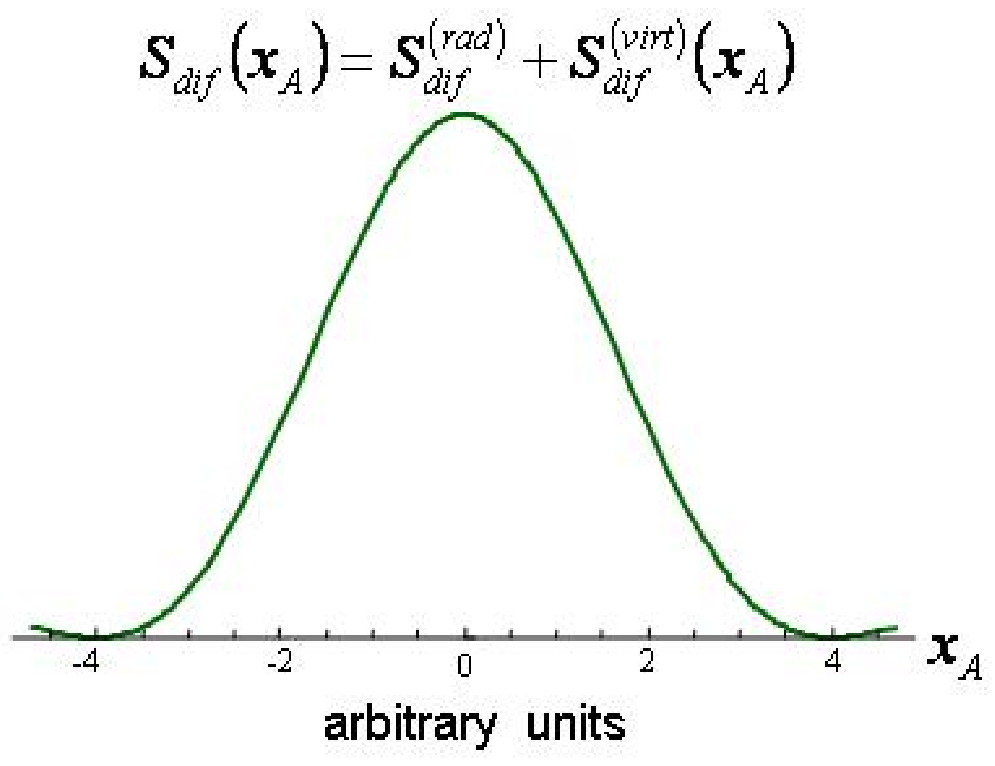} 
  & \includegraphics[width=0.25\textwidth]{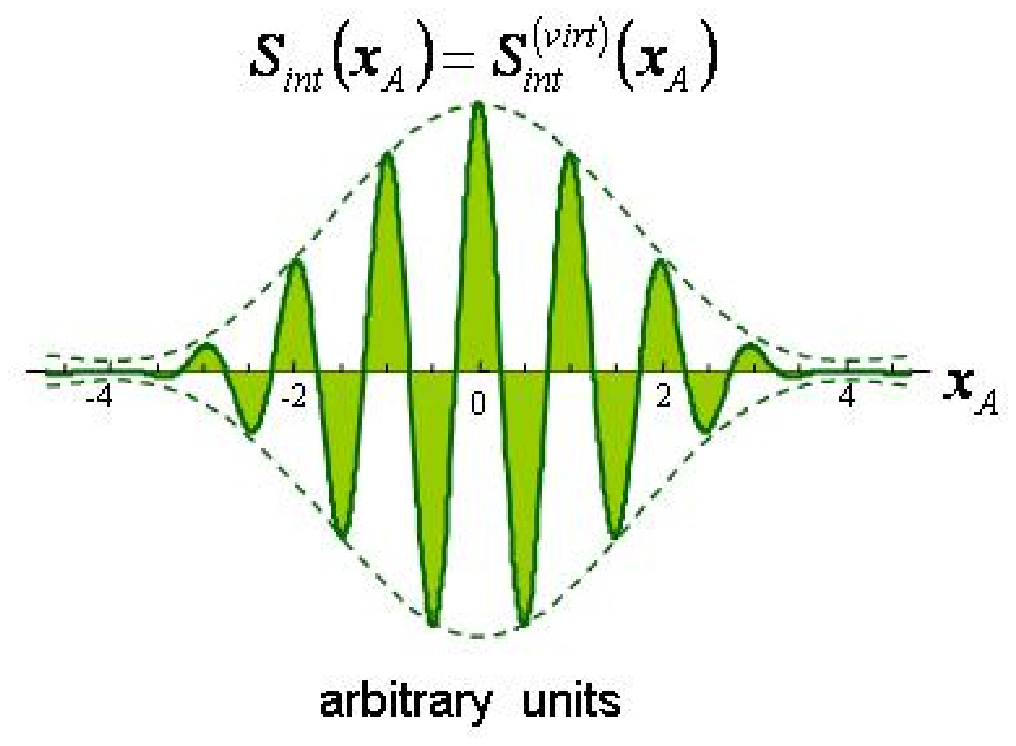} & \includegraphics[width=0.25\textwidth]{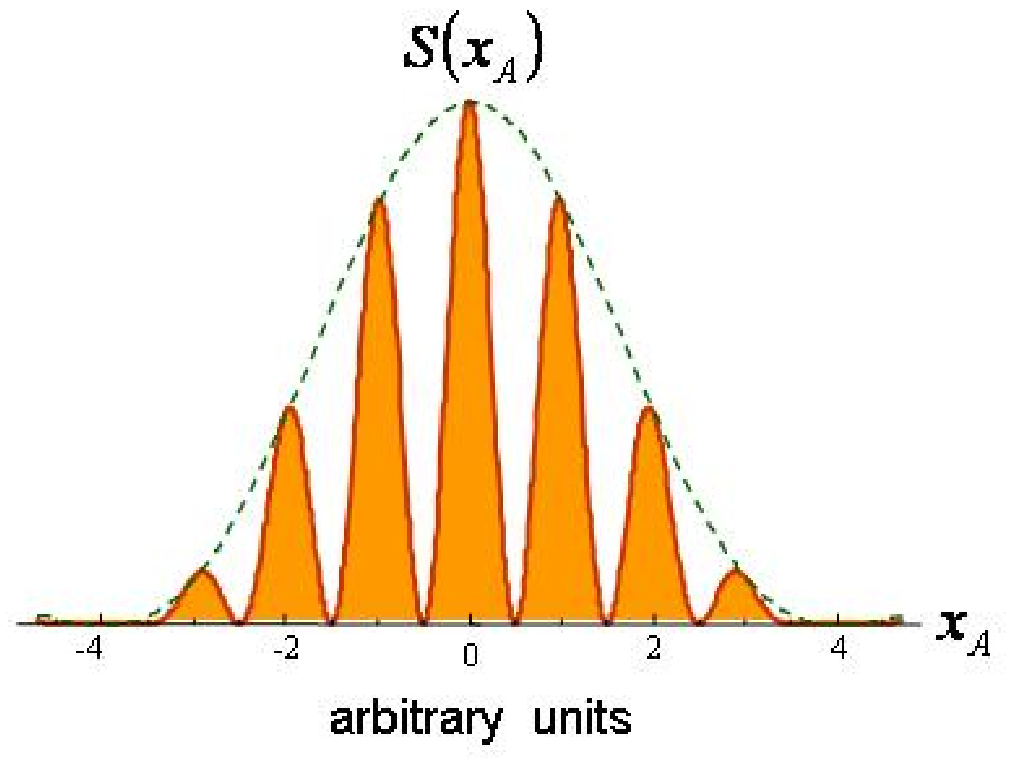}\\ 
  
  \textbf{(4)}& & & &\\ \hline   

  \includegraphics[width=0.22\textwidth]{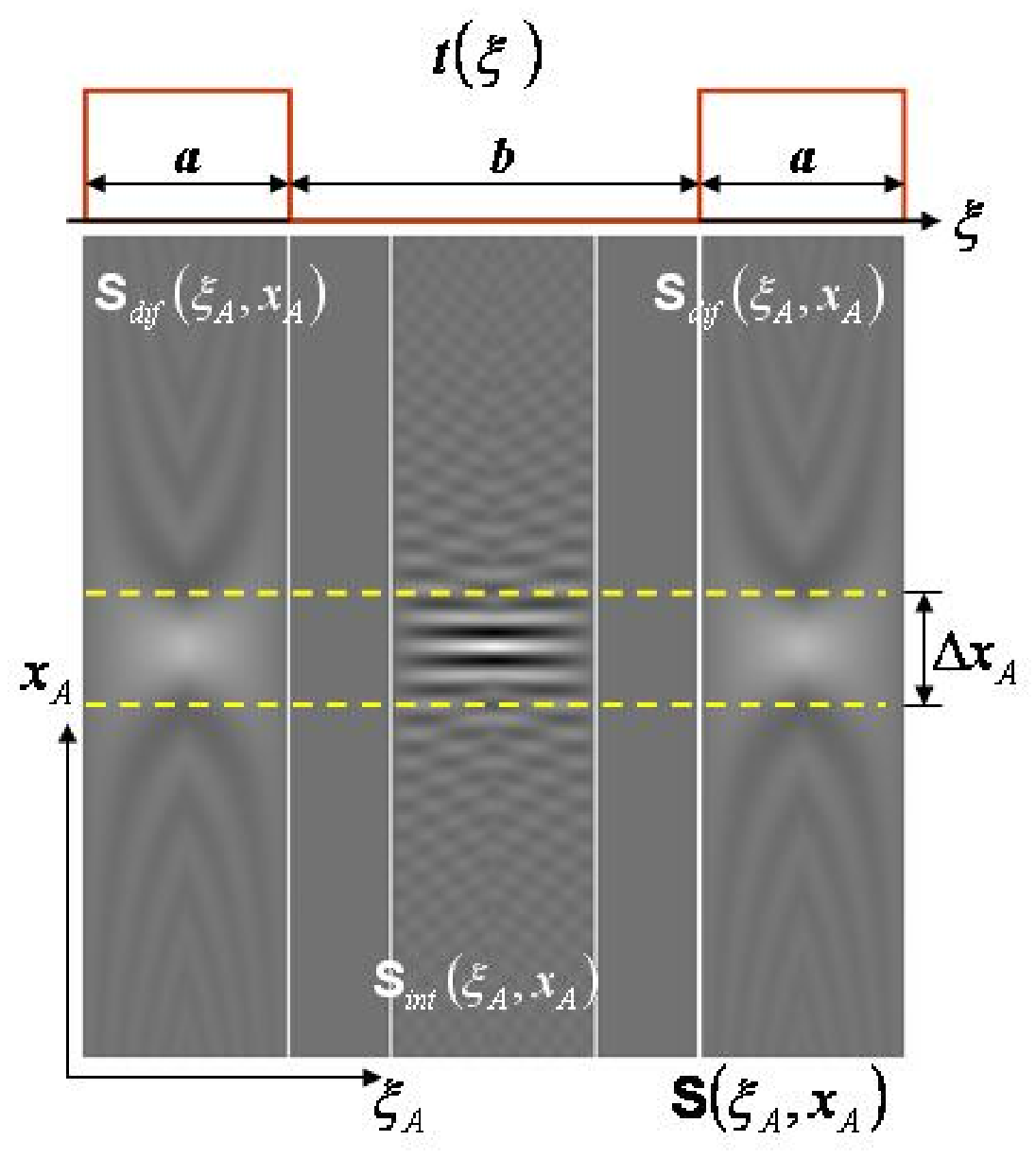}\label{fig9} & \includegraphics[width=0.25\textwidth]{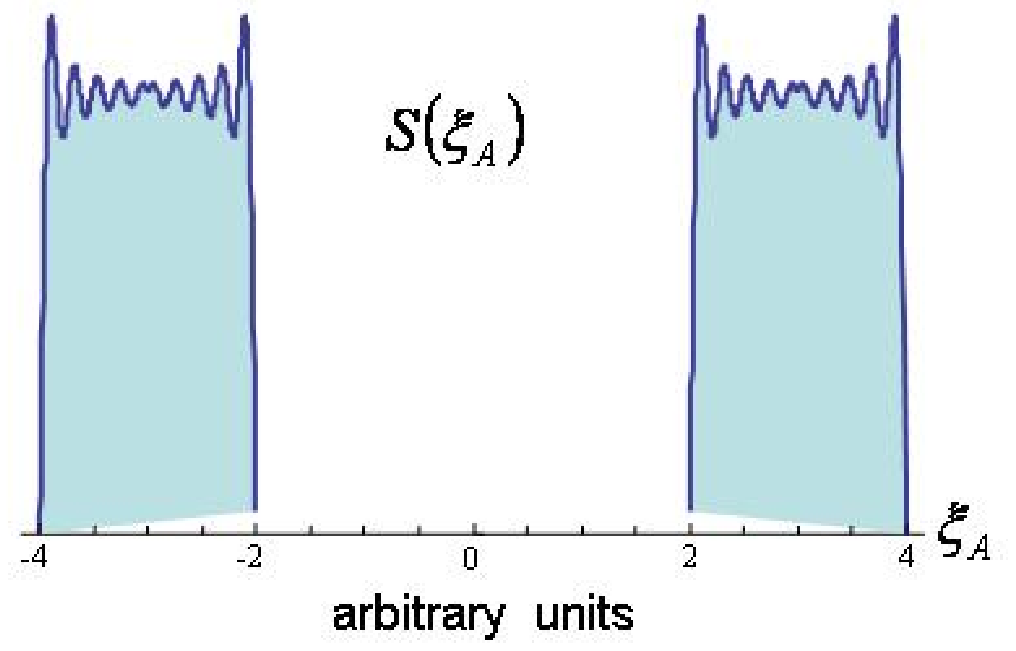} & \includegraphics[width=0.25\textwidth]{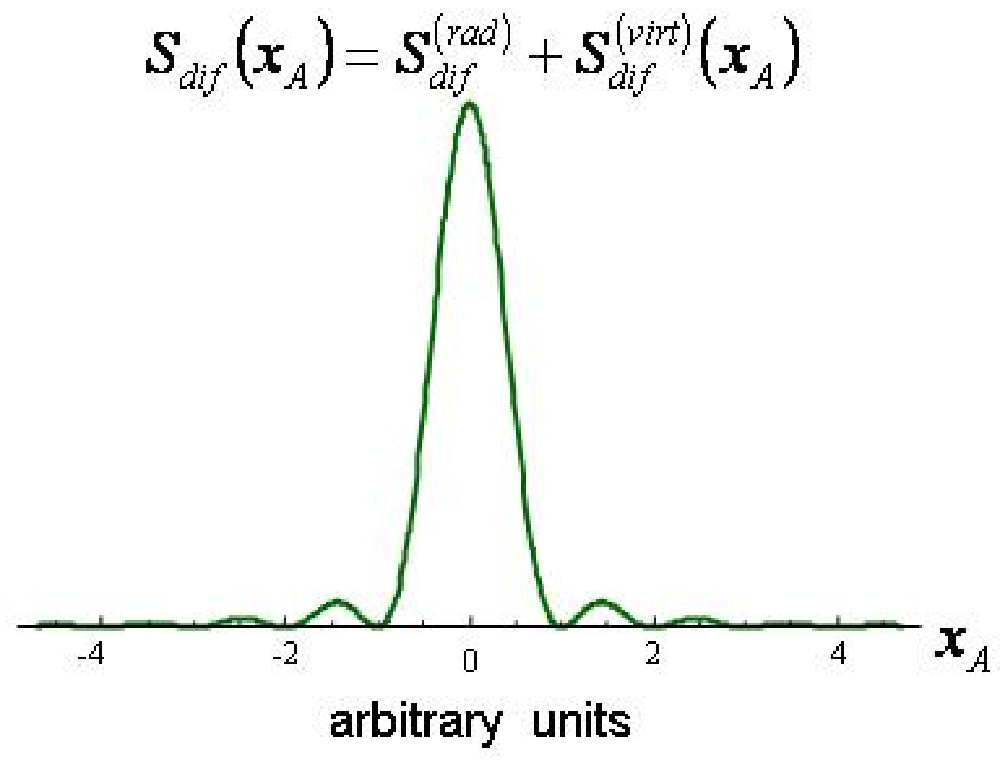} 
  & \includegraphics[width=0.25\textwidth]{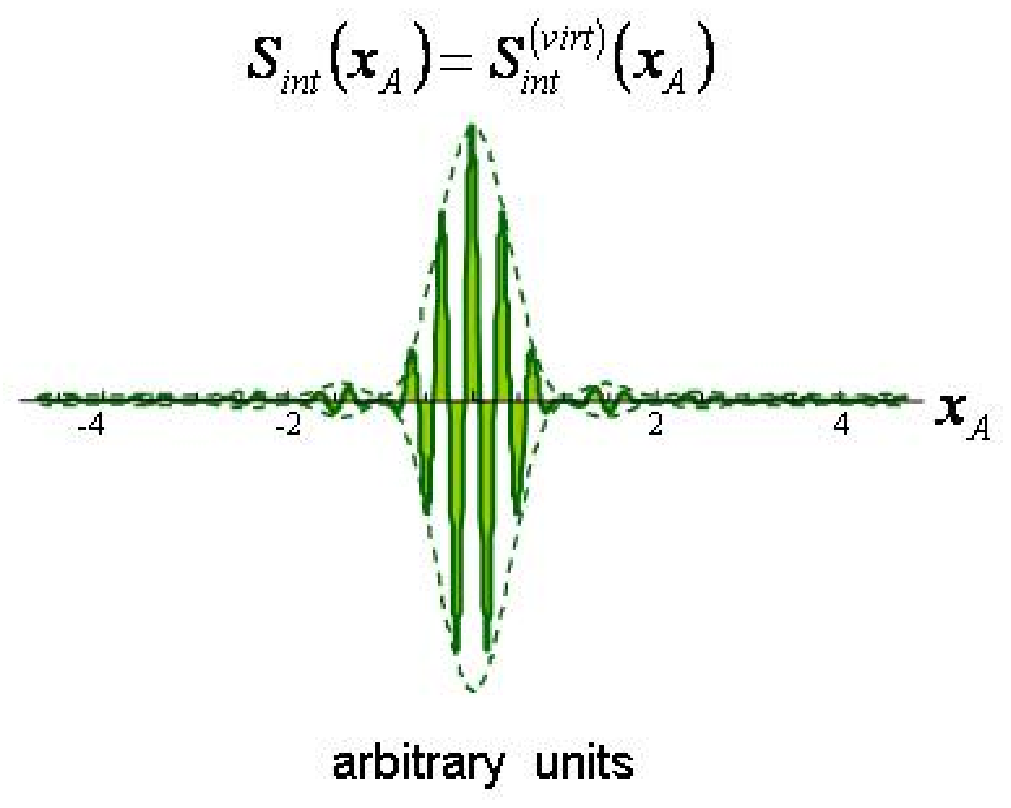} & \includegraphics[width=0.25\textwidth]{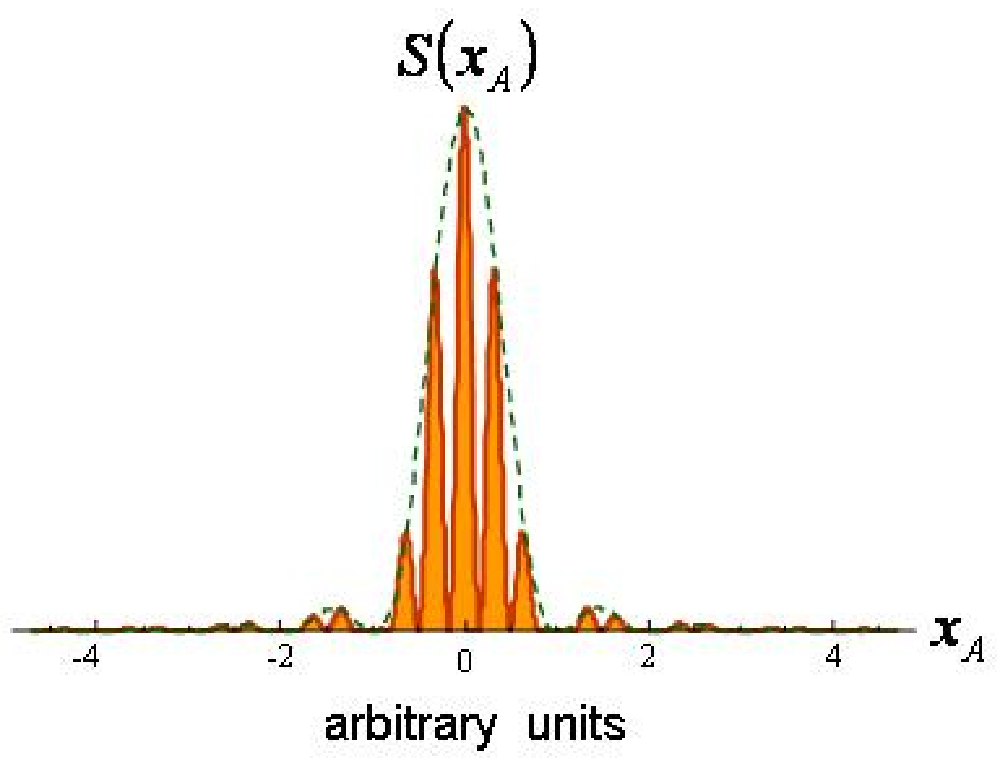} \\ 

  \textbf{(5)}& & & &\\ \hline

  \end{tabular}   
 \end{center}
\end{landscape}

\newpage
\begin{landscape}
 \begin{center}
  \begin{tabular}{|c|c|c|c|c|}\hline

  \includegraphics[width=0.23\textwidth]{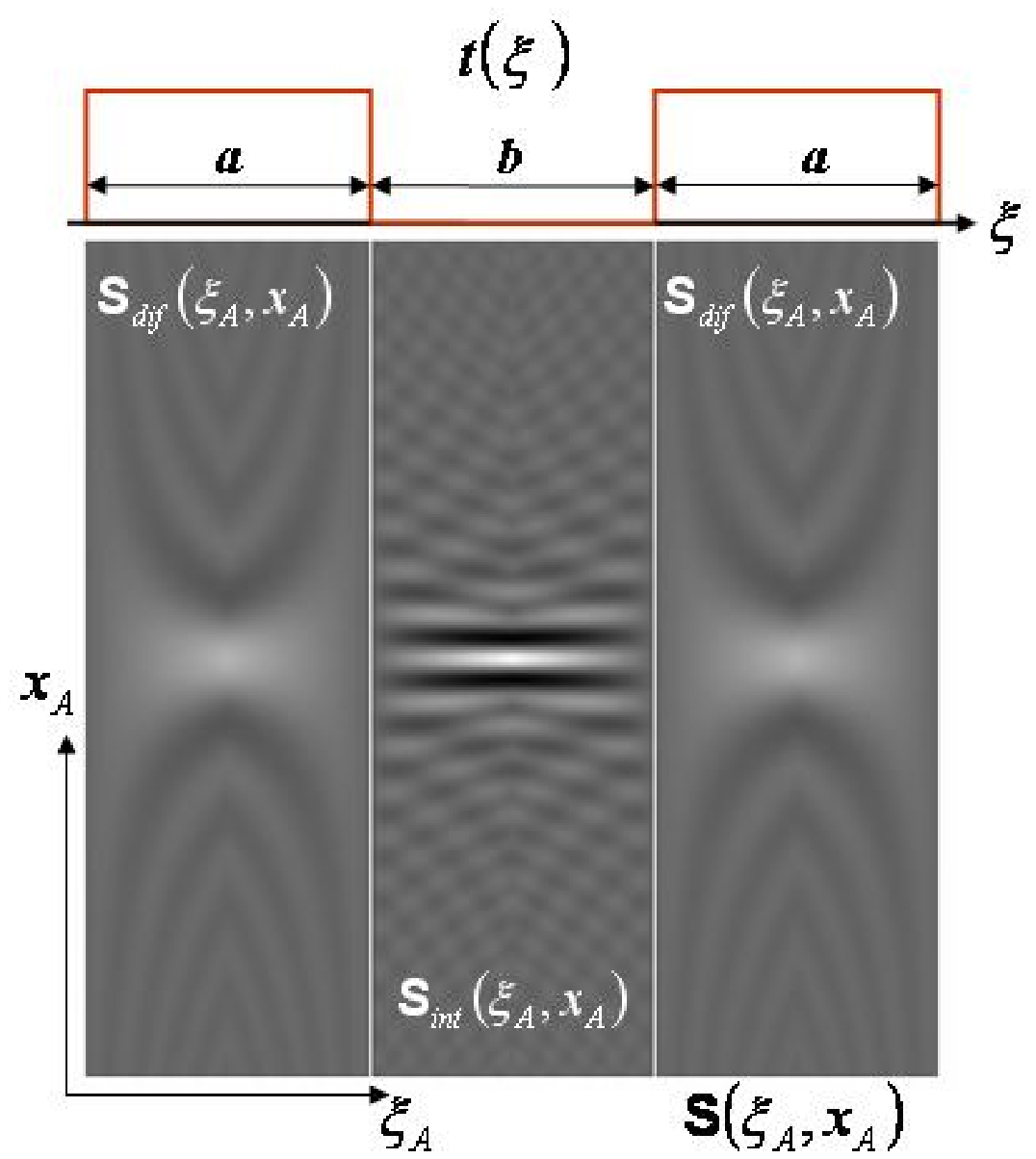}\label{fig10} & \includegraphics[width=0.25\textwidth]{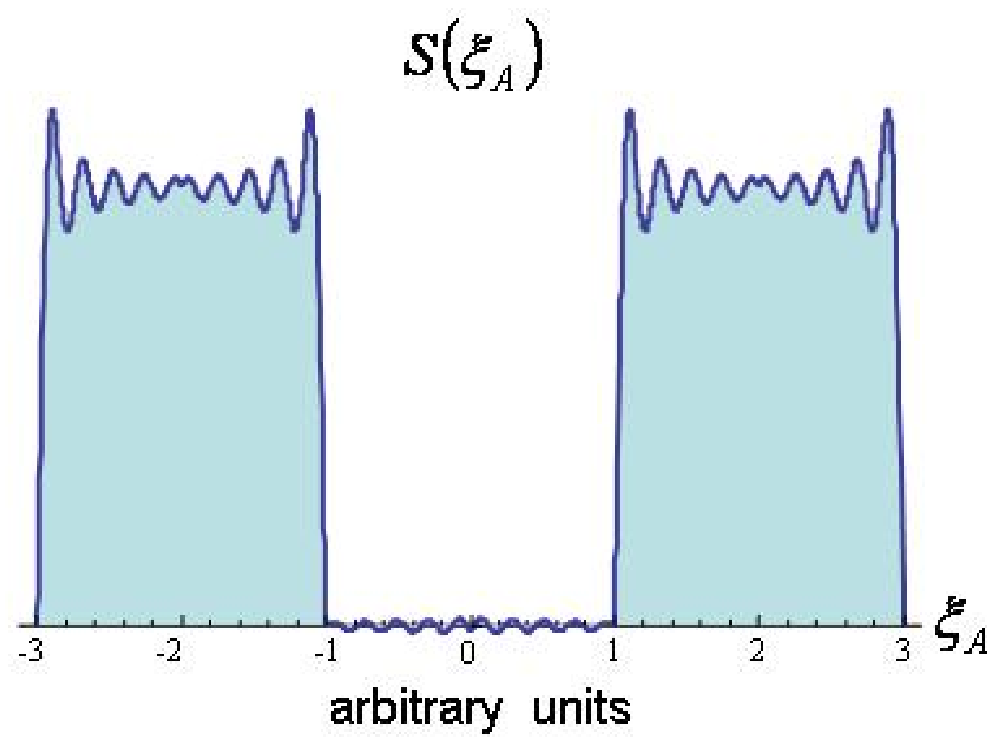} & \includegraphics[width=0.25\textwidth]{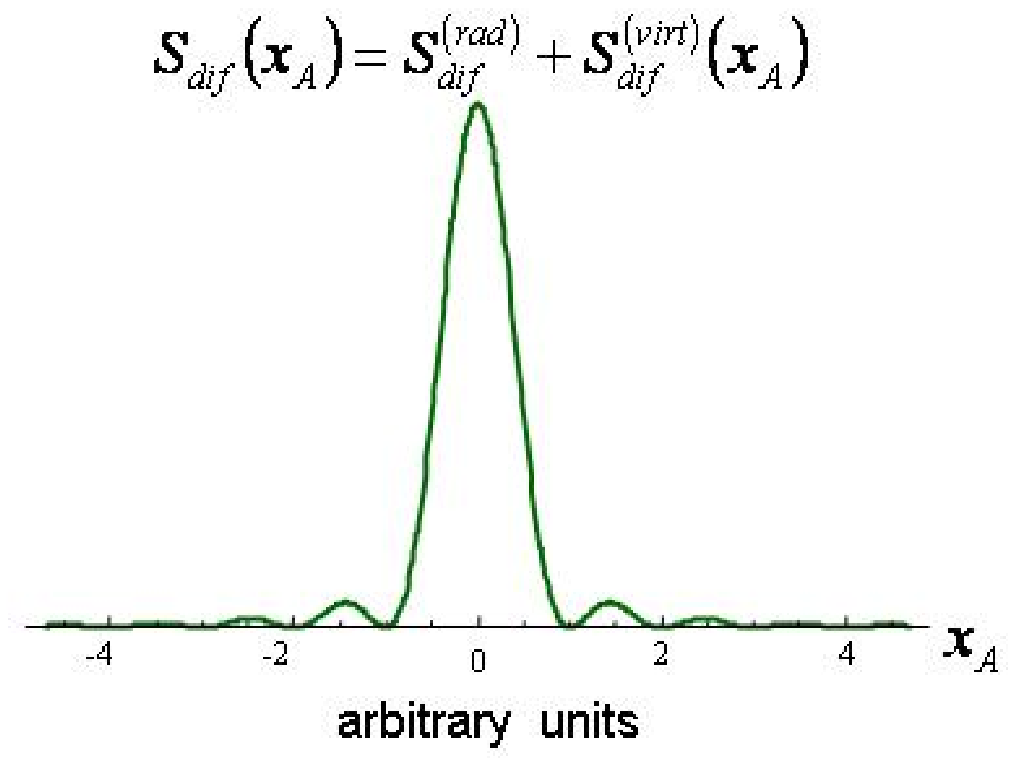} 
  & \includegraphics[width=0.25\textwidth]{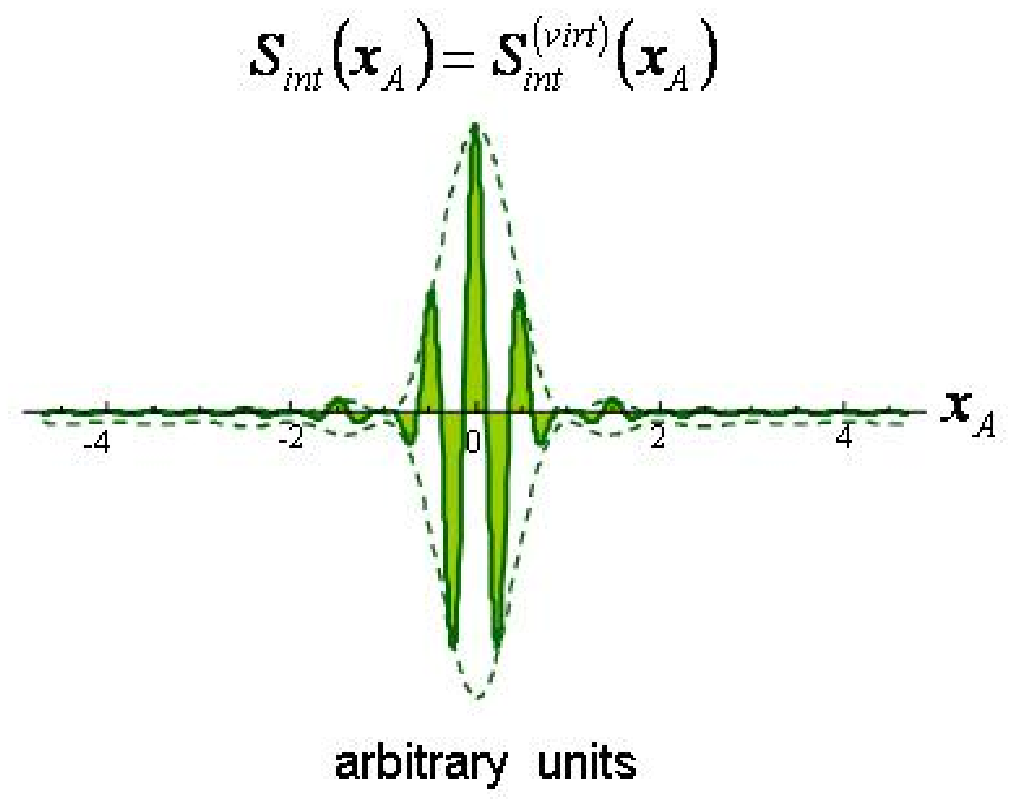} & \includegraphics[width=0.25\textwidth]{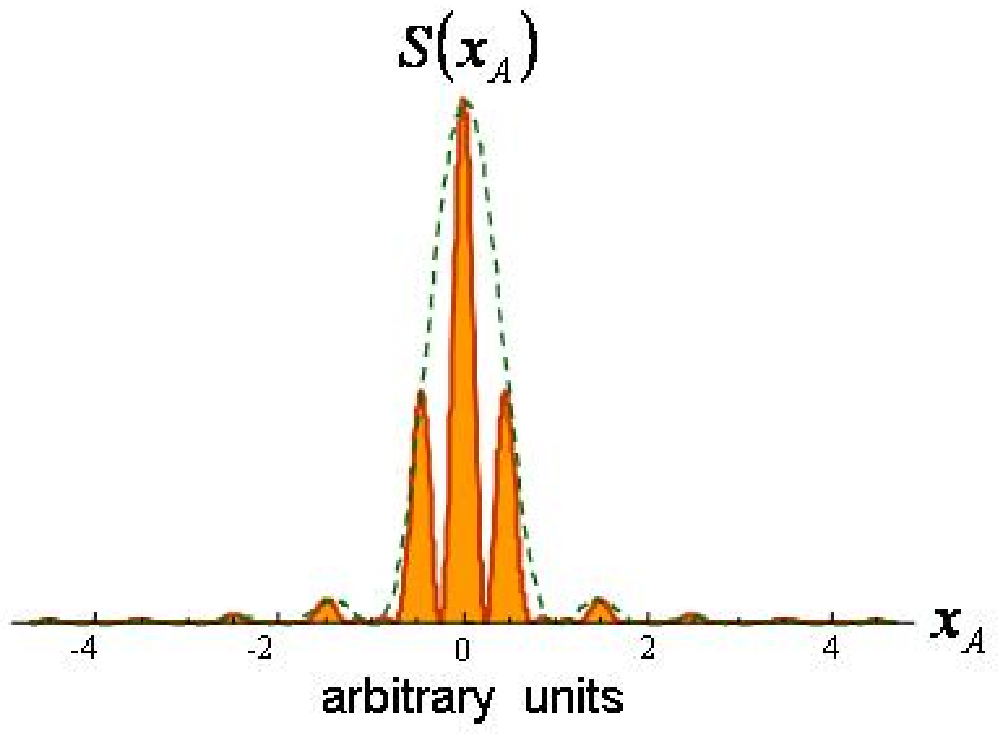}\\ 

  \textbf{(6)}& & & &\\ \hline 

  \includegraphics[width=0.23\textwidth]{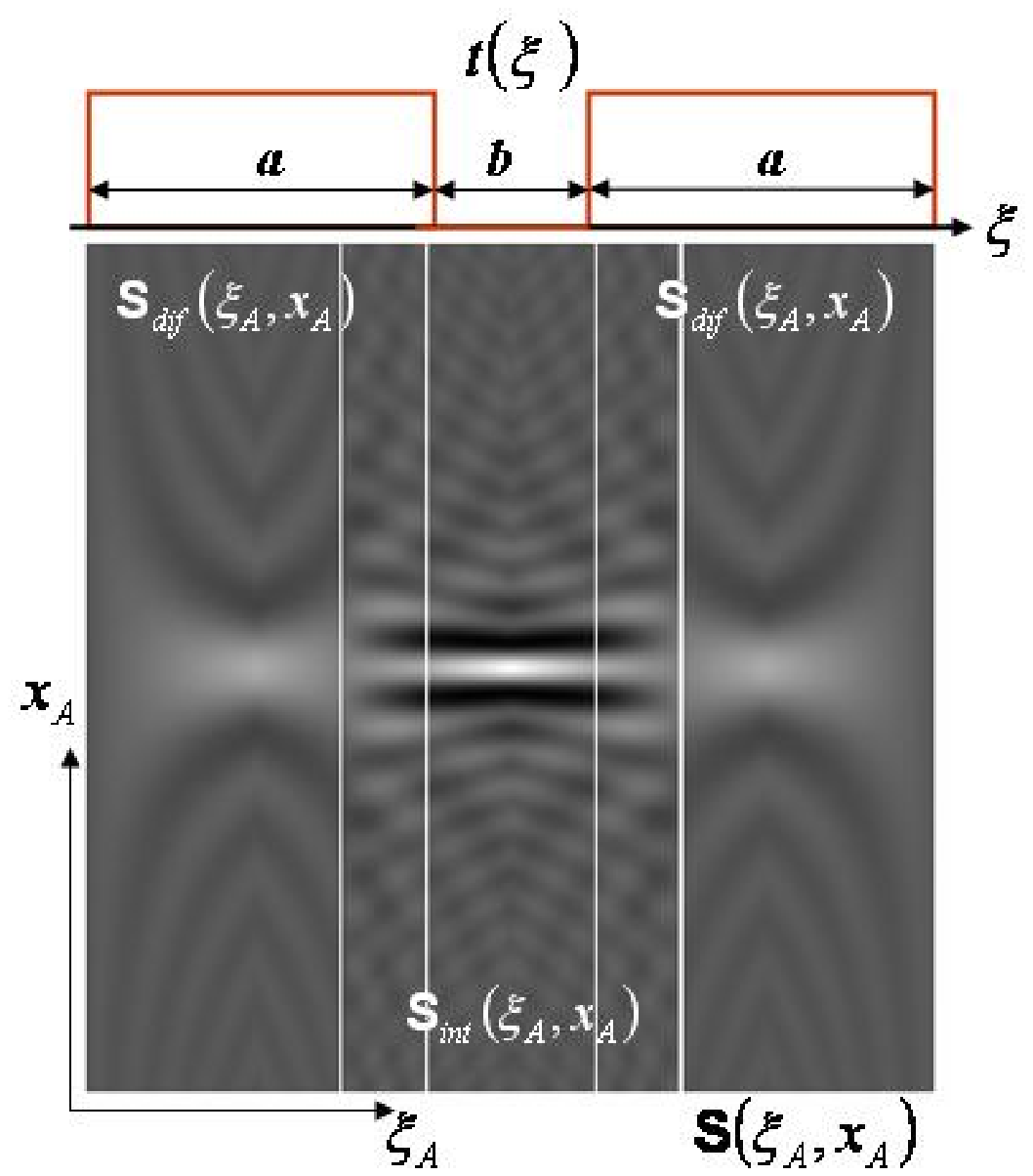}\label{fig11} & \includegraphics[width=0.245\textwidth]{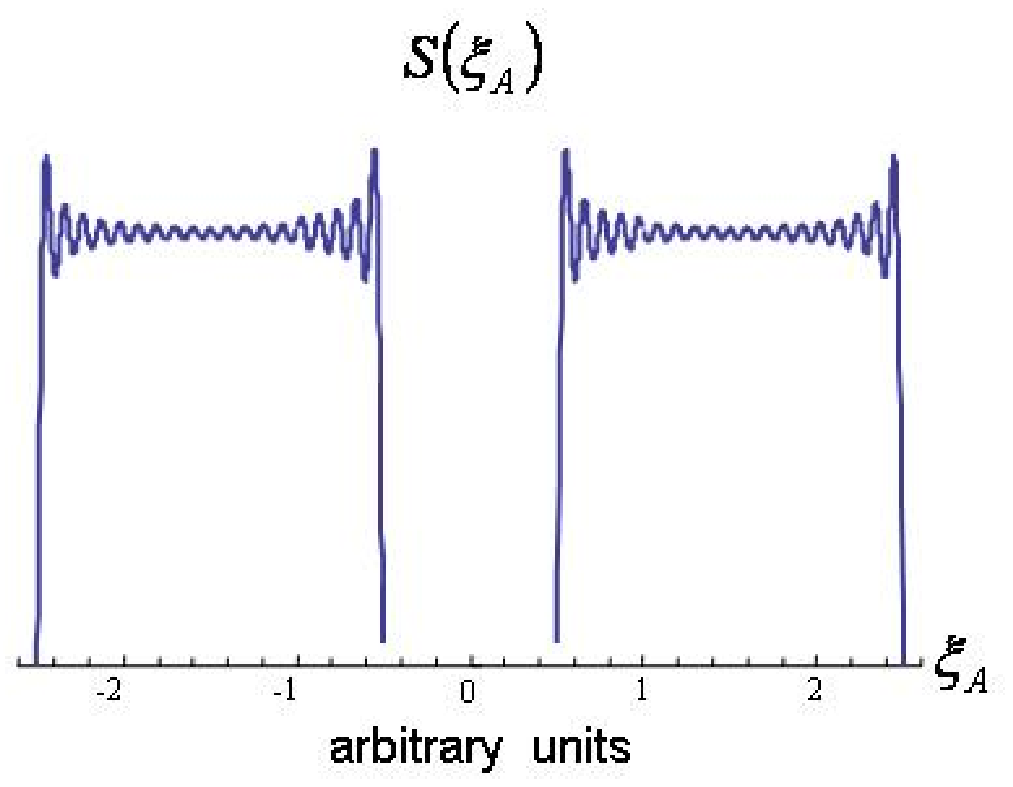} & \includegraphics[width=0.245\textwidth]{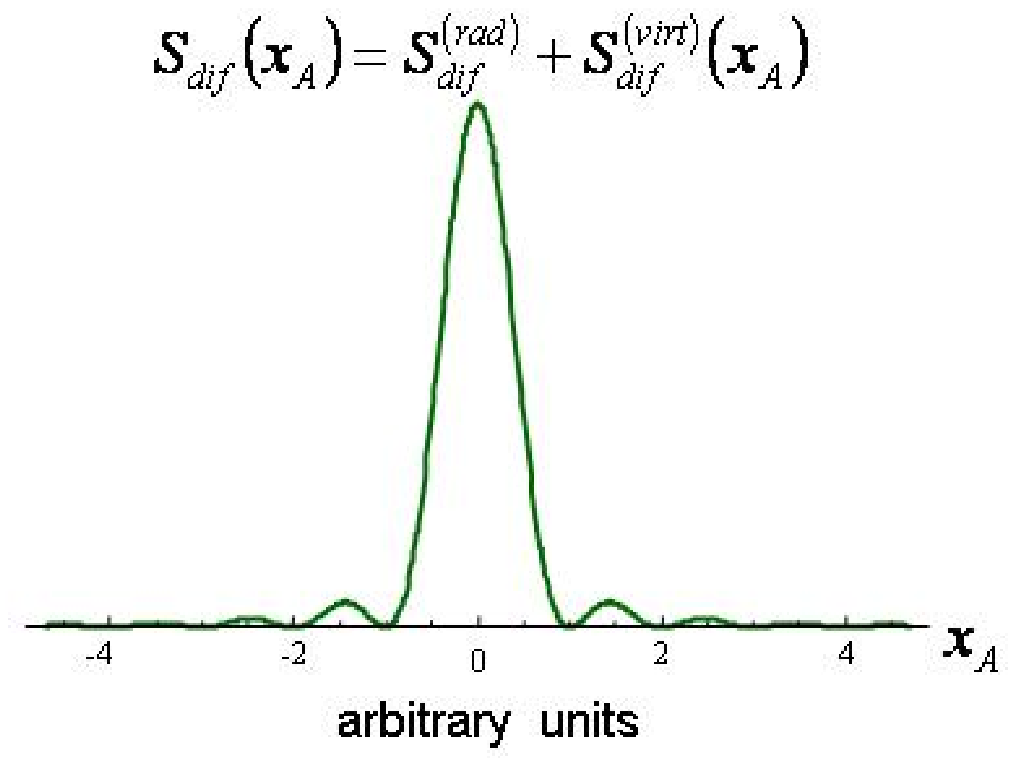} 
  & \includegraphics[width=0.245\textwidth]{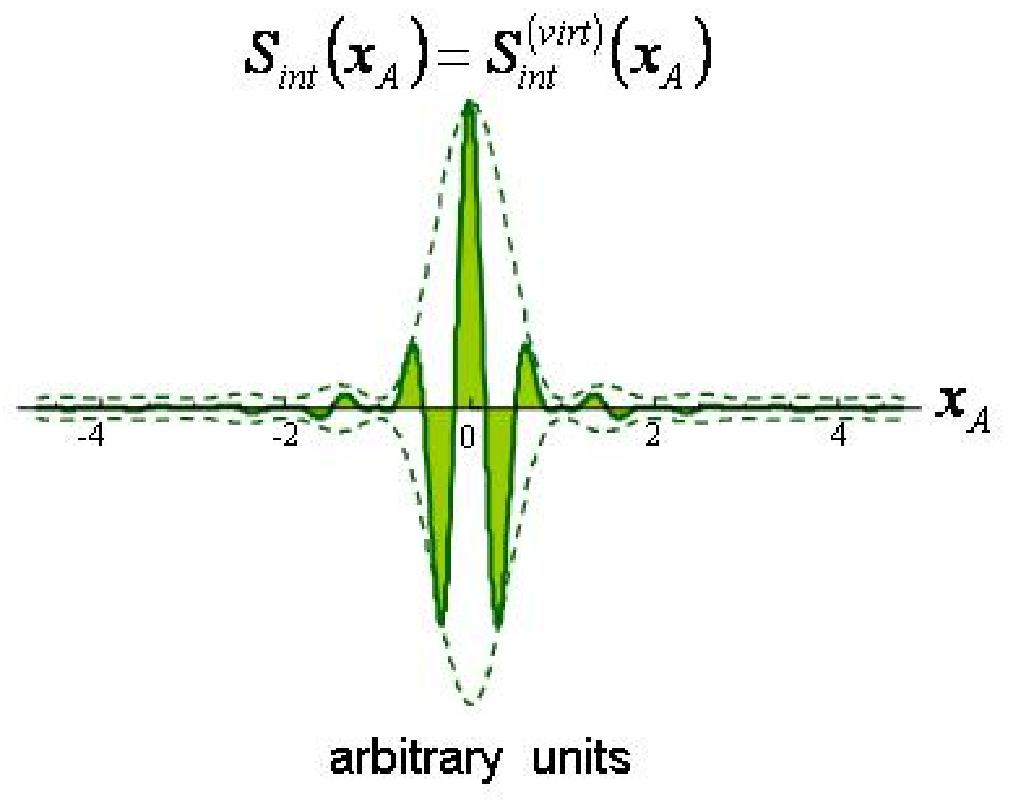} & \includegraphics[width=0.245\textwidth]{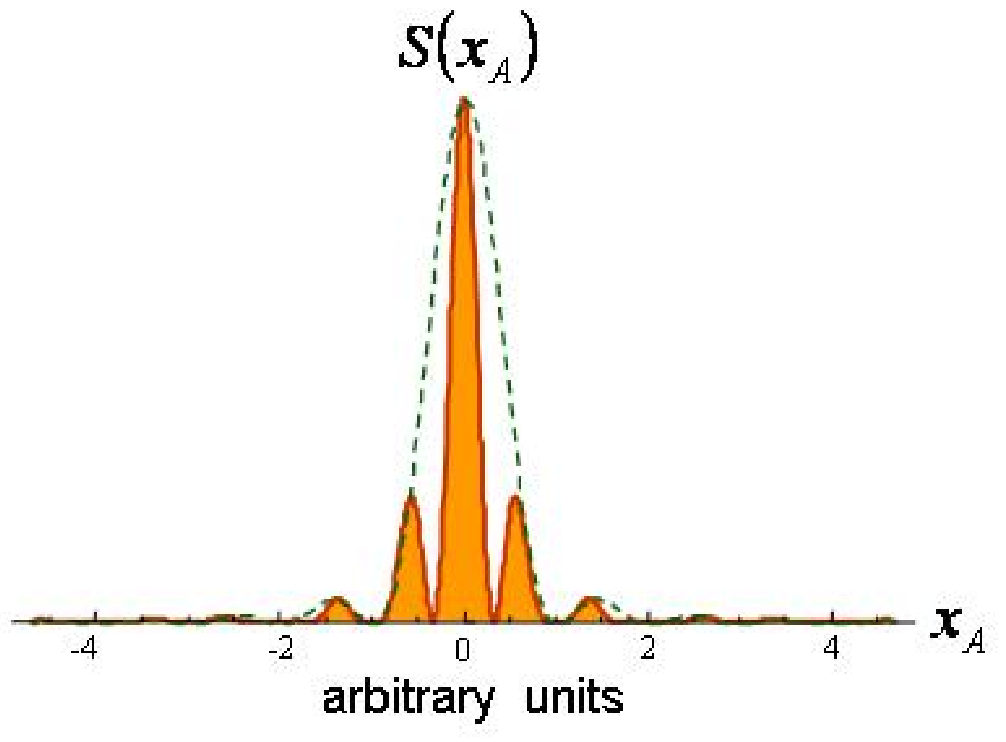}\\
  
  \textbf{(7)}& & & &\\ \hline
  
  \includegraphics[width=0.22\textwidth]{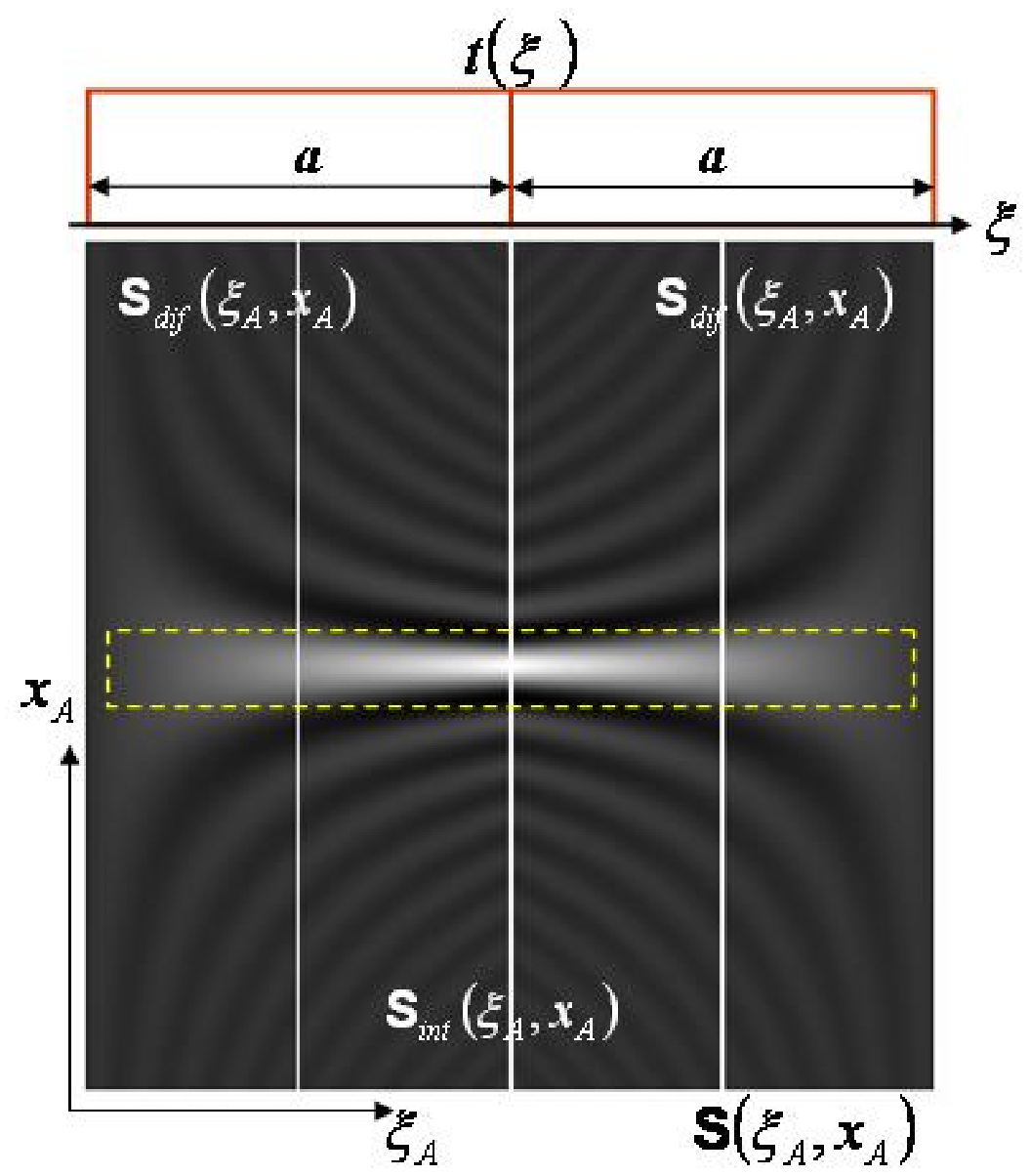}\label{fig12} & \includegraphics[width=0.25\textwidth]{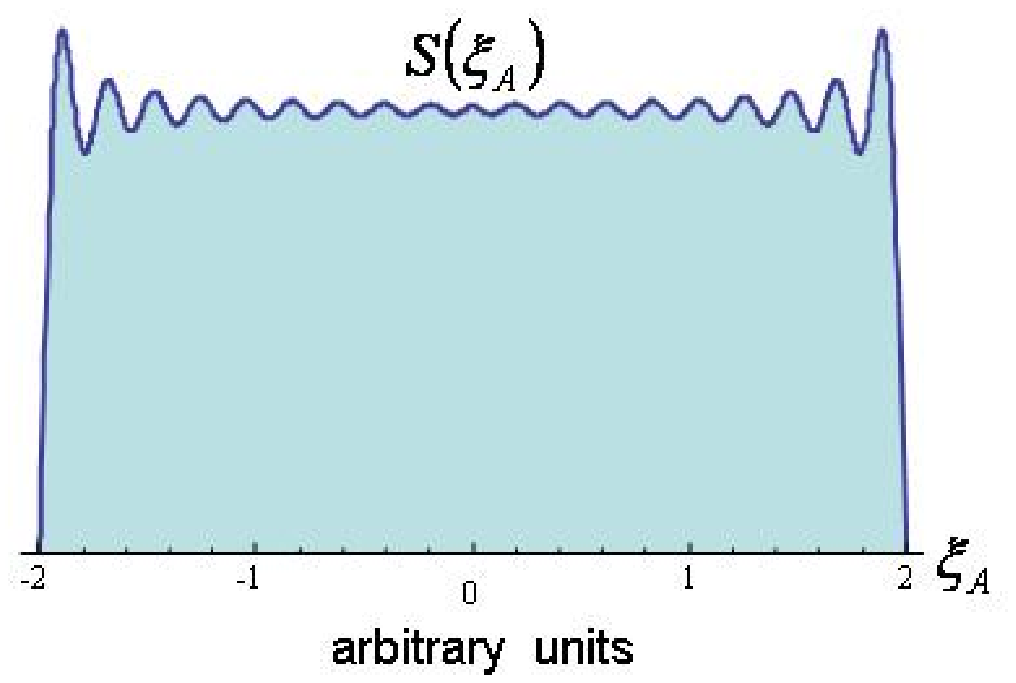} & \includegraphics[width=0.25\textwidth]{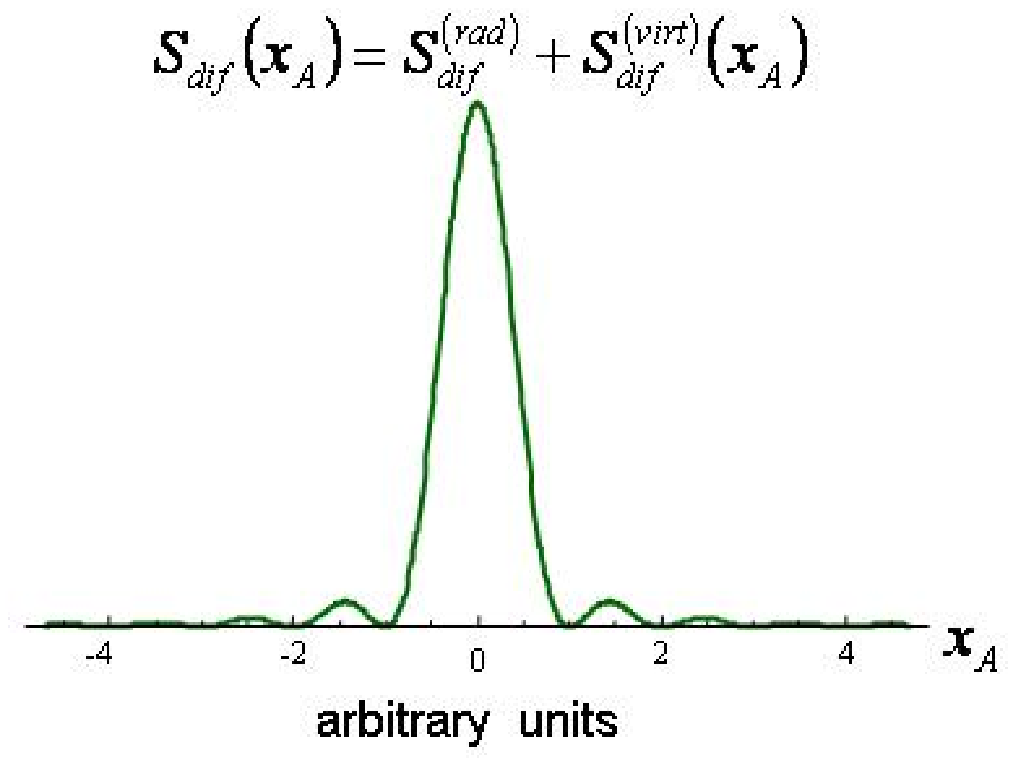} 
  & \includegraphics[width=0.25\textwidth]{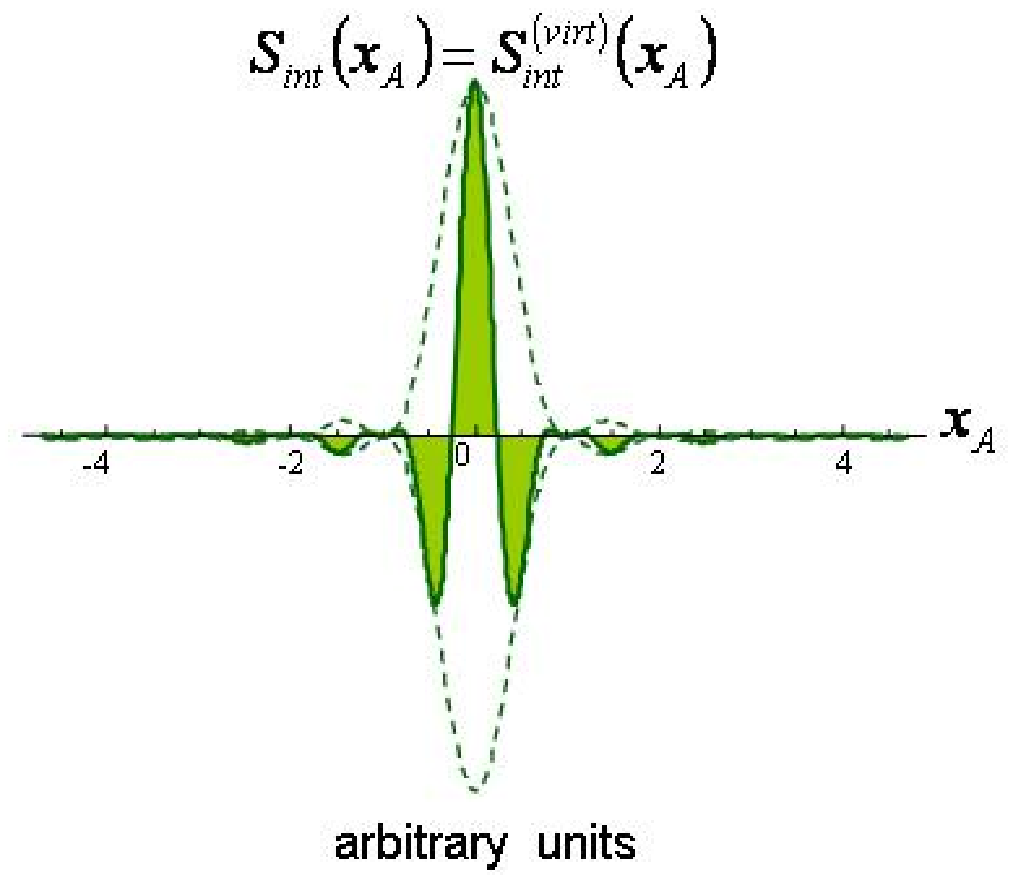} & \includegraphics[width=0.25\textwidth]{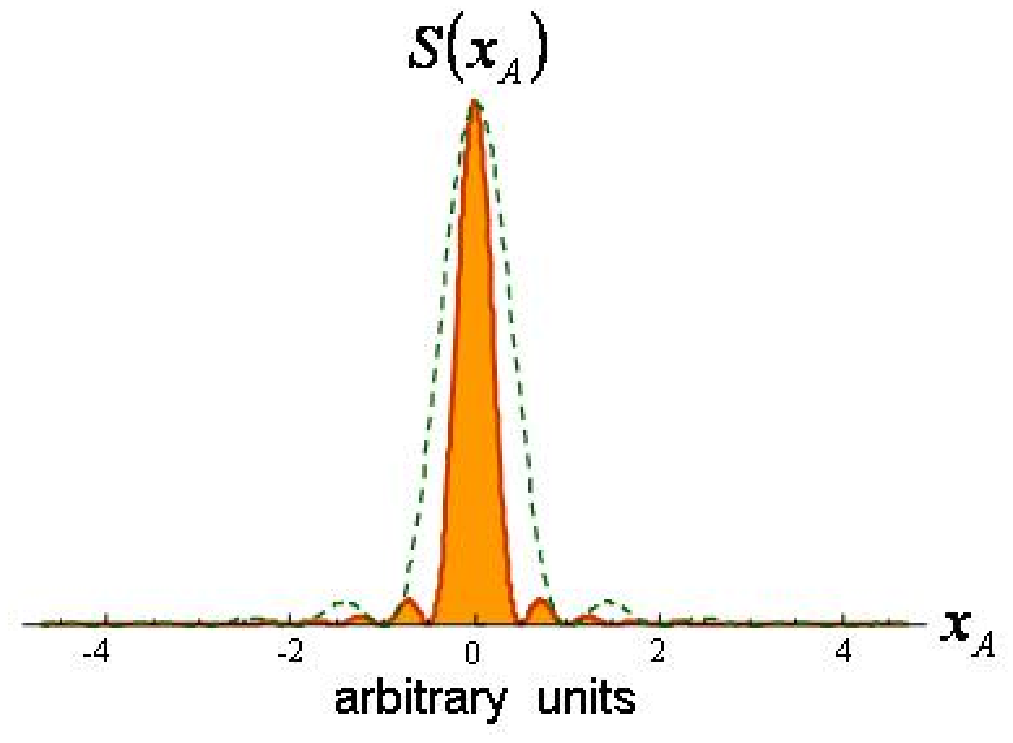}\\ 
  
  \textbf{(8)}& & & &\\ \hline

  \end{tabular}
 \end{center}
\end{landscape}

\newpage
\begin{landscape}
 \begin{center}
  \begin{tabular}{|c|c|c|c|c|}\hline
  
   \includegraphics[width=0.22\textwidth]{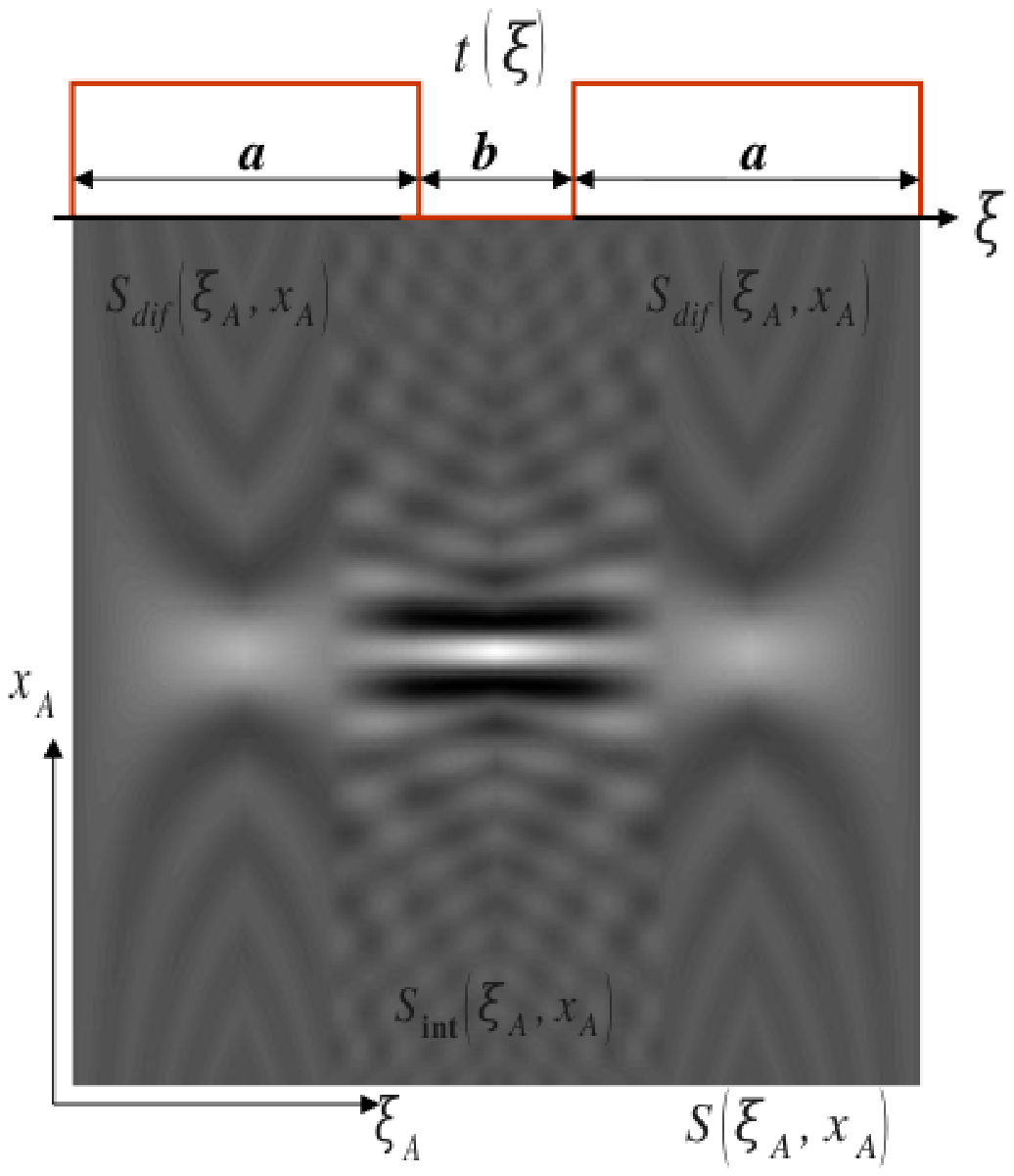}\label{fig13} & \includegraphics[width=0.25\textwidth]{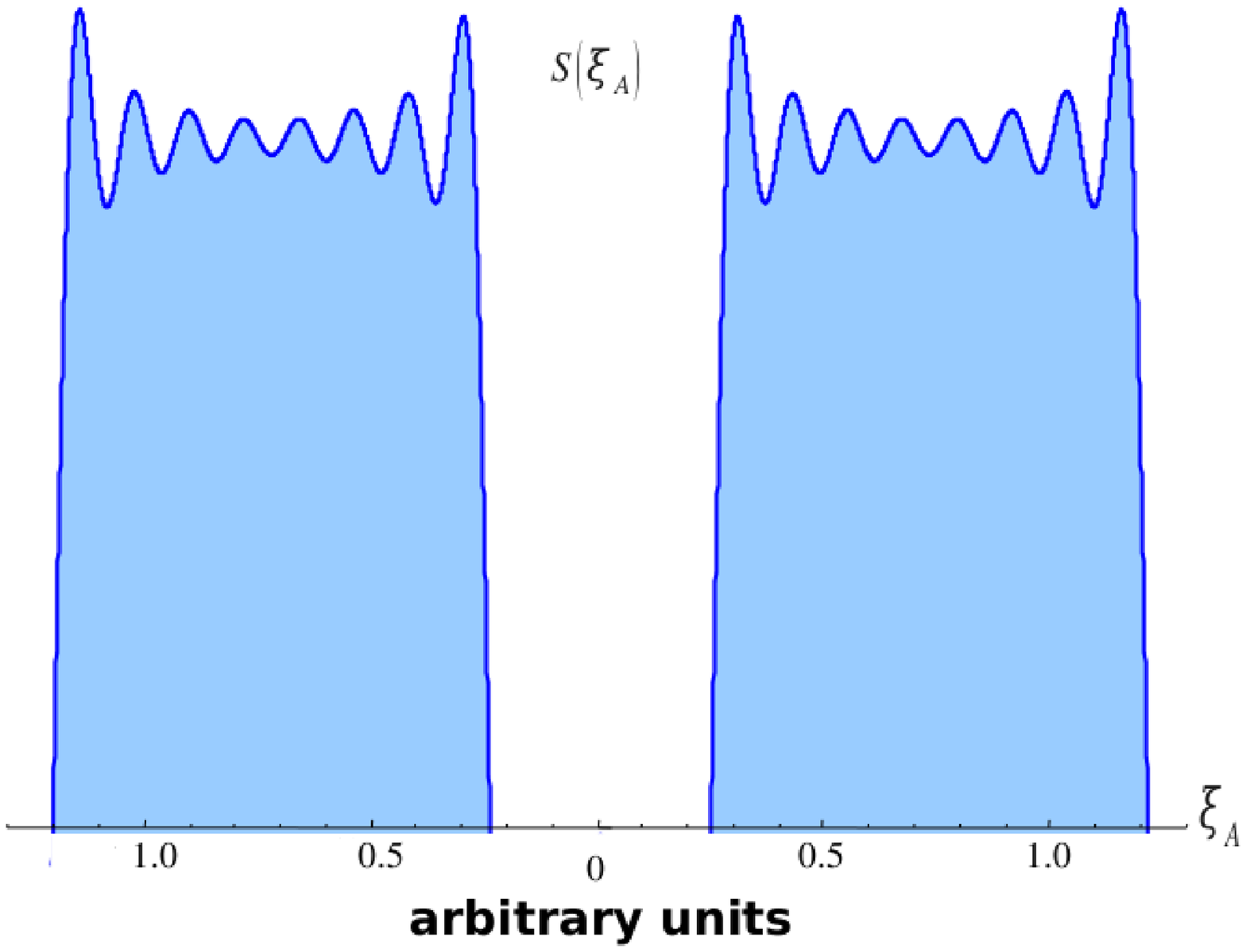} & \includegraphics[width=0.25\textwidth]{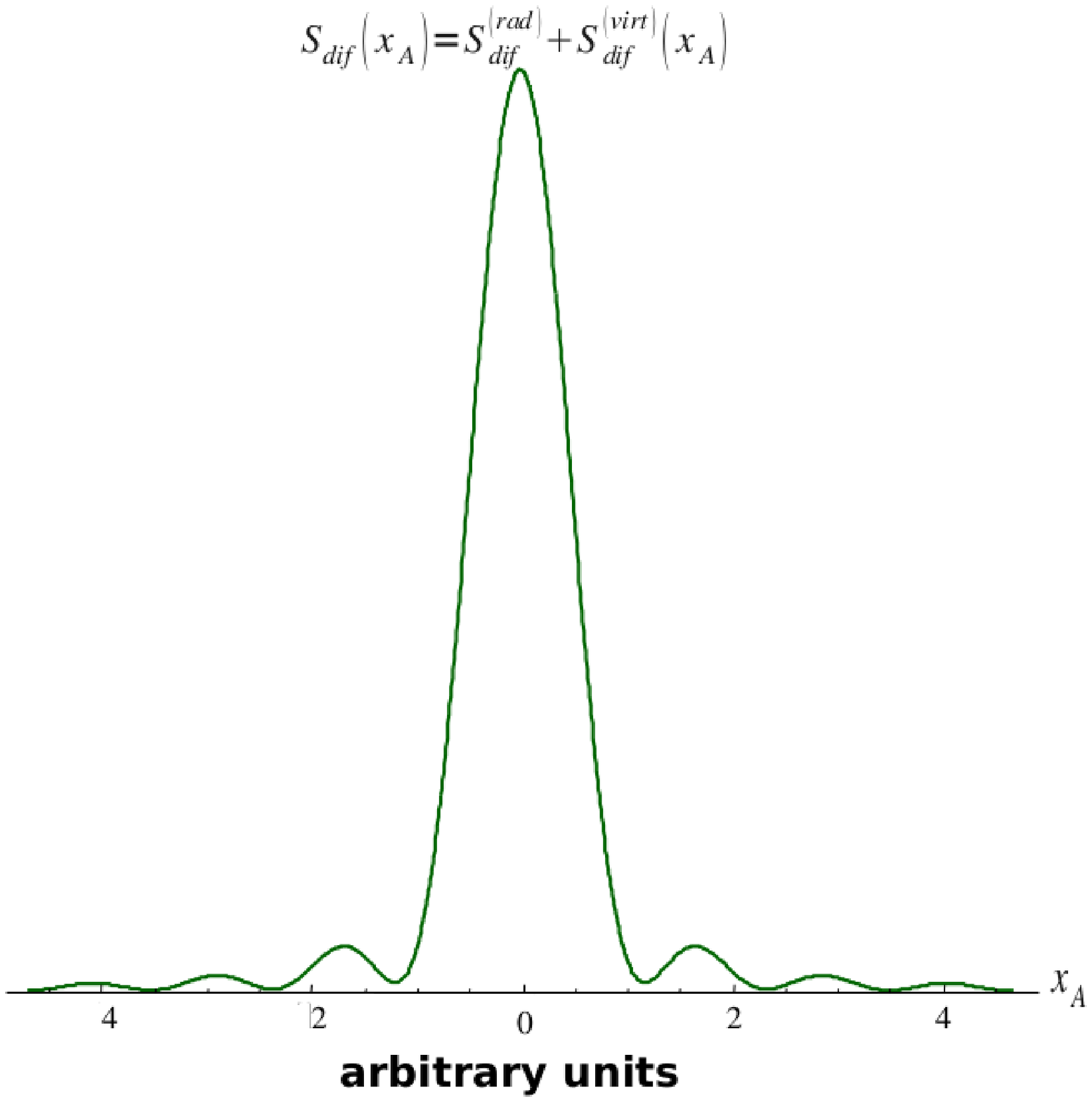} 
   & \includegraphics[width=0.25\textwidth]{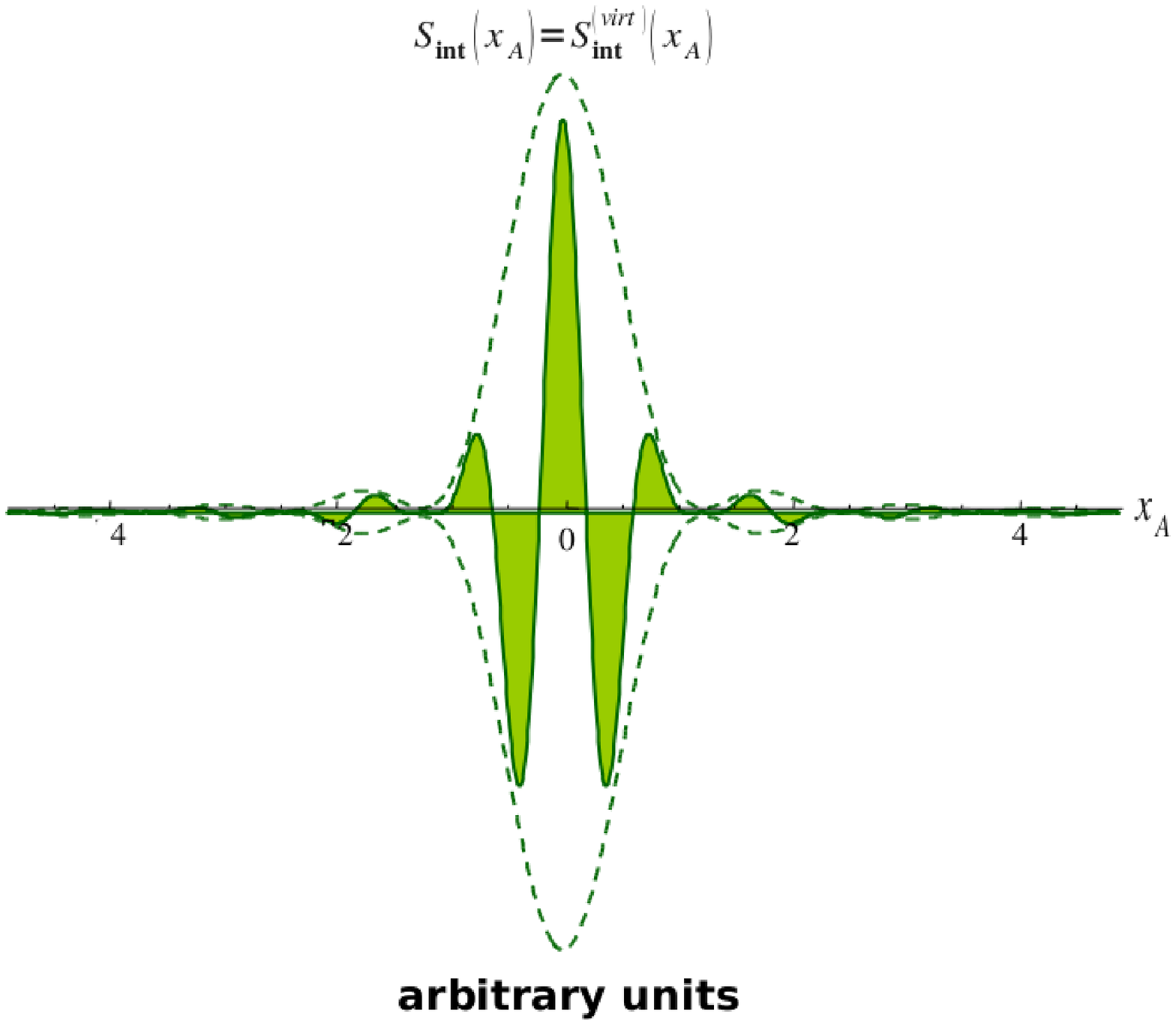} & \includegraphics[width=0.25\textwidth]{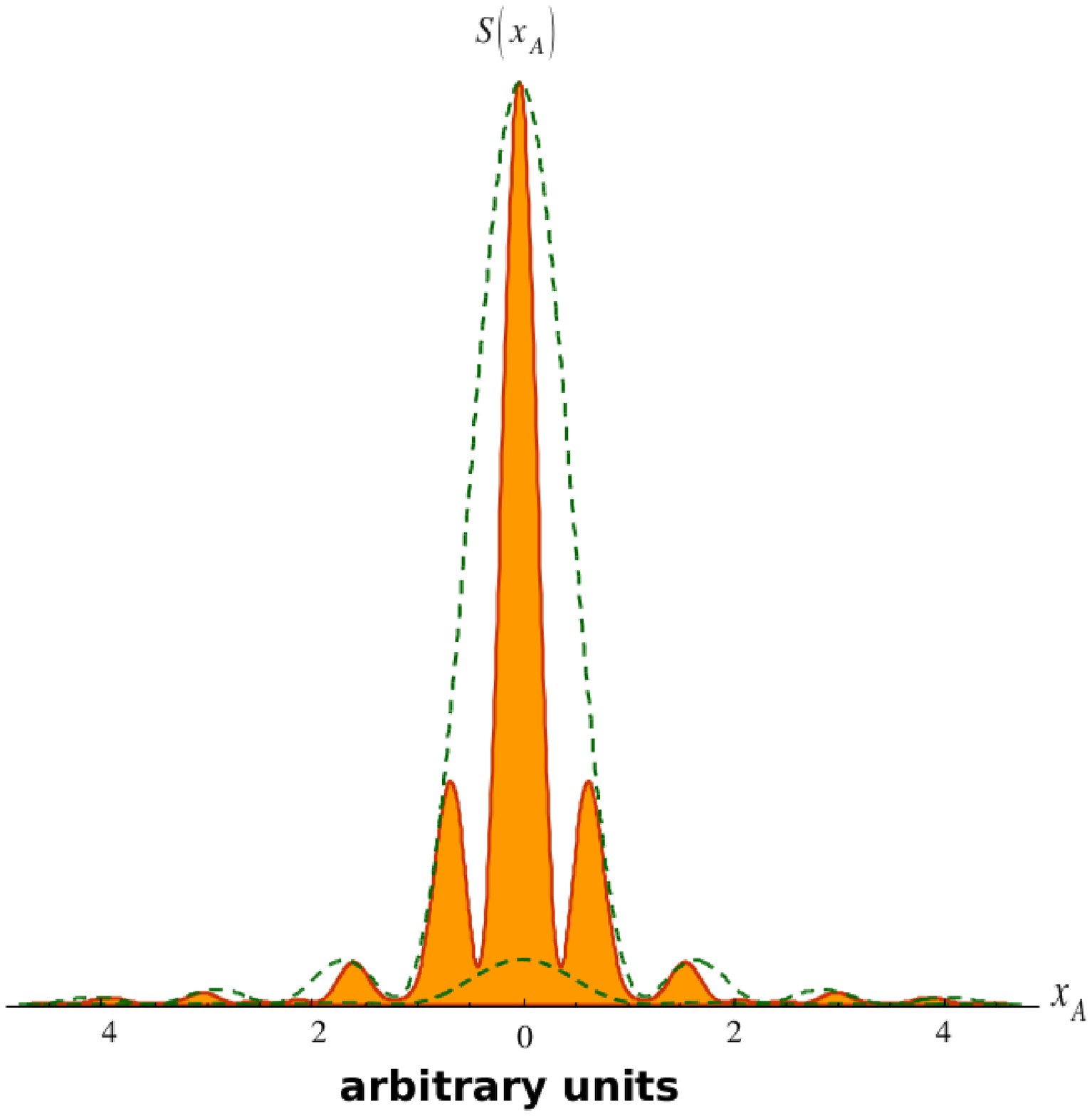}\\
 
   \textbf{(9)}& & & &\\ \hline

   \includegraphics[width=0.23\textwidth]{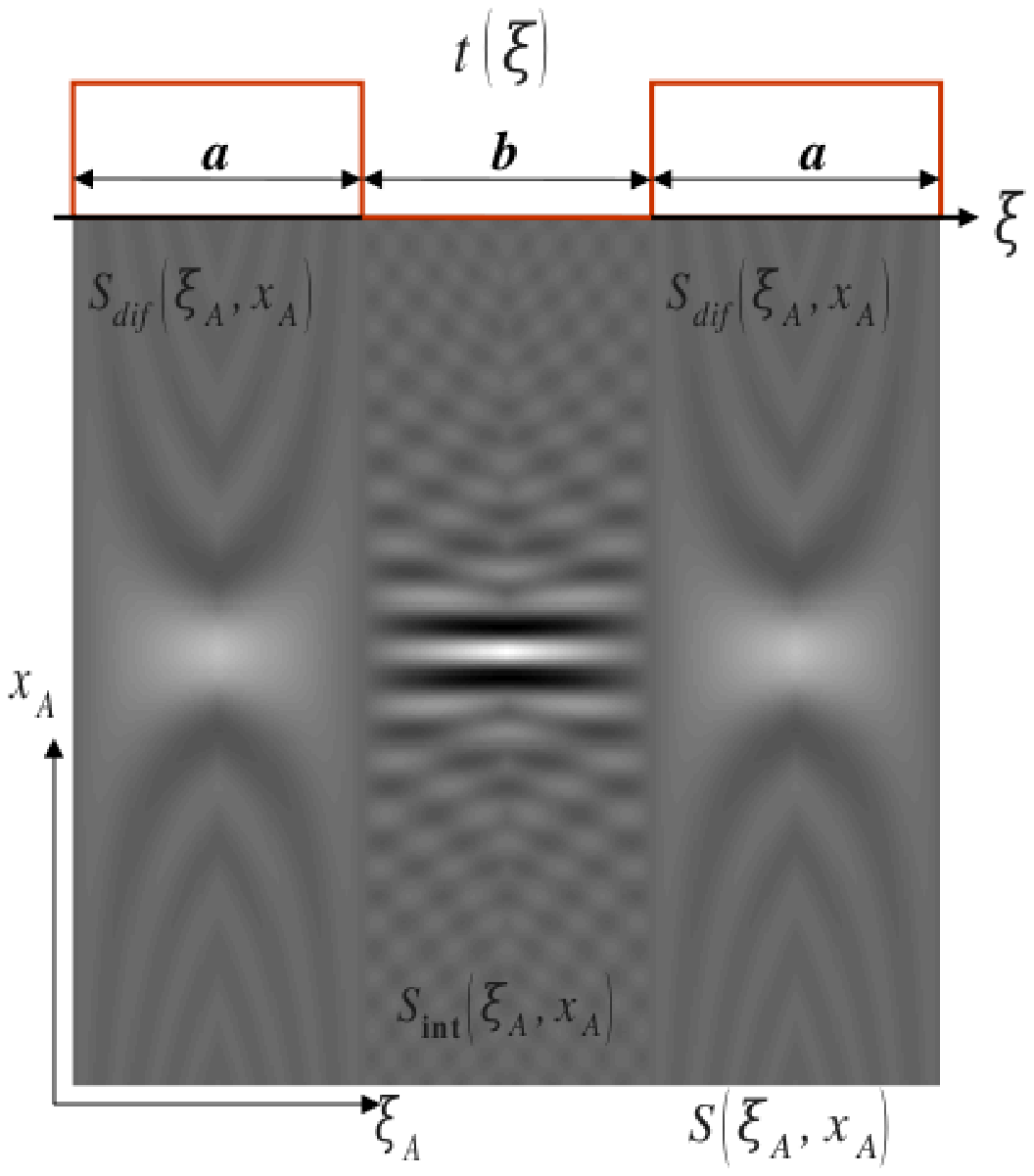}\label{fig14} & \includegraphics[width=0.25\textwidth]{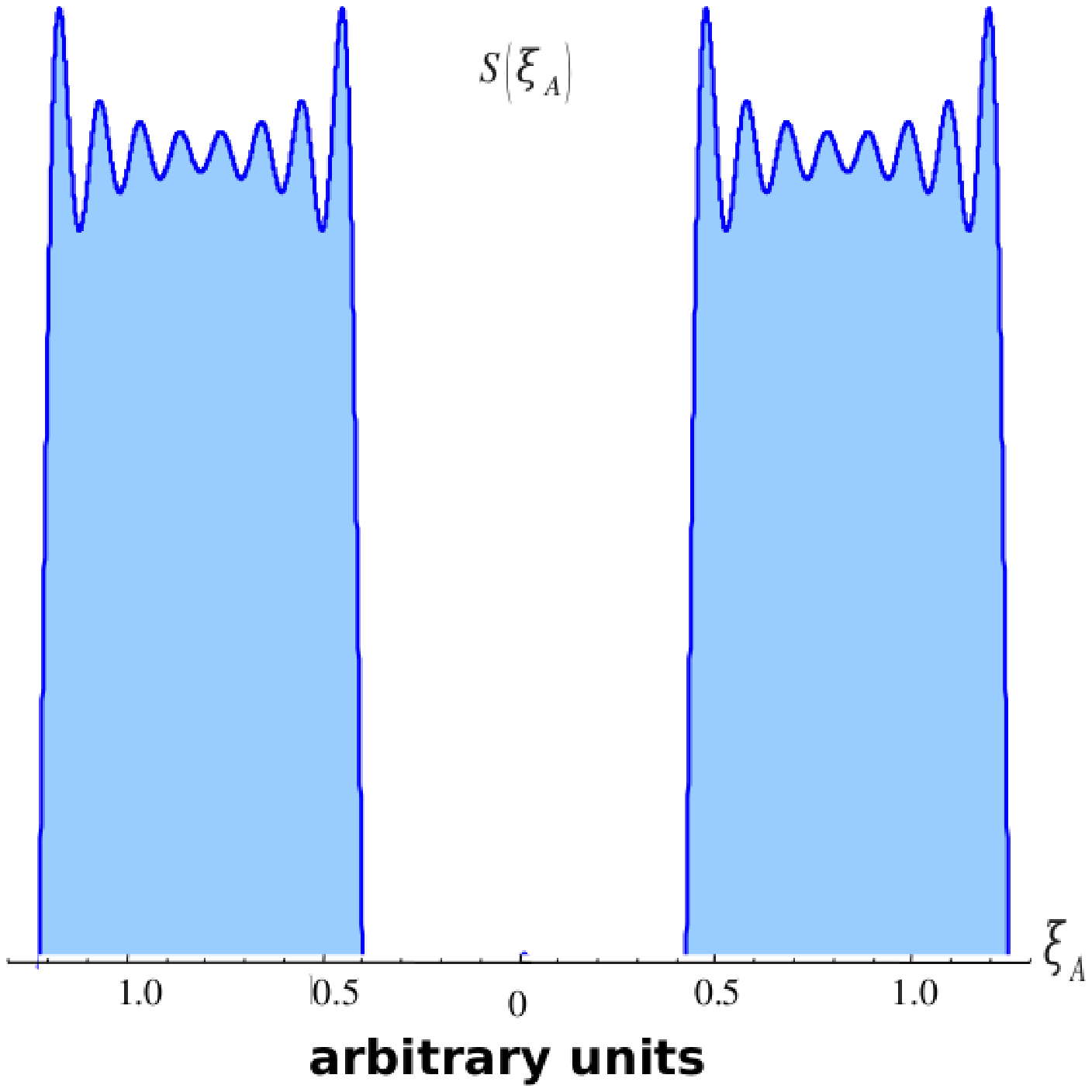} & \includegraphics[width=0.25\textwidth]{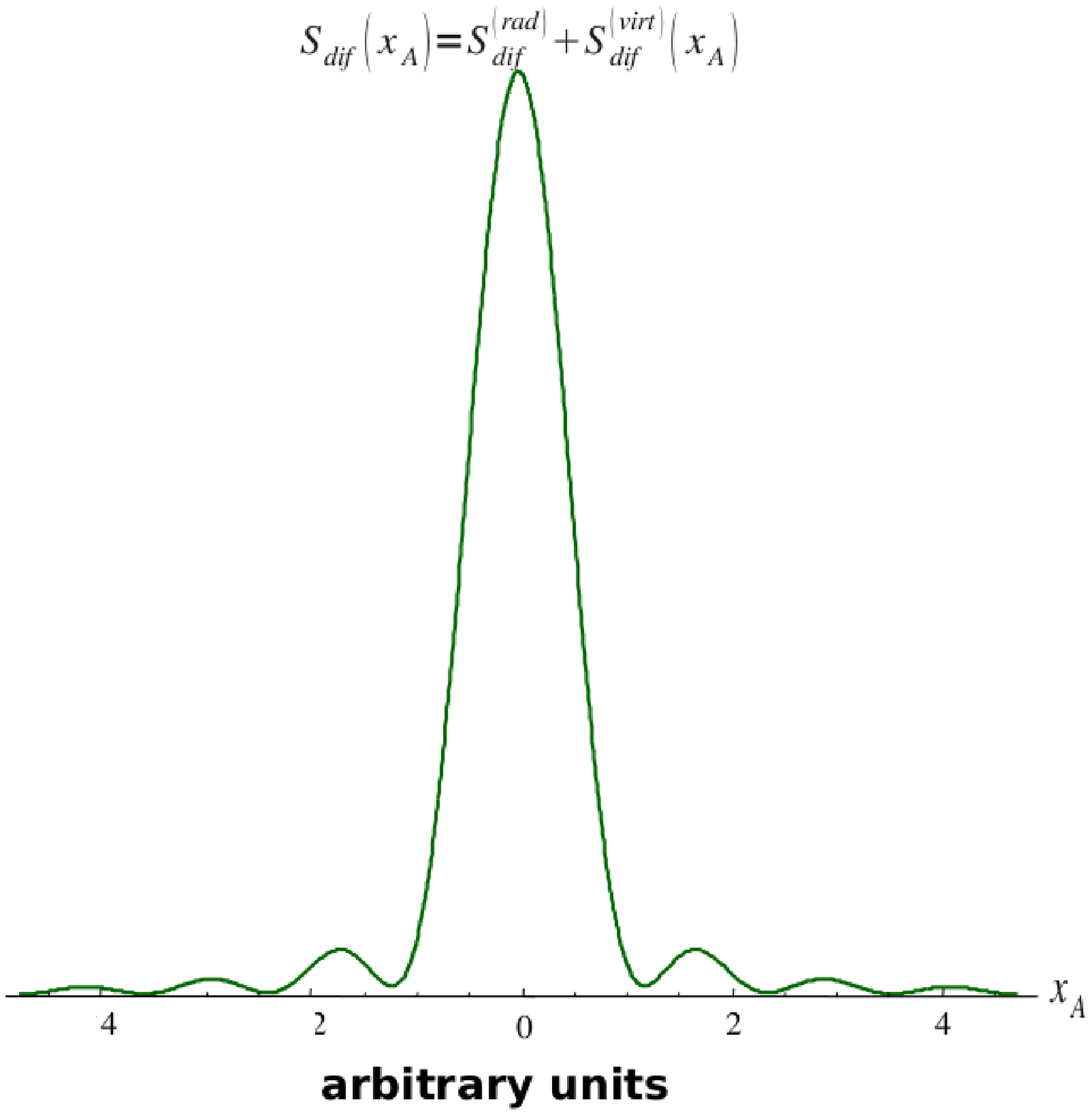} 
   & \includegraphics[width=0.25\textwidth]{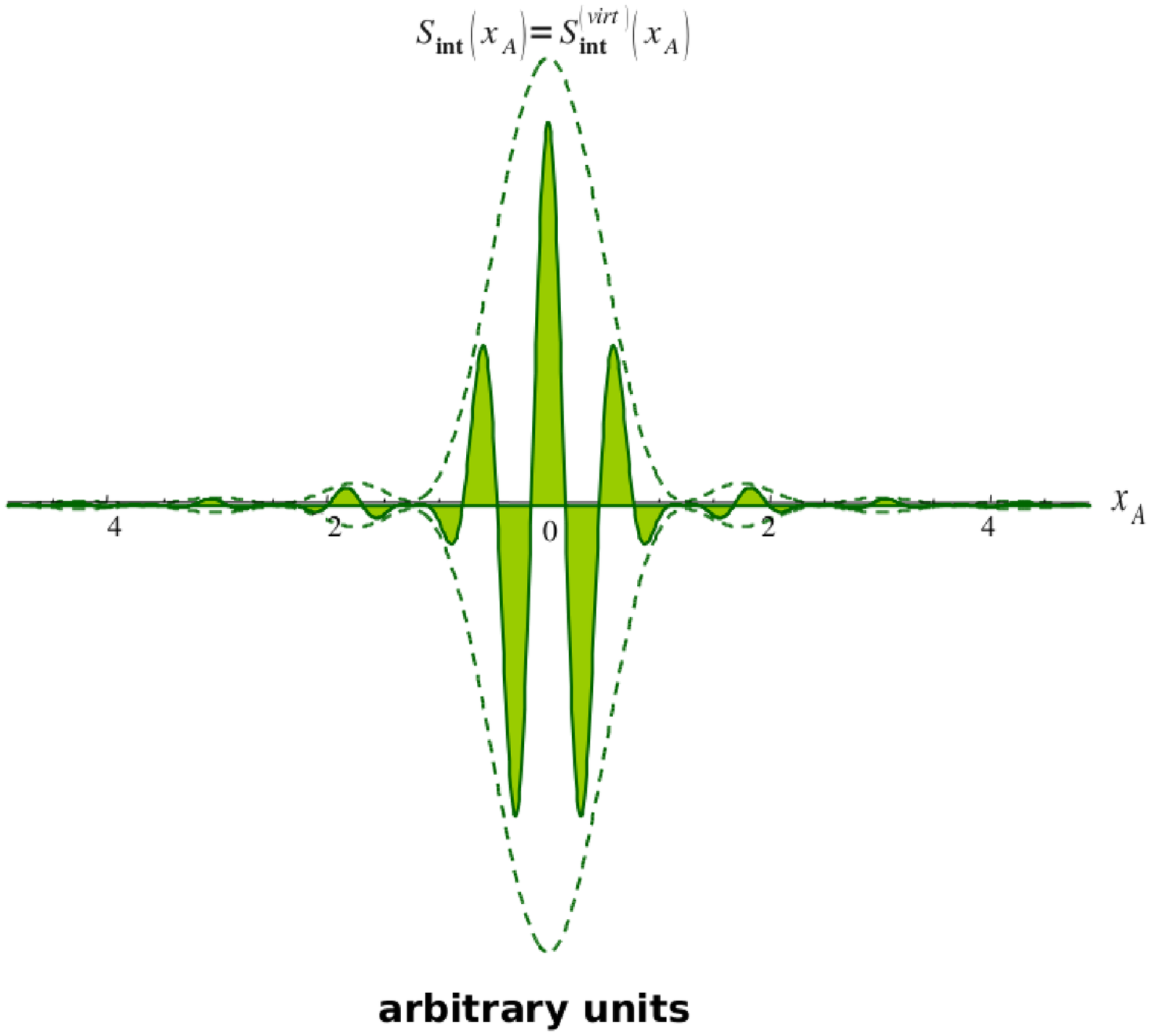} & \includegraphics[width=0.25\textwidth]{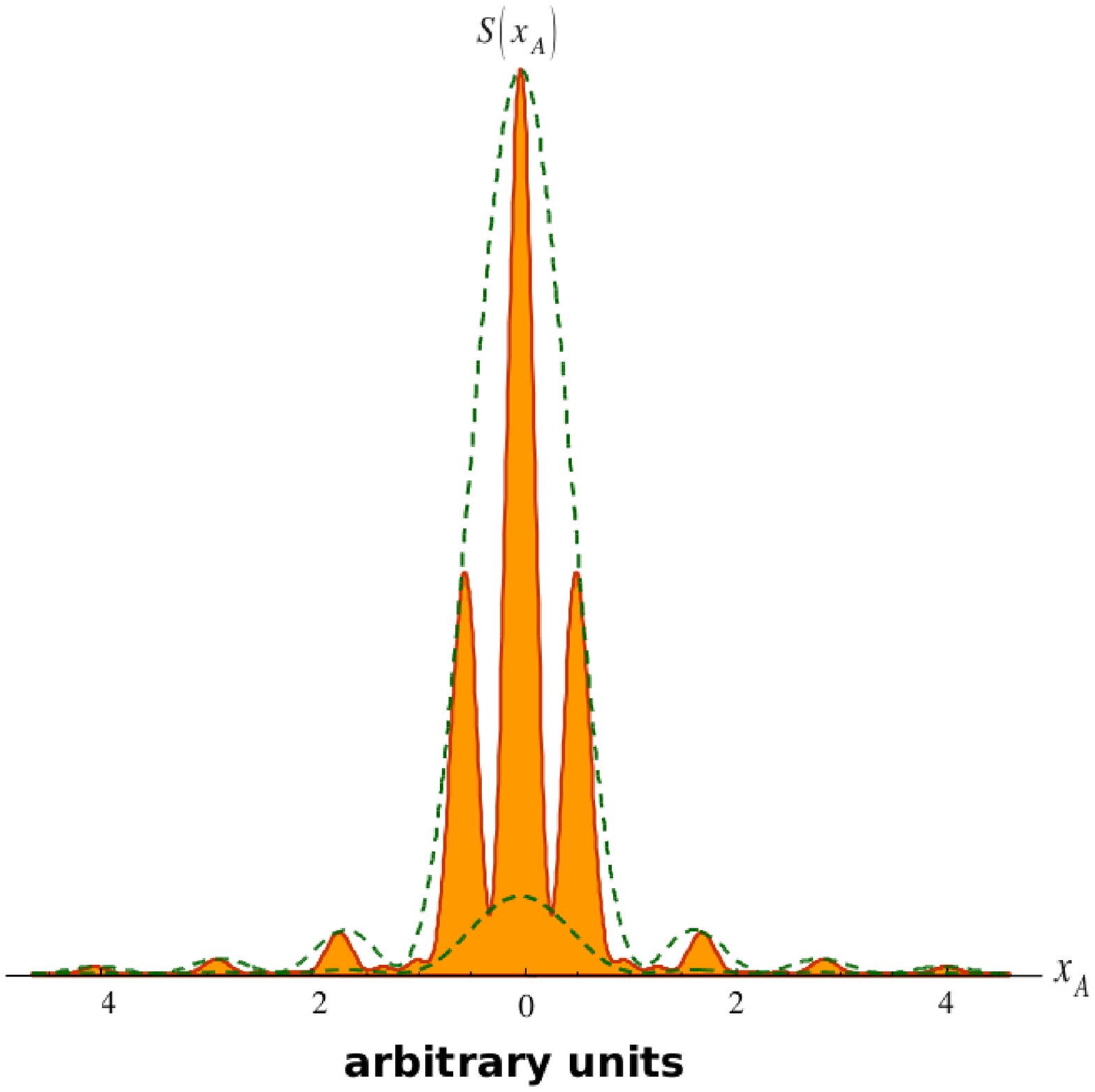}\\

   \textbf{(10)}& & & &\\ \hline

   \includegraphics[width=0.23\textwidth]{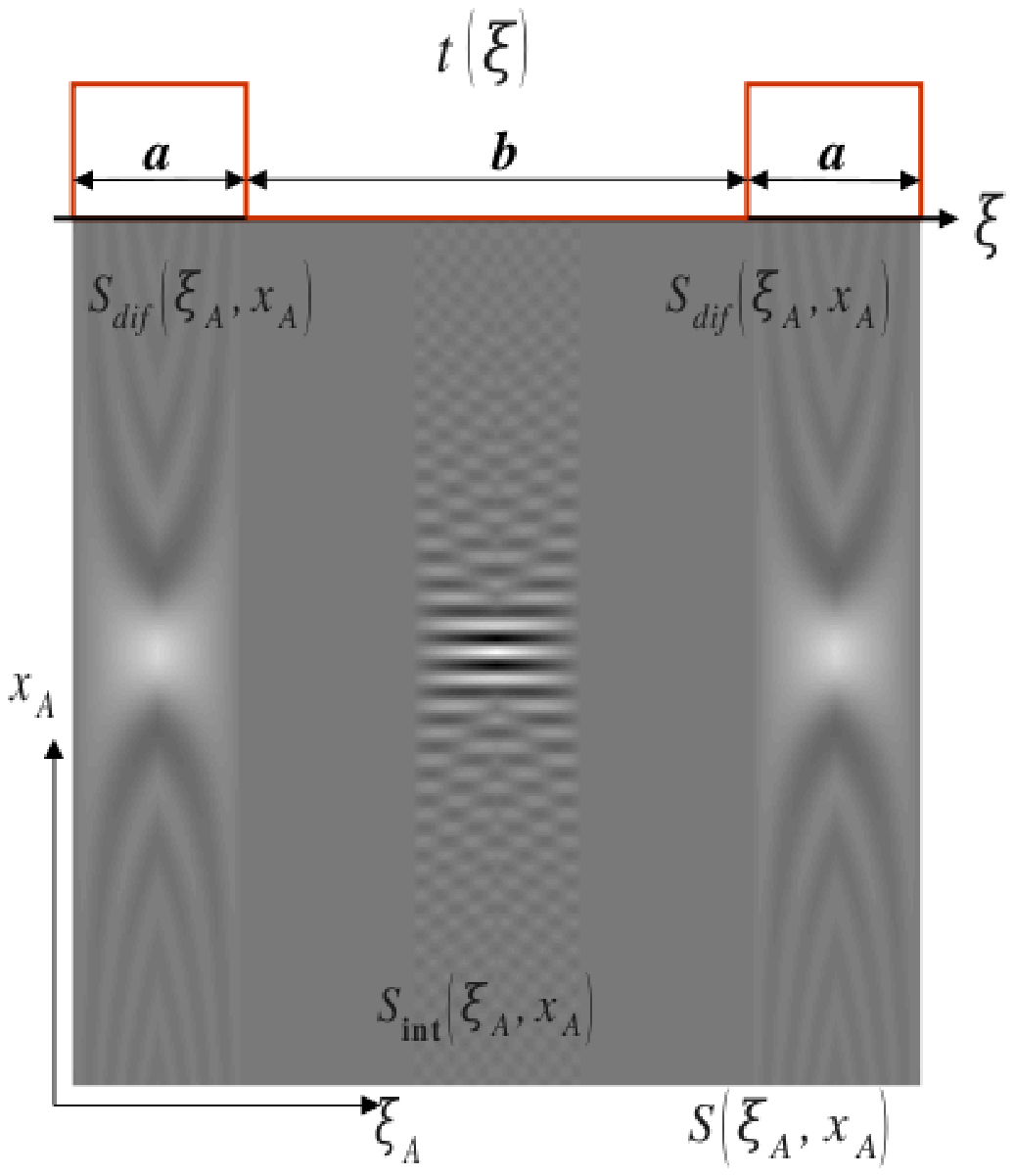}\label{fig15} & \includegraphics[width=0.25\textwidth]{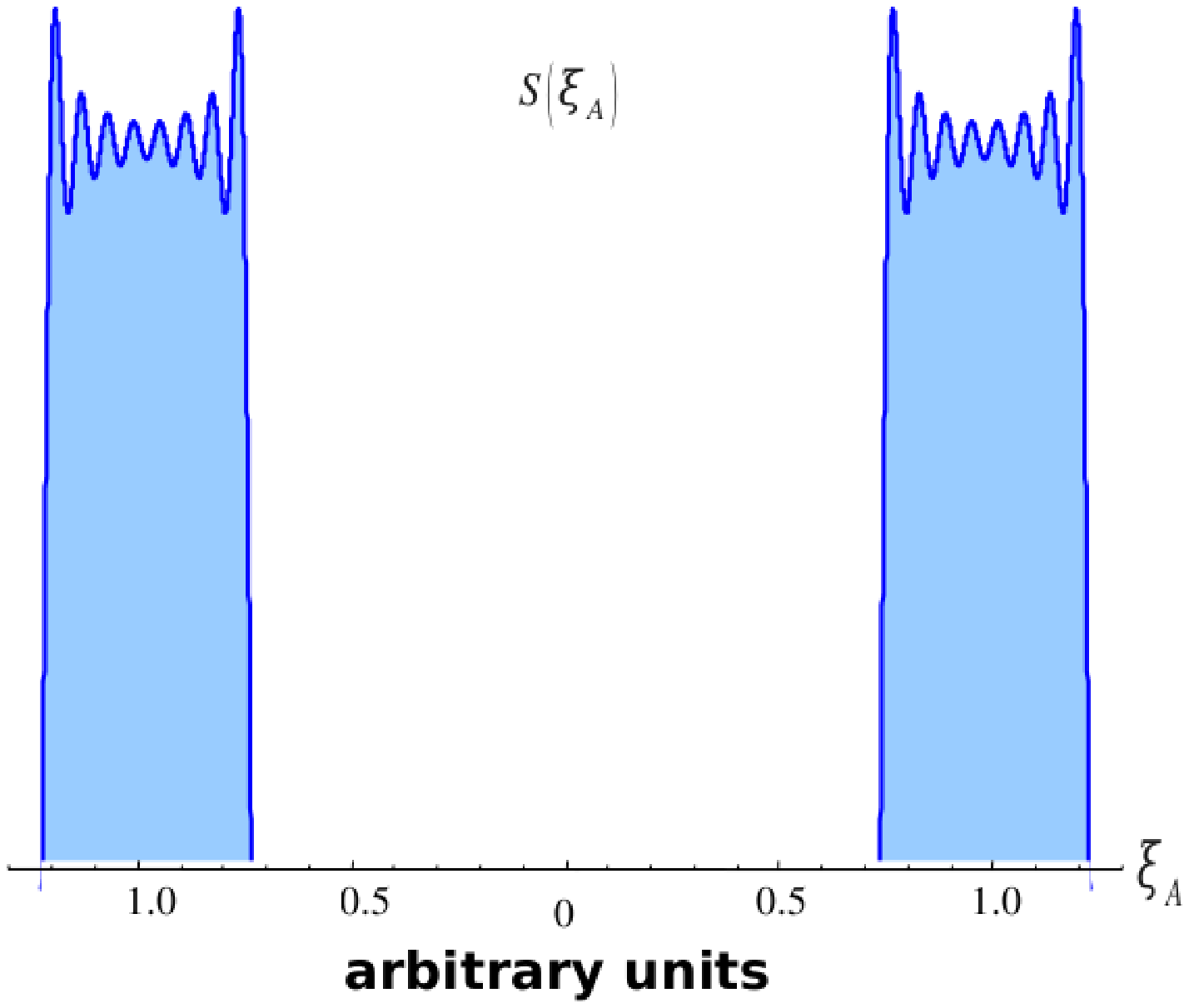} & \includegraphics[width=0.25\textwidth]{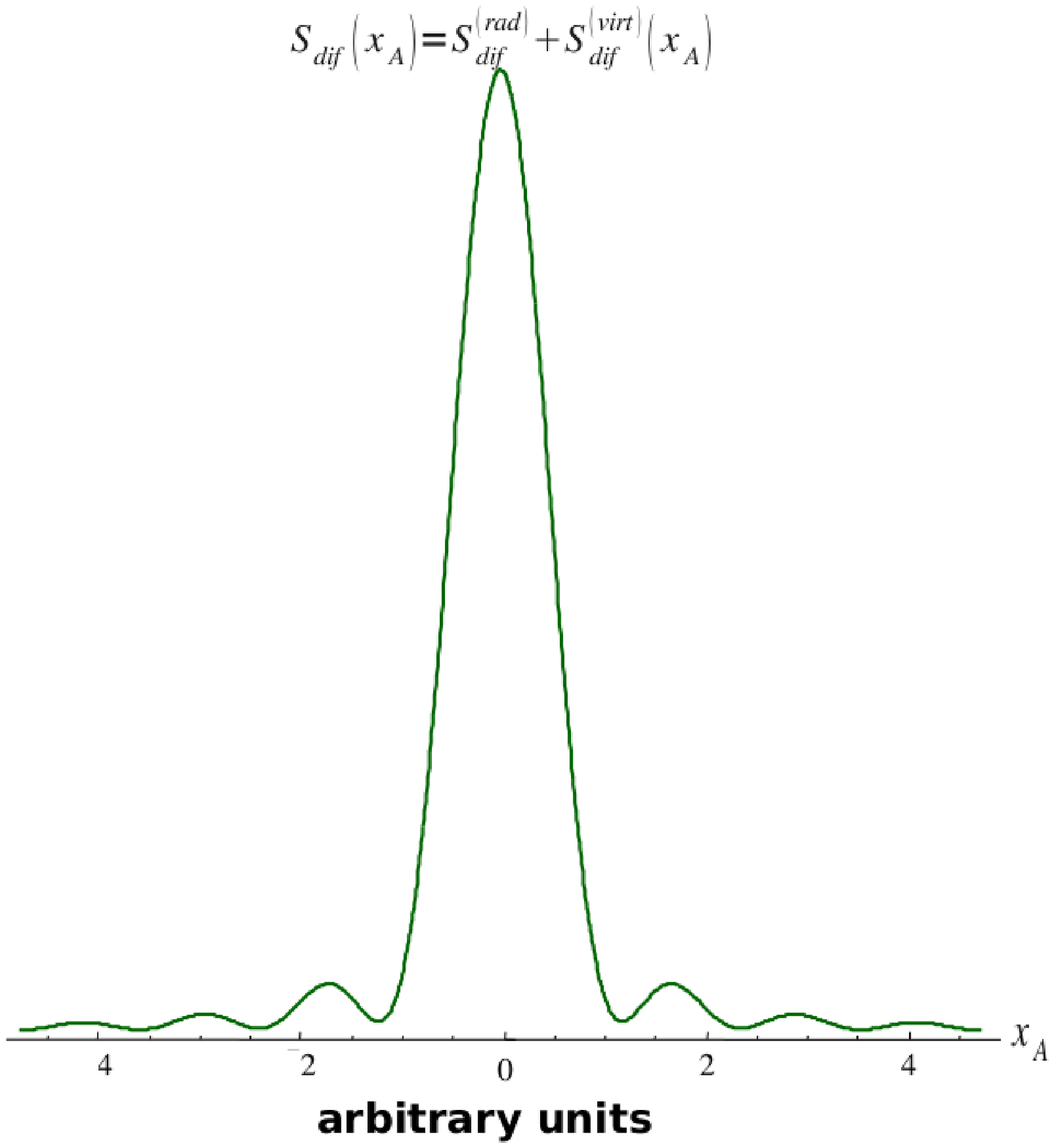} 
   & \includegraphics[width=0.25\textwidth]{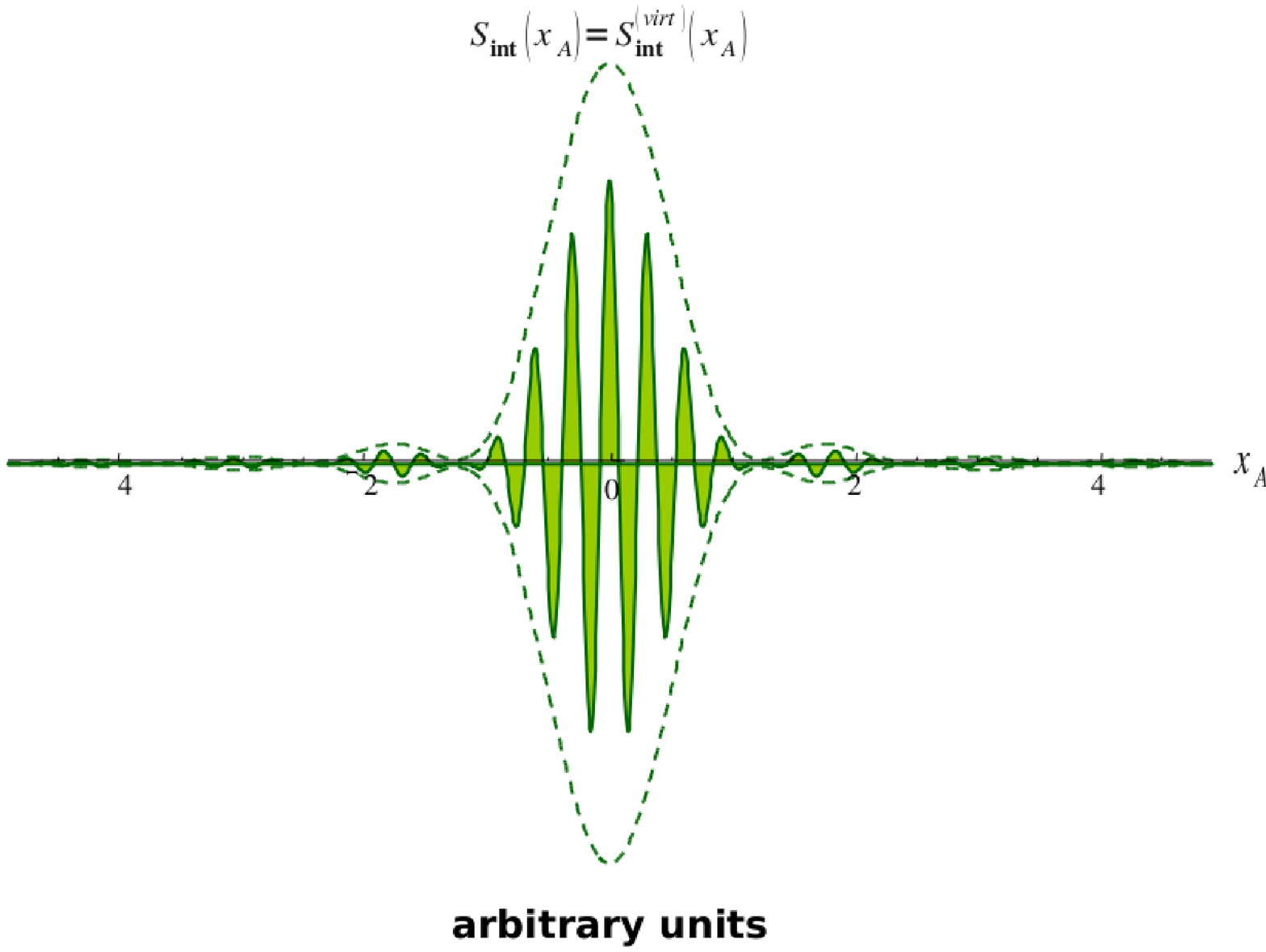} & \includegraphics[width=0.25\textwidth]{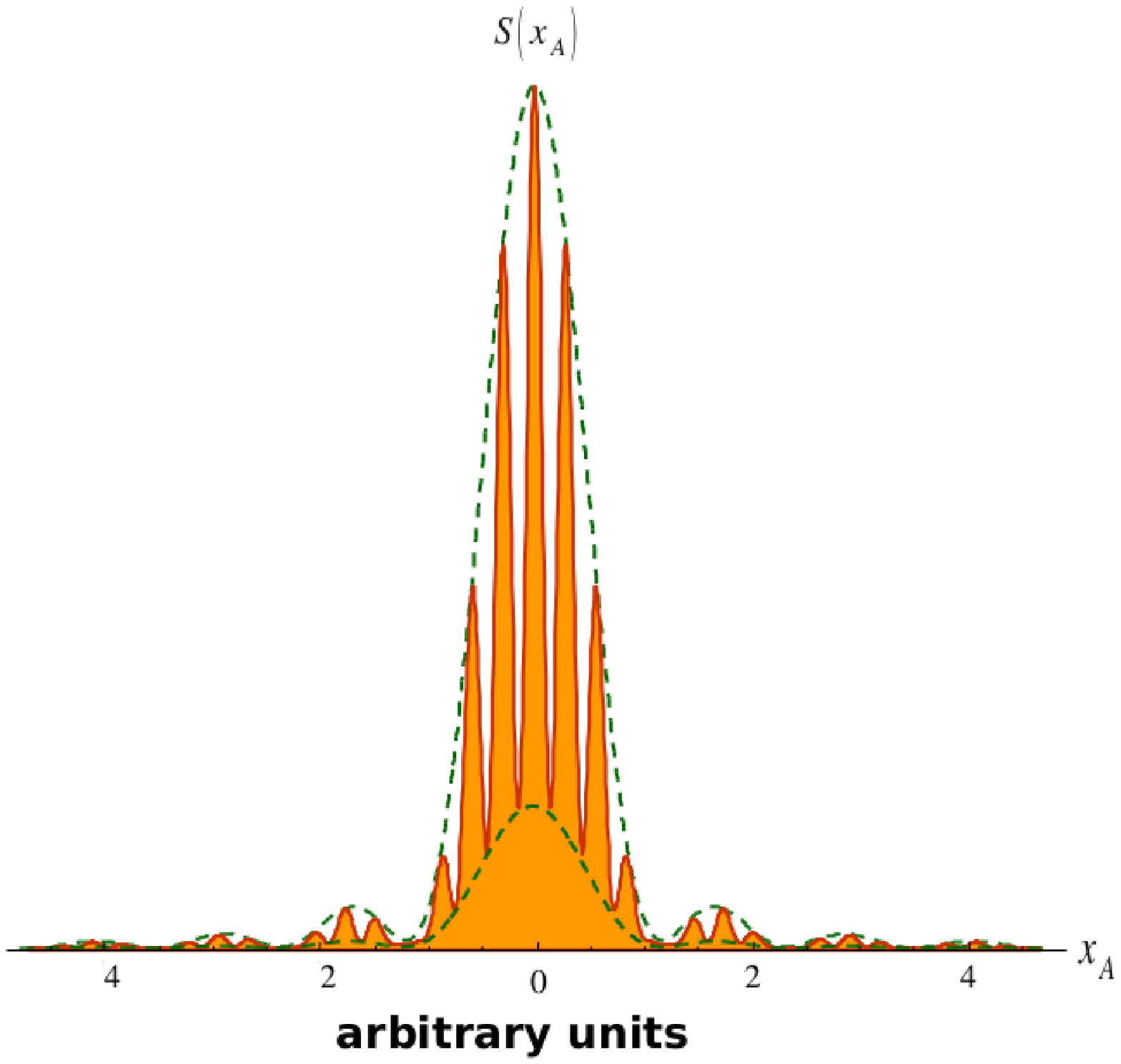}\\

   \textbf{(11)}& & & &\\ \hline
 
  \end{tabular} 
 \end{center}
\end{landscape}

\end{document}